\documentclass{aa}  
\usepackage{lscape}
\usepackage{lmodern}
\usepackage{graphicx}
\usepackage{color} 
\usepackage{longtable}
\usepackage{natbib,twoopt}
\usepackage[colorlinks,linkcolor=blue,urlcolor=blue,citecolor=black]{hyperref}
\makeatletter
\renewcommand*\aa@pageof{, page \thepage{} of \pageref*{LastPage}} 
\makeatother
\bibpunct{(}{)}{;}{a}{}{,} 
\usepackage{txfonts}
\defcitealias{2018A&A...615L...5I}{Paper~I}
\defcitealias{2018A&A...620A..48I}{Paper~II}

\begin{document} 

   \title{Hypervelocity stars in the \textit{Gaia} era}

   \subtitle{Revisiting the most extreme stars from the MMT HVS survey}

   \author{S.~Kreuzer
          \and 
          A.~Irrgang
          \and
          U.~Heber
          }

   \institute{Dr.~Karl~Remeis-Observatory \& ECAP, Astronomical Institute, Friedrich-Alexander University Erlangen-Nuremberg (FAU), Sternwartstr.~7, 96049 Bamberg, Germany\\
              \email{simon.kreuzer@fau.de}\label{remeis}
             }
             
   \date{Received / Accepted}

  \abstract
  {The hypervelocity star (HVS) survey conducted at the Multiple Mirror Telescope (MMT) identified 42 B-type stars in the Galactic halo whose radial velocity in the Galactic rest-frame exceeds $+275$\,km\,s${}^{-1}$. In order to unravel the nature and origin of those high-velocity outliers, their complete six-dimensional phase space information is needed. To this end, we complemented positions and proper motions from the second data release of {\it Gaia} with revised radial velocities and spectrophotometric distances that are based on a reanalysis of the available MMT spectra of 40 objects using state-of-the-art model spectra and a tailored analysis strategy. The resulting position and velocity vectors for 37 stars were then used as input for a subsequent kinematic investigation to obtain as complete a picture as possible. The combination of projected rotational velocity, position in the Kiel diagram, and kinematic properties suggests that all objects in the sample except two (B576, B598) are very likely to be main sequence stars. While the available data are still not precise enough to constrain the place of origin for 19 program stars, we identified eight objects that either come from the outer rim of the Galactic disk or not from the disk at all, along with ten that presumably stem from the Galactic disk. For almost all of those 18 targets with more or less well-constrained spatial origin, the Galactic center (GC) is disqualified as a possible place of origin. The most notable exception is B576, the origin of which coincides extremely well with the GC when assuming a blue horizontal branch (BHB) nature for it. HVS\,22 is by far the most extreme object in the sample. Although its origin is completely unconstrained, an ejection from the GC by the Hills mechanism is the most plausible explanation for its current Galactic rest-frame velocity of $1530^{+690}_{-560}$\,km\,s${}^{-1}$.
  }

          
   \maketitle
%

\section{Introduction}\label{sec:introduction}

The population of faint blue stars in the Galactic halo is dominated by white dwarf (WD), hot subdwarf, and blue horizontal branch (BHB) stars. However, now and then, some stars have turned out to be normal main sequence (MS) stars of spectral type B \citep{1966ApJ...144..496G, 1974ApJS...28..157G,1983MNRAS.205..435K,1987fbs..conf..149T}. These stars are not expected to be found in the halo because the clouds from which they formed are located in the Galactic disk rather than in the halo. Therefore, MS stars in the halo are believed to have been ejected from their place of birth in the Galactic disk and, accordingly, the term runaway stars was coined \citep{1957moas.book.....Z,1961BAN....15..265B}. The stars have to travel fast in order to reach their present-day positions in the halo within their comparatively short lifetimes. Every major survey for faint blue objects in the halo has continued to find MS B-type stars \citep[e.g., ][]{1997ApJ...491..172S,1997fbs..conf...87H,2004MNRAS.349..821L,2004MNRAS.353..633L}. \citet{2011MNRAS.411.2596S} list 96 bona-fide runaway B stars located beyond 1\,kpc from the Galactic disk with ejection velocities up to $\approx400$\,km\,s$^{-1}$. 

The most extreme runaway stars, the so-called hypervelocity stars \citep[HVS, for a review see][]{2015ARA&A..53...15B} were first discovered serendipitously by \citet{2005ApJ...622L..33B}, \citet{2005A&A...444L..61H} in the Sloan Digital Sky Survey \citep{2000AJ....120.1579Y}, and \citet{2005ApJ...634L.181E} in the Hamburg ESO survey \citep{1996A&AS..115..227W}, traveling with speeds that exceed their local Galactic escape velocity. HVSs have also been discovered among hot subdwarf stars \citep[e.g., HVS\,2, which is also known as US\,708,][]{2005A&A...444L..61H,2015Sci...347.1126G} and WDs \citep{2017Sci...357..680V}. 

These discoveries triggered a systematic search for more blue stars with high radial velocities in the Galactic halo. Since HVSs are distant objects and thus rather faint, it is not straightforward to obtain spectra with sufficient signal-to-noise ratio (S/N) to reveal their nature. The first HVS survey by \citet{2006ApJ...640L..35B} was later extended using the 6.5\,m Multiple Mirror Telescope (MMT) on Mount Hopkins, Arizona, yielding a moderate resolution survey  that covered 12\,000 square degrees of the sky, the so-called MMT HVS survey \citep{2009ApJ...690.1639B,2014ApJ...787...89B}. This survey finally led to the discovery of 21 HVSs and 16 runaway stars of late B spectral type \citep{2014ApJ...787...89B}. \citet{2015ApJ...804...49B} added five additional objects and studied the kinematics of 15 of them using proper motions measured with the Hubble Space Telescope (HST). Recent additions of four stars \citep{2014ApJ...785L..23Z,2017ApJ...847L...9H,2018AJ....156...87L} to the list of HVSs of A and B spectral type came from the LAMOST survey \citep{2012RAA....12.1197C}. The latest one \citep[S5-HVS1,][]{2020MNRAS.491.2465K} from the Southern Stellar Stream Spectroscopic Survey \citep[S$^{5}$,][]{2019MNRAS.490.3508L} is also the most extreme one, because its Galactic rest-frame velocity of 1700\,km\,s$^{-1}$ is record-high.

A classical ejection scenario for runaway stars is the binary supernova scenario developed by \citet{1961BAN....15..265B} in which the secondary star of a close binary system is ejected when the core-collapse of the more massive primary unbinds the binary. Alternatively, close encounters between (binary) stars in dense stellar clusters or the gravitational collapse of proto-stellar clusters may lead to the ejection of stars, most likely of the lightest one involved \citep[dynamical ejection scenario,][]{1967BOTT....4...86P}. Both of these classical scenarios are, however, not capable of ejecting stars beyond $\approx$ 500\,km\,s$^{-1}$ at the most \citep[e.g., ][]{2000ApJ...544..437P,2009MNRAS.396..570G,2012ApJ...751..133P,2015MNRAS.448L...6T,2016A&A...590A.107O}, see the discussion in \citet{2018A&A...615L...5I,2019A&A...628L...5I}. Runaway stars from those two channels could reach Galactic escape velocity only if their ejection happens to occur in the direction of Galactic rotation, which would give them an additional boost.

The only mechanism that is thought to be powerful enough to eject HVSs is the slingshot mechanism proposed by \cite{1988Natur.331..687H}. Via tidal interactions, a binary system may be disrupted during a close encounter with a supermassive black hole, leading to the ejection of one component with a velocity as large as 4000\,km\,s$^{-1}$. In this scenario, the place of origin must be the Galactic center (GC) as it is the only site in the Galaxy that hosts a supermassive black hole. The recent discovery of S5-HVS1 \citep{2020MNRAS.491.2465K}, whose trajectory points to an origin in the GC, may be considered as the smoking gun for the Hills mechanism. The origin of HVSs may, however, be extragalactic as well. The Large Magellanic Cloud has been proposed as a potential source of HVSs \citep{2016ApJ...825L...6B,2017MNRAS.469.2151B}. Indeed, the unique object HVS\,3 has been suggested to originate from the Large Magellanic Cloud \citep{2005ApJ...634L.181E}. This idea was supported by chemical tagging \citep{2008A&A...480L..37P} and finally confirmed with {\it Gaia} astrometry \citep{2018A&A...620A..48I,2019MNRAS.483.2007E}. Another extragalactic scenario involves the disruption of dwarf galaxies by the Milky Way \citep{2009ApJ...691L..63A}. However, simulations by \citet{2011A&A...535A..70P} rendered this possibility unlikely because the perturber needs to be unbound itself and the HVSs would travel along with the perturber.

ESA's {\it Gaia} space mission revolutionized astronomy by providing proper motions, parallaxes, and photometry of unprecedented precision for 1.3 billion objects. Its second data release \citep[{\it Gaia} DR2,][]{2018A&A...616A...1G} immediately triggered publications reporting spectacular discoveries, such as hypervelocity WDs as surviving companions of dynamically driven double-degenerate double-detonation Type Ia Supernovae \citep[D6 stars,][]{2018ApJ...865...15S}, and partly burnt runaway stellar remnants from peculiar thermonuclear supernovae leaving the Galaxy \citep{2019MNRAS.489.1489R}. The previously known B-type HVSs were readily studied from the new {\it Gaia} data \citep{2018A&A...620A..48I,2018ApJ...866...39B,2019MNRAS.483.2007E}. The nature of candidate HVSs of low mass was also clarified by {\it Gaia} data leading to the elimination of all but one of them \citep{2018MNRAS.479.2789B}. On the other hand, {\it Gaia} DR2 paved the way to search for new HVSs because it also provides radial velocities of cool stars, albeit limited to relatively bright ones. New cool nearby (10--15\,kpc) high speed stars have been reported \citep[e.g., ][]{2018ApJ...868...25B,2019MNRAS.490..157M,2019ApJS..244....4D}. However, \cite{2019MNRAS.486.2618B} pointed out that the radial velocities of a couple of those candidates may be flawed by blending with nearby stars as confirmed by independent radial-velocity measurements in one case. Additional ground-based spectroscopy is required to clarify their nature.

While proper motion measurements from {\it Gaia} are far superior to any ground-based ones, high-precision parallaxes are still limited to relatively nearby stars. To ease this problem, Bayesian statistical methods have been developed \citep{2015PASP..127..994B,2016ApJ...832..137A,2018AJ....156...58B}, which might be useful for distances of about 10\,kpc under the premise that an appropriate prior is available. The latter, however, is very difficult to derive in the case of ejected stars.

Because B-type HVSs are so far away, mostly beyond 30\,kpc, {\it Gaia} parallaxes are too uncertain to draw firm conclusions and will remain so even at the end of the {\it Gaia} mission. Therefore, spectrophotometric distances are and will be crucial to understand the kinematics of HVSs. This requires high-quality spectra, sophisticated model atmospheres and synthetic spectra, as well as an objective analysis strategy. Over the past years, those tools have been developed \citep{2011JPhCS.328a2015P,2014A&A...565A..63I}. Recently, we applied them to 14 out of the 42 highest velocity stars of the MMT HVS sample \citep{2018ApJ...866...39B} to derive their spectrophotometric distances \citep[][henceforth \citetalias{2018A&A...615L...5I}]{2018A&A...615L...5I}. Combining these distances with {\it Gaia} DR2 proper motions, we studied the kinematic properties and sites of origin of those 14 stars \citep[][henceforth \citetalias{2018A&A...620A..48I}]{2018A&A...620A..48I}. This particular subsample was chosen because HST proper motions are available that allowed a cross-check to be made, which showed that proper motions from both sources are consistent.

In this work, which we regard as Paper~III in this series, we extend the spectrophotometric analysis to 40 out of the 42 high-velocity outliers of the MMT HVS sample. We did not have access to the spectra of the two missing object. Complete astrometric data is available for 37 of them, allowing us to carry out a subsequent kinematic analysis as well. In Sect.~\ref{sec:models}, we describe our model atmospheres and synthetic spectra. Section~\ref{sec:spectral-analysis} illustrates how these models are used to fit the observed MMT spectra and derive atmospheric parameters as well as radial and rotational velocities. In Sect.~\ref{sec:spec-distances}, spectral energy distributions (SEDs) are constructed and spectrophotometric distances and stellar parameters are derived. Those are then used to perform the kinematic analysis presented in Sect.~\ref{sec:kinematics}. In Sect.~\ref{sec:results}, we discuss our results. Finally, we present our conclusions and our outlook in Sect.~\ref{sec:conclusion}.

\section{Model atmospheres and synthetic spectra}\label{sec:models}

We calculated a grid of synthetic spectra with solar chemical composition following the so-called ADS approach \citep{2011JPhCS.328a2015P}, which involves a sequence of the three codes {\sc Atlas12} \citep{1996ASPC..108..160K}, {\sc Detail} (\citealt{1981PhDT.......113G}; \citealt{detailsurface2}), and {\sc Surface} (\citealt{1981PhDT.......113G}; \citealt{detailsurface2}). Assuming local thermodynamic equilibrium (LTE), {\sc Atlas12} computes the initial atmospheric structure which is then used by {\sc Detail} to compute population numbers for specific chemical species in non-LTE. As we did previously in \citetalias{2018A&A...615L...5I}, we iteratively feed back the resulting population numbers for hydrogen and helium to {\sc Atlas12} to incorporate deviations from LTE also in the computation of the atmospheric structure. Once this iterative process has converged, a final synthetic spectrum with more sophisticated line-broadening data is calculated with {\sc Surface}. The resulting grid of synthetic spectra spans a range in effective temperature $T_\mathrm{eff}$ between 9000\,K and 16\,000\,K (in steps of 250\,K) and surface gravities $\log(g)$ between 3.0 and 4.8 (in steps of 0.2). 
Our recent improvements of all three codes, namely the implementation of the occupation probability formalism \citep{1988ApJ...331..794H} for hydrogen and ionized helium --~following the description given by \citet{1994A&A...282..151H}~-- as well as state-of-the-art line broadening tables for hydrogen \citep{2009ApJ...696.1755T}, are included as well. These changes result in a much more realistic representation for the region around the Balmer jump and are thus very important for the analysis of the available spectra, which cover exactly this region.

The list of spectral lines included in the ADS grid is tailored to B-type stars. Going to cooler temperatures, that is, to A-type stars, a rapidly increasing number of metal lines shows up in the spectrum that is not implemented in the grid. To cross-check whether this affects our analysis, we compared our results with those based on models computed with a combination of {\sc Atlas12} and the LTE spectrum synthesis code {\sc Synthe} \citep{1993sssp.book.....K}, which contains many more metal lines. The corresponding grid covers temperatures between 7200\,K and 11\,000\,K (in steps of 200\,K) and gravities between 3.0 and 4.6 (in steps of 0.2). The relatively large overlap between the two grids was chosen on purpose to enable us to analyze many stars with both sets of models in order to check whether our results are model dependent, which turned out not to be the case. In the following, we will therefore only refer to the results obtained with the ADS grid.

\begin{table*}
\caption{Results of the spectroscopic analysis. }\label{tab:atmos-results}
\begin{minipage}{.5\textwidth}
\setlength{\tabcolsep}{0.175cm}
\renewcommand{\arraystretch}{1.05}
\begin{center}

         \scriptsize 
\begin{tabular}{rrrrrrrr}
\hline\hline
     Object &    $T_\mathrm{eff}$ &    $\log(g)$ &     $\varv_\mathrm{rad}$ &       $\varv \sin(i)$ & $\chi^2_\mathrm{red}$ & S/N$_\mathrm{tot}$ & N  \\    
     \cline{4-5}&
      (K) & (cgs) & \multicolumn{2}{c}{(km\,s$^{-1}$)} & &  \\
      \hline
 HVS1 & $10\,290$ & $3.49$ & $832.5$ & $236$ &  $1.31$ &  $24.8$ &           5  \\ Stat. &             $^{+30}_{-40}$ & $^{+0.02}_{-0.02}$ &   $^{+2.7}_{-2.6}$ &                          $^{+9}_{-9}$  \\ Sys. & $^{+110}_{-110}$ & $^{+0.04}_{-0.04}$ & $^{+1.6}_{-1.6}$ &              $^{+4}_{-1}$ \\
 HVS4 & $13\,890$ & $3.97$ & $605.0$ & $122$ &  $1.29$ &  $24.0$ &           2  \\ Stat. &             $^{+60}_{-50}$ & $^{+0.03}_{-0.02}$ &   $^{+3.0}_{-2.9}$ &                        $^{+11}_{-12}$  \\ Sys. & $^{+140}_{-140}$ & $^{+0.04}_{-0.04}$ & $^{+0.6}_{-1.5}$ &              $^{+2}_{-1}$ \\
 HVS5 & $12\,550$ & $4.09$ & $541.5$ & $123$ &  $1.27$ &  $12.2$ &           1  \\ Stat. &             $^{+90}_{-80}$ & $^{+0.04}_{-0.04}$ &   $^{+5.9}_{-5.9}$ &                        $^{+19}_{-19}$  \\ Sys. & $^{+130}_{-130}$ & $^{+0.04}_{-0.04}$ & $^{+2.3}_{-1.5}$ &              $^{+1}_{-1}$ \\
 HVS6 & $12\,390$ & $4.30$ & $606.3$ &  $79$ &  $1.13$ &  $12.5$ &           1  \\ Stat. &             $^{+80}_{-80}$ & $^{+0.04}_{-0.04}$ &   $^{+5.3}_{-5.5}$ &                        $^{+23}_{-23}$  \\ Sys. & $^{+130}_{-130}$ & $^{+0.04}_{-0.04}$ & $^{+3.7}_{-2.8}$ &              $^{+9}_{-9}$ \\
 HVS7 & $12\,950$ & $3.96$ & $521.8$ &  $52$ &  $1.60$ &  $29.6$ &           1  \\ Stat. &             $^{+30}_{-30}$ & $^{+0.02}_{-0.02}$ &   $^{+2.2}_{-2.2}$ &                        $^{+12}_{-13}$  \\ Sys. & $^{+130}_{-130}$ & $^{+0.04}_{-0.04}$ & $^{+1.3}_{-1.1}$ &  $^{+10}_{-\phantom{0}6}$ \\
 HVS8 & $10\,880$ & $4.06$ & $498.9$ & $276$ &  $1.42$ &  $25.8$ &           1  \\ Stat. &             $^{+50}_{-40}$ & $^{+0.02}_{-0.02}$ &   $^{+3.3}_{-3.1}$ &                        $^{+12}_{-10}$  \\ Sys. & $^{+110}_{-110}$ & $^{+0.04}_{-0.04}$ & $^{+1.4}_{-1.3}$ &              $^{+7}_{-2}$ \\
 HVS9 & $10\,760$ & $3.44$ & $621.3$ & $353$ &  $1.43$ &  $20.2$ &           2  \\ Stat. &             $^{+40}_{-40}$ & $^{+0.02}_{-0.02}$ &   $^{+4.3}_{-3.9}$ &                        $^{+10}_{-10}$  \\ Sys. & $^{+110}_{-110}$ & $^{+0.04}_{-0.04}$ & $^{+0.5}_{-0.5}$ &              $^{+0}_{-9}$ \\
HVS10 & $10\,640$ & $3.94$ & $468.0$ &  $87$ &  $1.18$ &  $12.0$ &           1  \\ Stat. &             $^{+60}_{-60}$ & $^{+0.03}_{-0.03}$ &   $^{+4.4}_{-5.0}$ &                        $^{+18}_{-19}$  \\ Sys. & $^{+110}_{-110}$ & $^{+0.04}_{-0.04}$ & $^{+0.3}_{-0.3}$ &              $^{+1}_{-6}$ \\
HVS11 &  $9530$ & $4.12$ & $483.4$ & $187$ &  $1.26$ &  $15.9$ &           5  \\ Stat. &             $^{+50}_{-50}$ & $^{+0.03}_{-0.02}$ &   $^{+4.3}_{-5.1}$ &              $^{+17}_{-\phantom{0}9}$  \\ Sys. & $^{+100}_{-100}$ & $^{+0.04}_{-0.04}$ & $^{+0.5}_{-0.5}$ &              $^{+2}_{-1}$ \\
HVS12 & $11\,020$ & $4.08$ & $545.1$ &  $11$ &  $1.19$ &  $14.0$ &           3  \\ Stat. &             $^{+50}_{-50}$ & $^{+0.03}_{-0.03}$ &   $^{+3.8}_{-3.4}$ &           $^{+32}_{-11}$  \\ Sys. & $^{+120}_{-120}$ & $^{+0.04}_{-0.04}$ & $^{+0.3}_{-0.2}$ &            $^{+13}_{-10}$ \\
HVS13 & $10\,810$ & $3.92$ & $570.8$ & $187$ &  $1.24$ &  $14.7$ &           3  \\ Stat. &             $^{+60}_{-60}$ & $^{+0.03}_{-0.03}$ &   $^{+5.2}_{-4.9}$ &                        $^{+16}_{-35}$  \\ Sys. & $^{+110}_{-110}$ & $^{+0.04}_{-0.04}$ & $^{+1.5}_{-0.5}$ &              $^{+1}_{-1}$ \\
HVS14 & $11\,420$ & $4.01$ & $540.0$ & $148$ &  $1.28$ &  $16.1$ &           2  \\ Stat. &             $^{+60}_{-60}$ & $^{+0.03}_{-0.02}$ &   $^{+4.5}_{-4.5}$ &                        $^{+19}_{-19}$  \\ Sys. & $^{+120}_{-120}$ & $^{+0.04}_{-0.04}$ & $^{+2.2}_{-4.0}$ &              $^{+6}_{-2}$ \\
HVS15 & $11\,000$ & $4.07$ & $464.0$ & $127$ &  $1.33$ &  $11.4$ &           2  \\ Stat. &             $^{+50}_{-90}$ & $^{+0.04}_{-0.04}$ &   $^{+6.2}_{-6.2}$ &                        $^{+27}_{-26}$  \\ Sys. & $^{+110}_{-110}$ & $^{+0.04}_{-0.04}$ & $^{+2.6}_{-2.3}$ &              $^{+2}_{-1}$ \\
HVS16 & $10\,550$ & $4.06$ & $423.7$ & $165$ &  $1.21$ &  $13.3$ &           2  \\ Stat. &             $^{+60}_{-60}$ & $^{+0.04}_{-0.03}$ &   $^{+5.3}_{-5.4}$ &                        $^{+20}_{-19}$  \\ Sys. & $^{+110}_{-110}$ & $^{+0.04}_{-0.04}$ & $^{+0.4}_{-0.3}$ &              $^{+0}_{-0}$ \\
HVS17 & $12\,620$ & $4.09$ & $255.5$ & $129$ &  $1.30$ &  $23.5$ &           2  \\ Stat. &             $^{+40}_{-40}$ & $^{+0.02}_{-0.02}$ &   $^{+3.0}_{-3.0}$ &                        $^{+13}_{-15}$  \\ Sys. & $^{+130}_{-130}$ & $^{+0.04}_{-0.04}$ & $^{+1.5}_{-2.2}$ &              $^{+1}_{-5}$ \\
HVS18 & $11\,600$ & $3.88$ & $239.4$ & $132$ &  $1.31$ &  $21.6$ &           4  \\ Stat. &             $^{+40}_{-50}$ & $^{+0.02}_{-0.02}$ &   $^{+3.4}_{-3.2}$ &                        $^{+18}_{-16}$  \\ Sys. & $^{+120}_{-120}$ & $^{+0.04}_{-0.04}$ & $^{+2.0}_{-1.5}$ &              $^{+5}_{-9}$ \\
HVS19 & $11\,480$ & $4.17$ & $593.4$ & $224$ &  $1.23$ &   $8.3$ &           3  \\ Stat. & $^{+\phantom{0}90}_{-120}$ & $^{+0.04}_{-0.06}$ &   $^{+9.1}_{-9.0}$ &                        $^{+27}_{-43}$  \\ Sys. & $^{+120}_{-120}$ & $^{+0.04}_{-0.04}$ & $^{+2.2}_{-0.7}$ &              $^{+1}_{-1}$ \\
HVS20 & $10\,160$ & $3.78$ & $510.5$ & $316$ &  $1.28$ &  $14.3$ &           4  \\ Stat. &             $^{+90}_{-90}$ & $^{+0.06}_{-0.03}$ &   $^{+5.4}_{-5.5}$ &                        $^{+20}_{-11}$  \\ Sys. & $^{+110}_{-110}$ & $^{+0.04}_{-0.04}$ & $^{+2.9}_{-2.9}$ &              $^{+0}_{-0}$ \\
HVS21 & $12\,760$ & $4.10$ & $357.4$ &  $47$ &  $1.30$ &  $11.1$ &           1  \\ Stat. &             $^{+90}_{-70}$ & $^{+0.04}_{-0.04}$ &   $^{+5.5}_{-5.4}$ &           $^{+62}_{-47}$  \\ Sys. & $^{+130}_{-130}$ & $^{+0.04}_{-0.04}$ & $^{+2.8}_{-3.9}$ &            $^{+50}_{-29}$ \\
HVS22 & $10\,350$ & $3.94$ & $596.4$ & $156$ &  $1.21$ &   $9.0$ &           3  \\ Stat. & $^{+100}_{-\phantom{0}80}$ & $^{+0.06}_{-0.04}$ &   $^{+7.4}_{-7.7}$ &                        $^{+32}_{-31}$  \\ Sys. & $^{+110}_{-110}$ & $^{+0.04}_{-0.04}$ & $^{+0.5}_{-0.3}$ &              $^{+1}_{-0}$ \\
\hline
     \end{tabular}\\

\end{center}
\end{minipage}
\begin{minipage}{.5\textwidth}
\setlength{\tabcolsep}{0.175cm}
\renewcommand{\arraystretch}{1.05}
\begin{center}
         \scriptsize 
\begin{tabular}{rrrrrrrr}
\hline\hline
     Object &    $T_\mathrm{eff}$ &    $\log(g)$ &     $\varv_\mathrm{rad}$ &       $\varv \sin(i)$ & $\chi^2_\mathrm{red}$ & S/N$_\mathrm{tot}$ & N  \\    
     \cline{4-5}&
      (K) & (cgs) & \multicolumn{2}{c}{(km\,s$^{-1}$)} & &  \\
      \hline
HVS23 & $10\,400$ & $3.60$ & $248.9$ & $112$ &  $1.32$ &   $4.3$ &           1  \\ Stat. &           $^{+160}_{-230}$ & $^{+0.09}_{-0.12}$ & $^{+14.4}_{-14.3}$ &                        $^{+51}_{-59}$  \\ Sys. & $^{+110}_{-110}$ & $^{+0.04}_{-0.04}$ & $^{+0.2}_{-0.1}$ &              $^{+2}_{-0}$ \\
HVS24 & $10\,900$ & $3.95$ & $496.0$ & $213$ &  $1.24$ &  $16.6$ &           2  \\ Stat. &             $^{+60}_{-60}$ & $^{+0.03}_{-0.03}$ &   $^{+4.2}_{-4.4}$ &                        $^{+17}_{-19}$  \\ Sys. & $^{+110}_{-110}$ & $^{+0.04}_{-0.04}$ & $^{+1.5}_{-1.5}$ &              $^{+1}_{-2}$ \\
 B095 &  $9850$ & $4.17$ & $206.8$ &  $89$ &  $1.30$ &  $15.1$ &           6  \\ Stat. &             $^{+60}_{-50}$ & $^{+0.05}_{-0.03}$ &   $^{+4.2}_{-4.1}$ &                        $^{+14}_{-12}$  \\ Sys. & $^{+100}_{-100}$ & $^{+0.04}_{-0.04}$ & $^{+0.5}_{-0.5}$ &              $^{+4}_{-2}$ \\
  B129 & $10\,720$ & $3.56$ & $351.7$ & $177$ &  $1.12$ &  $12.4$ &           1  \\ Stat. &             $^{+70}_{-70}$ & $^{+0.03}_{-0.04}$ &   $^{+5.1}_{-4.3}$ &                        $^{+19}_{-20}$  \\ Sys. & $^{+110}_{-110}$ & $^{+0.04}_{-0.04}$ & $^{+2.5}_{-1.5}$ &              $^{+1}_{-2}$ \\
  B143 & $10\,910$ & $4.03$ & $217.6$ & $269$ &  $1.30$ &  $15.7$ &           1  \\ Stat. &             $^{+70}_{-50}$ & $^{+0.03}_{-0.03}$ &   $^{+4.8}_{-4.8}$ &                        $^{+21}_{-18}$  \\ Sys. & $^{+110}_{-110}$ & $^{+0.04}_{-0.04}$ & $^{+2.4}_{-0.9}$ &              $^{+1}_{-3}$ \\
  B167 & $11\,250$ & $4.20$ & $297.9$ &   $0$ &  $1.19$ &  $19.4$ &           1  \\ Stat. &             $^{+40}_{-40}$ & $^{+0.02}_{-0.02}$ &   $^{+2.6}_{-2.8}$ & $^{+34}_{-\phantom{0}0}$ \\ Sys. & $^{+120}_{-120}$ & $^{+0.04}_{-0.04}$ & $^{+0.5}_{-1.1}$ & $^{+2}_{-0}$ \\
  B329 & $10\,800$ & $3.85$ & $213.7$ &  $58$ &  $1.24$ &   $8.0$ &           1  \\ Stat. & $^{+100}_{-\phantom{0}80}$ & $^{+0.06}_{-0.05}$ &   $^{+6.8}_{-7.2}$ &                        $^{+34}_{-48}$  \\ Sys. & $^{+110}_{-110}$ & $^{+0.04}_{-0.04}$ & $^{+1.4}_{-0.4}$ &              $^{+7}_{-7}$ \\
  B434 & $10\,140$ & $3.84$ & $441.3$ &  $89$ &  $1.33$ &  $35.0$ &           3  \\ Stat. &             $^{+40}_{-40}$ & $^{+0.02}_{-0.02}$ &   $^{+1.9}_{-1.9}$ &                          $^{+9}_{-8}$  \\ Sys. & $^{+110}_{-110}$ & $^{+0.04}_{-0.04}$ & $^{+2.2}_{-2.1}$ &  $^{+\phantom{0}7}_{-13}$ \\   
 B458 &  $9810$ & $3.80$ & $454.0$ & $104$ &  $1.23$ &   $6.1$ &           1  \\ Stat. & $^{+120}_{-\phantom{0}90}$ & $^{+0.05}_{-0.05}$ &   $^{+9.6}_{-8.5}$ &                        $^{+26}_{-23}$  \\ Sys. & $^{+100}_{-100}$ & $^{+0.04}_{-0.04}$ & $^{+1.4}_{-0.3}$ &              $^{+2}_{-1}$ \\    
 B481 & $10\,300$ & $3.61$ & $133.1$ & $190$ &  $1.28$ &   $7.4$ &           1  \\ Stat. &           $^{+120}_{-160}$ & $^{+0.06}_{-0.06}$ &   $^{+8.9}_{-8.3}$ &                        $^{+24}_{-30}$  \\ Sys. & $^{+110}_{-110}$ & $^{+0.04}_{-0.04}$ & $^{+2.1}_{-1.8}$ &              $^{+1}_{-1}$ \\
 B485 & $15\,710$ & $3.92$ & $422.7$ &  $81$ &  $1.51$ &  $40.1$ &           1  \\ Stat. &             $^{+70}_{-50}$ & $^{+0.02}_{-0.02}$ &   $^{+1.8}_{-1.8}$ &                          $^{+9}_{-8}$  \\ Sys. & $^{+160}_{-160}$ & $^{+0.04}_{-0.04}$ & $^{+1.0}_{-1.9}$ &  $^{+\phantom{0}7}_{-11}$ \\ 
 B537 & $11\,760$ & $3.75$ & $150.9$ & $181$ &  $1.23$ &   $7.5$ &           1  \\ Stat. &           $^{+120}_{-140}$ & $^{+0.06}_{-0.07}$ &   $^{+9.3}_{-9.2}$ &                        $^{+37}_{-29}$  \\ Sys. & $^{+120}_{-120}$ & $^{+0.04}_{-0.04}$ & $^{+1.7}_{-2.7}$ &  $^{+15}_{-\phantom{0}1}$ \\
 B572 & $10\,920$ & $4.23$ & $130.0$ & $138$ &  $1.27$ &   $6.4$ &           1  \\ Stat. &           $^{+140}_{-140}$ & $^{+0.06}_{-0.06}$ & $^{+10.7}_{-12.2}$ &                        $^{+40}_{-38}$  \\ Sys. & $^{+110}_{-110}$ & $^{+0.04}_{-0.04}$ & $^{+2.3}_{-3.7}$ &              $^{+1}_{-1}$ \\
 B576 & $11\,400$ & $3.70$ & $216.1$ &  $47$ &  $1.33$ &  $27.9$ &           1  \\ Stat. &             $^{+30}_{-30}$ & $^{+0.02}_{-0.02}$ &   $^{+1.9}_{-1.7}$ &                        $^{+12}_{-13}$  \\ Sys. & $^{+120}_{-120}$ & $^{+0.04}_{-0.04}$ & $^{+0.4}_{-0.3}$ &              $^{+8}_{-8}$ \\
 B598 & $10\,730$ & $4.52$ & $282.5$ & $192$ &  $1.13$ &  $17.0$ &           1  \\ Stat. &             $^{+60}_{-70}$ & $^{+0.03}_{-0.03}$ &   $^{+4.6}_{-4.8}$ &                        $^{+18}_{-15}$  \\ Sys. & $^{+110}_{-110}$ & $^{+0.04}_{-0.04}$ & $^{+1.0}_{-1.7}$ &              $^{+1}_{-1}$ \\
 B711 & $10\,410$ & $3.98$ & $267.6$ &  $17$ &  $2.54$ &  $36.4$ &           1  \\ Stat. &             $^{+30}_{-40}$ & $^{+0.02}_{-0.02}$ &   $^{+1.6}_{-1.7}$ &           $^{+16}_{-17}$  \\ Sys. & $^{+110}_{-110}$ & $^{+0.04}_{-0.04}$ & $^{+1.5}_{-1.8}$ &            $^{+12}_{-17}$ \\
 B733 & $10\,280$ & $4.04$ & $351.4$ & $278$ &  $2.83$ &  $53.2$ &           1  \\ Stat. &             $^{+30}_{-30}$ & $^{+0.02}_{-0.01}$ &   $^{+2.2}_{-2.0}$ &                          $^{+7}_{-7}$  \\ Sys. & $^{+110}_{-110}$ & $^{+0.04}_{-0.04}$ & $^{+2.4}_{-1.8}$ &              $^{+9}_{-6}$ \\
B1080 & $10\,700$ & $3.89$ & $501.5$ & $150$ &  $1.21$ &  $14.3$ &           2  \\ Stat. &             $^{+70}_{-70}$ & $^{+0.03}_{-0.03}$ &   $^{+4.6}_{-4.5}$ &                        $^{+21}_{-22}$  \\ Sys. & $^{+110}_{-110}$ & $^{+0.04}_{-0.04}$ & $^{+2.2}_{-1.2}$ &              $^{+1}_{-1}$ \\
B1085 & $11\,020$ & $3.79$ & $483.9$ & $319$ &  $1.27$ &  $23.9$ &           2  \\ Stat. &             $^{+40}_{-40}$ & $^{+0.03}_{-0.03}$ &   $^{+3.7}_{-2.5}$ &                        $^{+10}_{-12}$  \\ Sys. & $^{+120}_{-120}$ & $^{+0.04}_{-0.04}$ & $^{+0.9}_{-0.4}$ &              $^{+0}_{-9}$ \\
B1139 & $11\,660$ & $4.26$ &  $65.7$ &  $30$ &  $1.37$ &   $8.0$ &           1  \\ Stat. &           $^{+120}_{-120}$ & $^{+0.06}_{-0.05}$ &   $^{+7.4}_{-8.2}$ &           $^{+45}_{-30}$  \\ Sys. & $^{+120}_{-120}$ & $^{+0.04}_{-0.04}$ & $^{+3.4}_{-4.3}$ &            $^{+15}_{-24}$ \\
\hline
     \end{tabular}\\

\end{center}
\end{minipage}
\tablefoot{Statistical uncertainties (\textit{``Stat.''}) are $1\sigma$ confidence limits based on $\chi^2$ statistics. Systematic uncertainties (\textit{``Sys.''}) cover only the effects induced by additional variations of $1\%$ in $T_{\mathrm{eff}}$ and $0.04$ in $\log(g)$ and are formally taken to be $1\sigma$ confidence limits (see \citealt{2014A&A...565A..63I} for details). The quantity $\chi^2_\mathrm{red}$ is the reduced $\chi^2$ at the best fit. The total number of available spectra is denoted in the 'N' column, whereas the total S/N is given as quadratically added S/N of all available spectra.}
\end{table*}

\section{Spectral analysis}\label{sec:spectral-analysis}

The basic strategy of the spectral analysis is very similar to \citetalias{2018A&A...615L...5I}. However, there is one significant improvement, namely that flux-calibrated instead of normalized spectra are considered here. Consequently, we also re-analyze the objects from \citetalias{2018A&A...615L...5I} to have a homogeneously studied sample.

  \begin{figure*}
   \centering
   \includegraphics[width=0.47\linewidth]{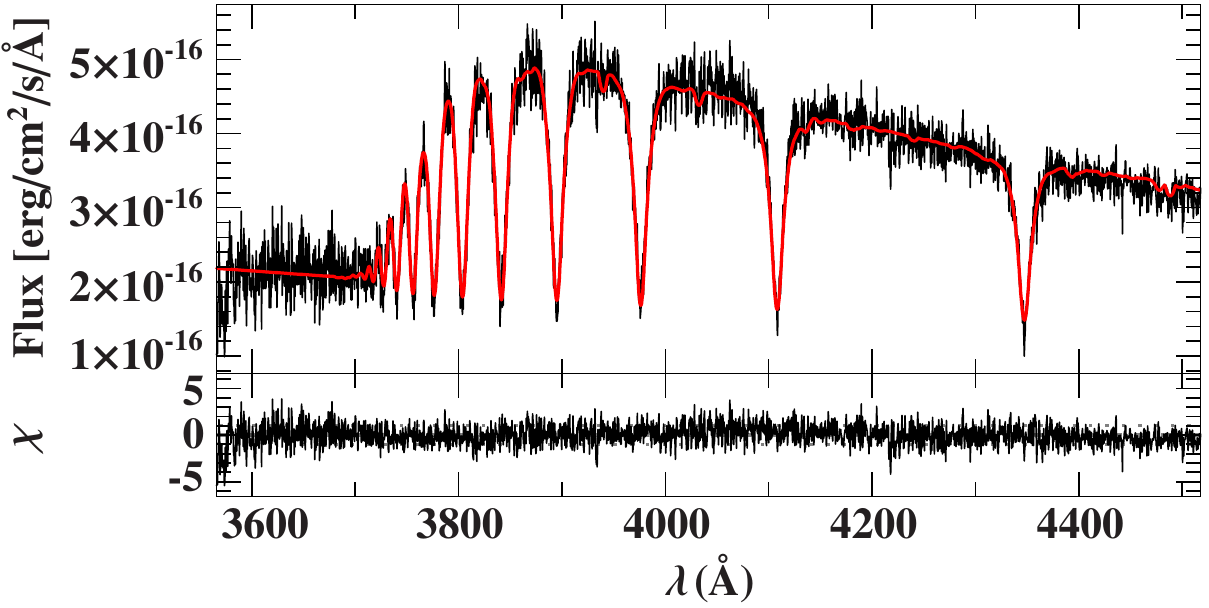}
   \includegraphics[width=0.49\linewidth]{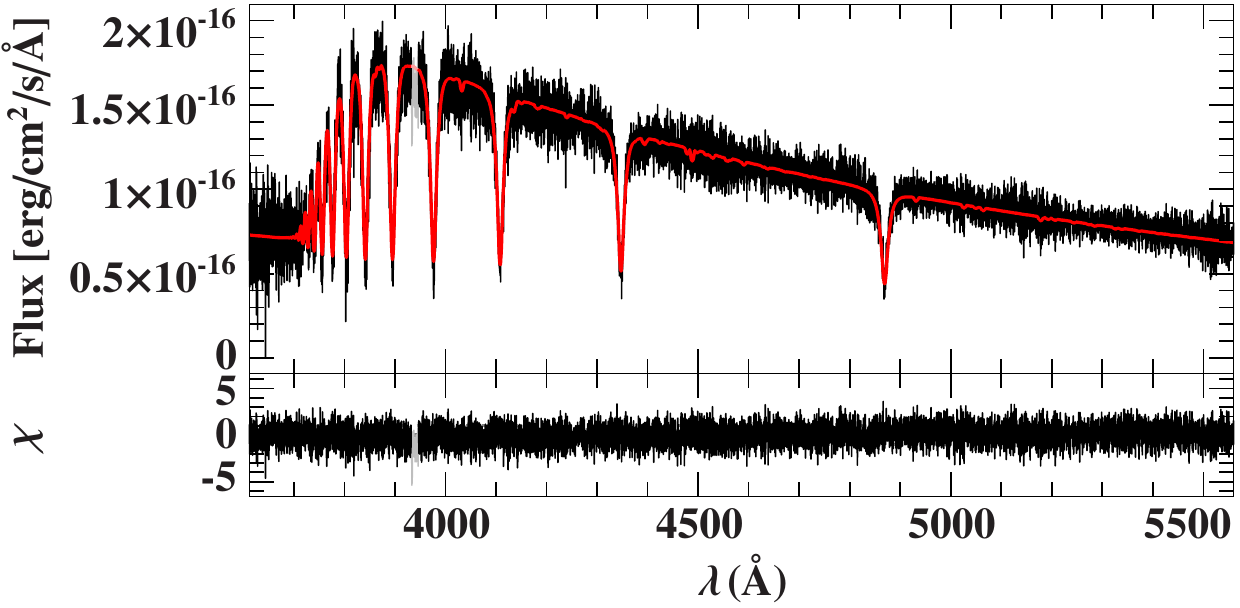}   
   \caption{Comparison of best-fitting model spectrum (red) with observation (black) in the case of B1085. Residuals $\chi$ are shown in the lower panels. The \textit{left figure} shows one of the available MMT spectra, the quality of which is typical for the sample. The \textit{right figure} shows an exemplary flux-calibrated X-shooter UVB spectrum. Contrary to the absolute fluxes, the calibration of the relative fluxes is almost identical, demonstrating that our procedure for relative flux-calibration works. Combined with synthetic spectra that properly account for the Balmer jump, this enables us to derive reliable effective temperatures and surface gravities.}
   \label{fig:B1085-xshooter-spectrum}
   \end{figure*}

\subsection{MMT survey data and relative flux calibration}

The spectra analyzed in this work were taken during the MMT HVS survey and kindly provided by Warren Brown. A prime goal of that survey was spectral classification rather than a high-precision quantitative analysis. Consequently, the average S/N of the co-added --~for some of the stars, more than one spectrum is available~-- spectra is only of the order of 10--30 because most of the targets are quite distant and thus very faint, even for a 6.5\,m telescope. Table~\ref{tab:atmos-results} lists the number of individual observations as well as the wavelength-averaged S/N of the co-added spectra.

The low S/N and the relatively small wavelength coverage of the MMT spectra make it crucial to use as much information as possible to determine accurate temperatures and surface gravities. Consequently and in contrast to \citetalias{2018A&A...615L...5I}, we fitted flux-calibrated rather than normalized spectra to also exploit the information contained in the slope of the continuum as well as in the shape of the Balmer jump. The effective temperature mainly affects the height of the Balmer jump while the surface gravity primarily its slope. By using this approach, the derived values for $T_\mathrm{eff}$ and $\log(g)$ are more accurate and less uncertain, which is important for the spectrophotometric distance estimation where both parameters contribute significantly to the error budget. Our relative flux calibration followed the typical procedure, that is, we corrected for Rayleigh scattering, aerosols \citep[see e.g., ][]{2011A&A...527A..91P}, telluric absorption features \citep{2014A&A...568A...9M}, and then use a standard star to calibrate the flux. MMT spectra for the flux standards were taken from the same night whenever possible, otherwise from the previous or following one. The flux-calibrated reference spectra of the standards were available in the HST CALSPEC database \citep{2014PASP..126..711B}.

\subsection{Fit method}

The spectral analysis strategy basically followed \citet{2014A&A...565A..63I}. The underlying idea was to simultaneously fit all individual spectra of a star over their entire spectral range using the concept of $\chi^2$ minimization. Given the limited quality of the available spectra, it was not possible to determine abundances of individual chemical elements. Therefore, a solar chemical composition was assumed and the microturbulence was kept fixed at 2\,km\,s$^{-1}$, which is characteristic of late B-type MS stars. We were thus left with four fitting parameters for the stellar spectrum: the effective temperature $T_\mathrm{eff}$, the surface gravity $\log(g)$, the projected rotational velocity $\varv\sin(i)$, and the radial velocity $\varv_{\mathrm{rad}}$. Because we dealt with flux-calibrated spectra, we also had to consider interstellar reddening. Using the extinction law by \cite{1999PASP..111...63F}, three additional parameters were introduced: a distance scaling parameter, the color excess $E(B-V)$, and the extinction coefficient $R_V$, which was kept fixed at its typical value for the interstellar medium, that is, $R_V=3.1$.

\subsection{Cross-checks against medium- and high-resolution spectra}

\begin{table}
\centering
\renewcommand{\arraystretch}{1.25}
\caption{Atmospheric and derived stellar parameters for B1085 based on two different sets of spectra.
\label{tab:B1085}}
\begin{tabular}{lrr}
\hline\hline
 & MMT & X-shooter\\
 $T_\mathrm{eff}$ (K) & $11\,020\pm120$ &  $10\,890\pm120$\\
 $\log(g)$ (cgs) & $3.79\pm0.05$&  $3.85\pm0.05$\\
 $\varv_{\mathrm{rad}}$ (km s$^{-1}$) & $484^{+4}_{-3}$ & $491\pm2$ \\
 $\varv\sin(i)$ (km s$^{-1}$) & $319^{+10}_{-15}$ & $290\pm1$\\
 \hline
 $M$ ($M_\odot$) &   $3.38^{+0.10}_{-0.11}$&  $3.14\pm0.1$\\
 $\tau$ (Myr) & $224^{+24}_{-13}$ & $260^{+\phantom{0}9}_{-11}$\\
 $d$ (kpc) & $42^{+3}_{-4}$ & $37\pm3$\\
 \hline
\end{tabular}
\tablefoot{Uncertainties are $1\sigma$ and cover statistical as well as systematic effects. The derivations of stellar parameters and distances are outlined in Sect.~\ref{subsection:atmos_stellar_params} and \ref{sec:distance_estimation}.}
\end{table}

   \begin{figure*}
   \centering
   \includegraphics[width=0.49\linewidth]{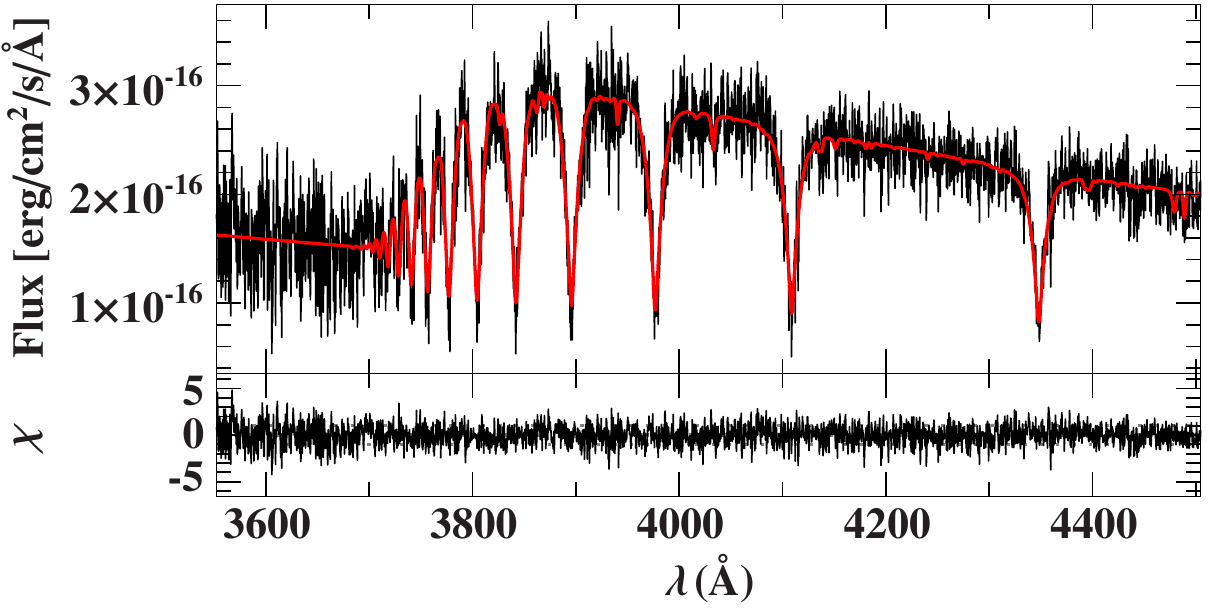}
   \hspace{.5cm}
   \includegraphics[width=0.45\linewidth]{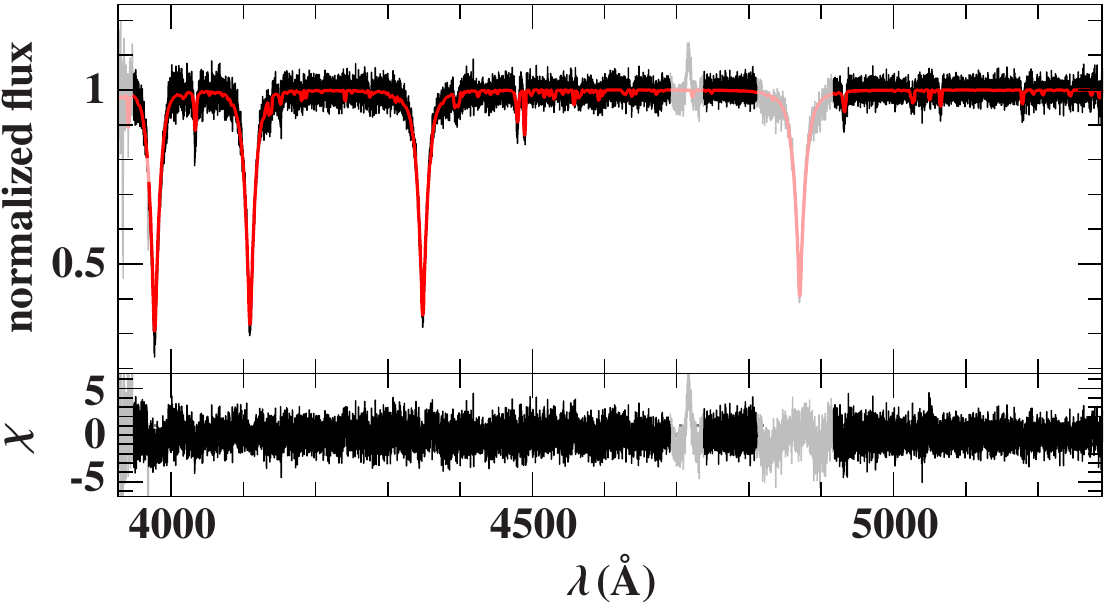}
   \caption{Same as Fig.~\ref{fig:B1085-xshooter-spectrum} but for HVS\,5. The \textit{left figure} shows a flux-calibrated MMT spectrum while the normalized co-added HIRES spectrum is shown in the \textit{right figure}. Light colors mark regions that have been excluded from fitting, e.g., due to data reduction problems.}
   \label{fig:hvs5-spectrum}
   \end{figure*}

Medium- and high-resolution spectra with larger spectral coverage are available for a few objects of the sample. By comparing the results based on those spectra with the ones based on the MMT spectra, we are able to validate our approach. The first test case is B1085, for which two MMT spectra with exposure times of 120\,s and 660\,s are available. In addition, we obtained four flux-calibrated X-shooter \citep{2011A&A...536A.105V} spectra with individual exposure times of 1200\,s in the UVB and the VIS channel, which together span a range of 3600--9400\,{\small \AA}. The second test case is HVS\,5, for which we downloaded HIRES \citep{1994SPIE.2198..362V} spectra from the KOA archive, which have already been analyzed by \cite{2012ApJ...754L...2B}. We reduced the data anew, performed a continuum normalization, and co-added the blue channel of all nine exposures. The corresponding spectral fits for both targets are shown in Figs.~\ref{fig:B1085-xshooter-spectrum} and \ref{fig:hvs5-spectrum} and the resulting atmospheric and derived stellar parameters are contrasted in Tables~\ref{tab:B1085} and \ref{tab:hvs5}. The good agreement between results based on MMT spectra and medium- to high-resolution spectra of different spectral coverage is very reassuring, showing that we can derive accurate parameters from the flux-calibrated MMT spectra.

\begin{table}
\centering
\setlength{\tabcolsep}{0.17cm}
\renewcommand{\arraystretch}{1.25}
\caption{Atmospheric and derived stellar parameters for HVS\,5 based on two different sets of spectra.
\label{tab:hvs5}}
\begin{tabular}{lrrr}
\hline\hline
 & MMT & HIRES & Brown+ (2012)\\
 $T_\mathrm{eff}$ (K) & $12\,550^{+160}_{-150}$ &  $12\,190^{+290}_{-260}$ &  $12\,000 \pm 350$\\
 $\log(g)$ (cgs) & $4.09\pm0.06$ & $4.11^{+0.11}_{-0.12}$& $3.89 \pm 0.13$\\
 $\varv_{\mathrm{rad}}$ (km s$^{-1}$) & $542^{+7}_{-6}$ &  $552\pm3$ &  $552 \pm 3$\\
 $\varv\sin(i)$ (km s$^{-1}$) &$123 \pm 19$ & $132^{+3}_{-2}$&$133 \pm 7$\\
 \hline
 $M$ ($M_\odot$) &  $3.40^{+0.10}_{-0.11}$ & $3.23^{+0.09}_{-0.07}$&$3.62 \pm 0.11$\\
 $\tau$ (Myr) &$149^{+20}_{-26}$  &$160^{+60}_{-70}$ &$170 \pm 17$\\
 $d$ (kpc) &  $37\pm4$ &$34^{+7}_{-5}$ & $44 \pm 4$\\
 \hline
\end{tabular}
\tablefoot{Same as Table~\ref{tab:B1085}. Values obtained by \cite{2012ApJ...754L...2B} from the same HIRES data are listed in the third column.}
\end{table}

 \subsection{Atmospheric and stellar parameters}\label{subsection:atmos_stellar_params}

   \begin{figure}
   \centering
   \includegraphics[width=\linewidth]{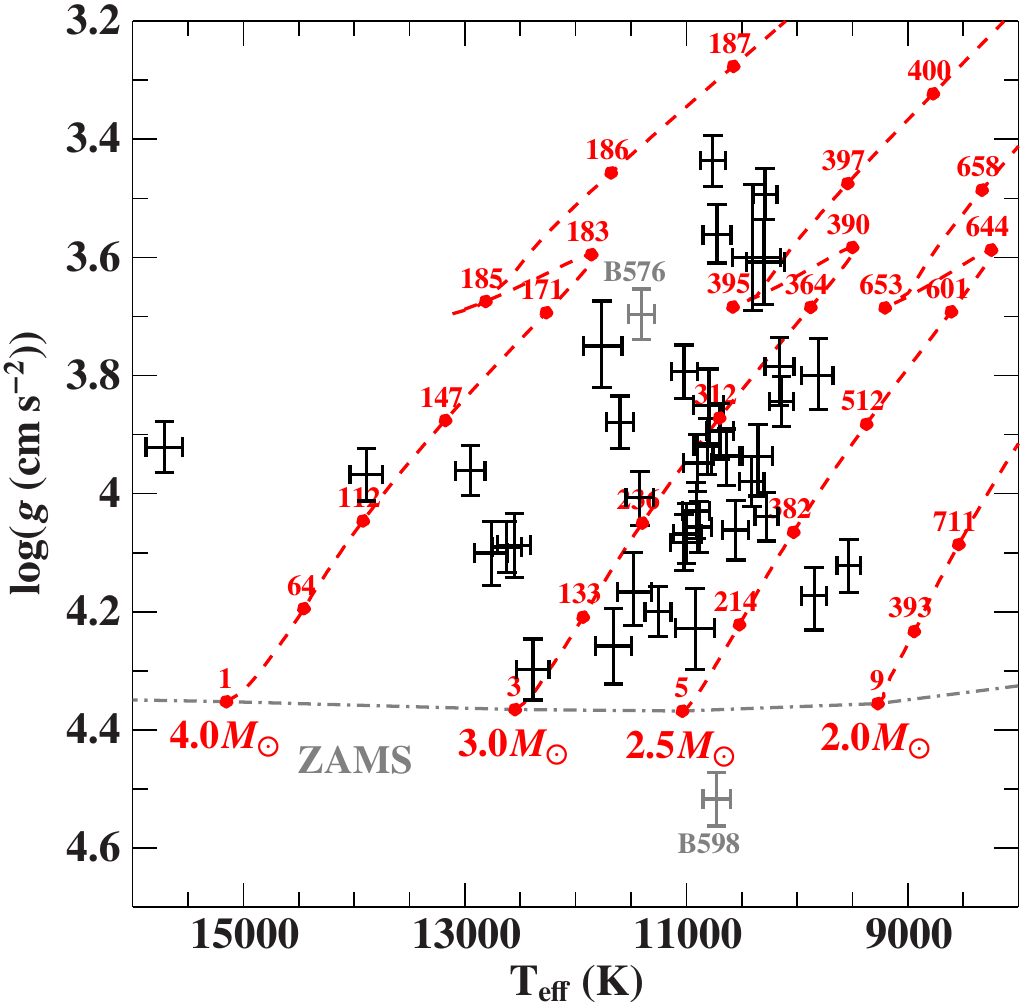}
      \caption{Position of the program stars in the Kiel diagram. Evolutionary tracks for rotating ($\Omega/\Omega_\mathrm{crit} = 0.4$) MS stars of solar metallicity and different initial masses \citep{2013A&A...553A..24G} are overlaid in red. Red filled circles and numbers mark the age in Myr. The locus of the zero-age MS (ZAMS) is indicated as a gray dashed line. Error bars are $1\sigma$ and cover statistical and systematic uncertainties. The two objects (B576, B598) for which a MS nature is unlikely are marked in gray (see Sect.~\ref{sec:B598}).}
         \label{fig:evol-tracks}
   \end{figure}
   
      \begin{figure}
   \centering
   \includegraphics[width=\linewidth]{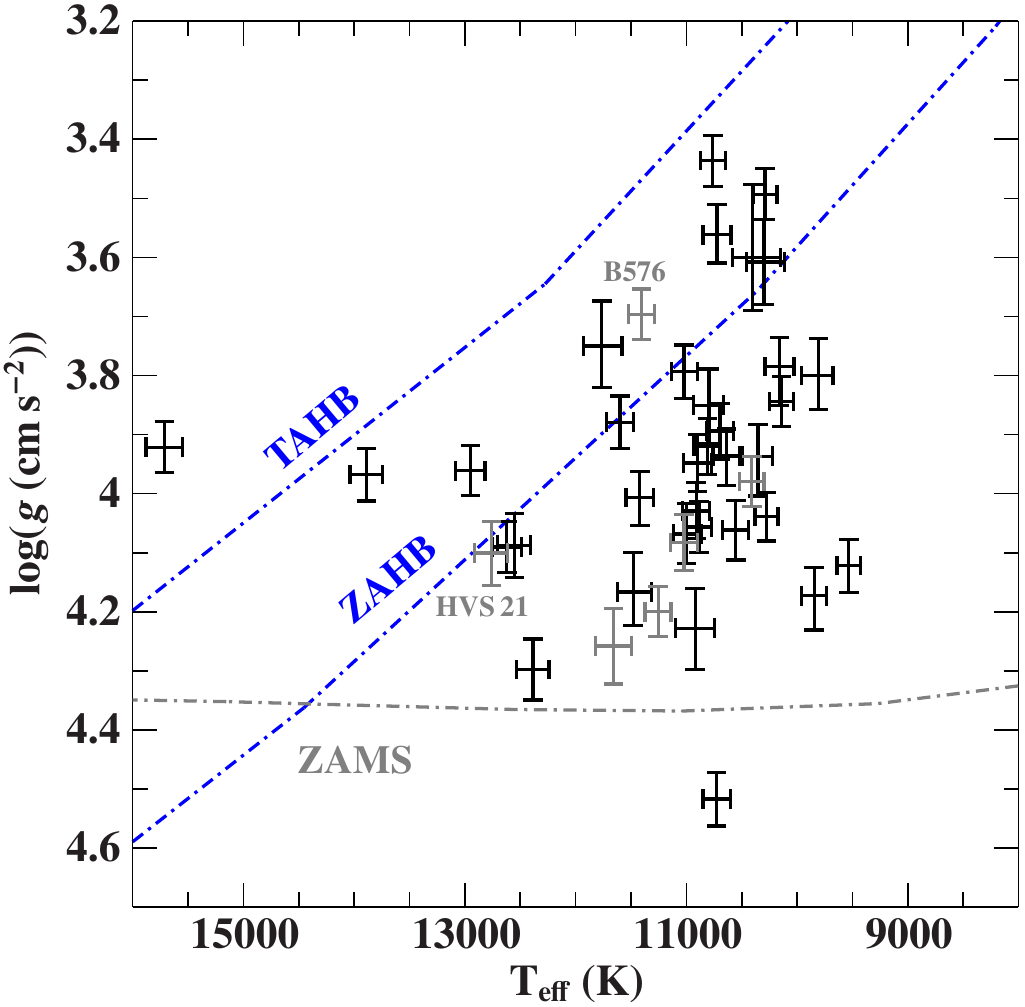}
      \caption{Same as Fig.~\ref{fig:evol-tracks} but MS tracks are replaced by loci for the zero- and terminal-age horizontal branch (ZAHB, TAHB) for a helium abundance of 0.247 and $[\mathrm{Fe}/\mathrm{H}] = -1.48$ from \cite{1993ApJ...419..596D}. Stars with $\varv\sin(i)< 50$\,km\,s$^{-1}$ are displayed in gray.}
         \label{fig:evol-tracks-dorman}
   \end{figure}

\begin{figure}
   \centering
   \includegraphics[width=\linewidth]{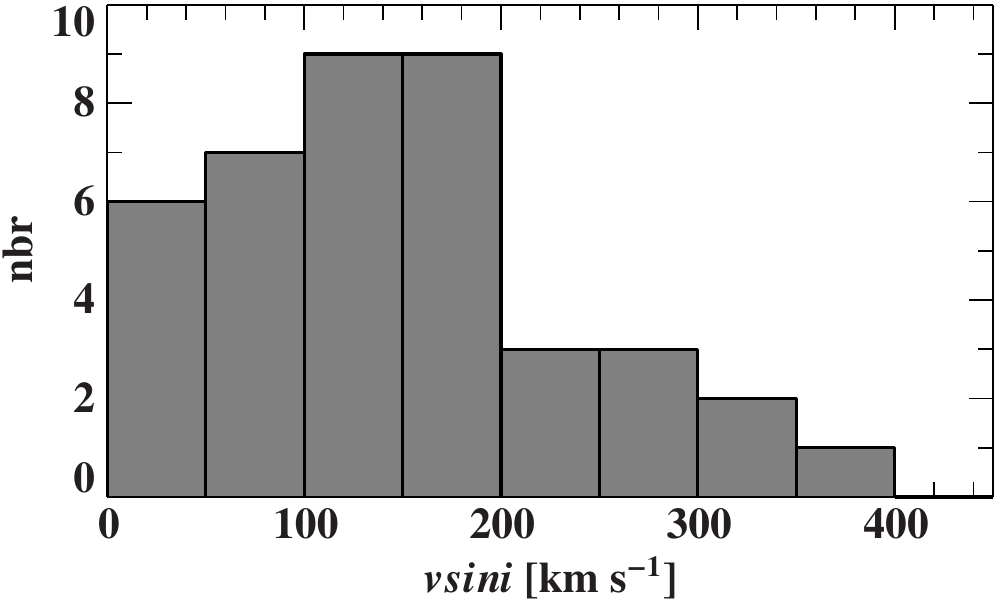}
      \caption{Histogram showing the distribution of $\varv\sin(i)$ values for all program stars. Typical individual uncertainties (see Table~\ref{tab:atmos-results}) are smaller than the bin size. Most stars rotate fast and are thus very likely MS stars.}
      \label{fig:hist_vsini}
   \end{figure}

The results of the spectral analysis are summarized in Table~\ref{tab:atmos-results} and visualized in Fig.~\ref{fig:evol-tracks}. All program stars except one (B598) have surface gravities and effective temperatures that are perfectly consistent with a MS nature. The surface gravities derived here are on average lower than those reported by \cite{2014ApJ...787...89B}, naturally solving the issue that some stars such as HVS\,11, HVS\,12, and HVS\,19 were lying below the ZAMS, a result that was barely compatible with a MS HVS nature. The atmospheric parameters of some stars would also be consistent with BHB stars, that is, evolved stars of lower mass; see Fig.~\ref{fig:evol-tracks-dorman}. An important criterion to differentiate between the two cases is stellar rotation. BHB stars tend to rotate slowly \citep[less than a few tens of km\,s$^{-1}$,][]{2003ApJS..149..101B} while MS B-type stars typically rotate fast (hundreds of km\,s$^{-1}$). Indeed, most program stars rotate rapidly (see Table~\ref{tab:atmos-results} and Fig.~\ref{fig:hist_vsini}), predominantly with $\varv\sin(i)$ = 50--200\,km\,s$^{-1}$. While three stars rotate even faster than 300\,km\,s$^{-1}$, only a handful of targets shows low projected rotation velocities (upper limits of $\approx50$\,km\,s$^{-1}$). All slowly rotating stars except B576 are located outside the region enclosed by the ZAHB and the TAHB (see Fig.~\ref{fig:evol-tracks-dorman}) indicating that they do not belong to an old population of low mass BHB stars. Low $\varv\sin(i)$ values of MS stars could be explained by low inclinations $i$, that is, by seeing the object almost pole-on.

We conclude that most of the program stars are likely MS stars. Under this assumption, stellar masses $M$, ages $\tau$, radii $R$, luminosities $L$, and ratios of actual angular velocity to critical velocity $\Omega/\Omega_\mathrm{crit}$ can be derived by comparing the stars' position in the Kiel diagram with theoretical predictions; see Fig.~\ref{fig:evol-tracks}. The outcome of this exercise is tabulated in Table~\ref{tab:stellar_params}. 

The nature of B598 remains unclear for the moment because its effective temperature places the star below the ZAMS, although it is rapidly rotating. Two evolutionary scenarios can explain stars in this region of the Kiel diagram. On the one hand, we note that the locus of the ZAMS is a function of metallicity. The lower the metallicity, the more compact the stars, the higher their surface gravity. Hence, B598 could be a low metallicity MS star of an old stellar population. Such stars are classified as blue stragglers, and correspondingly B598 may be a rejuvenated 2--3\,$M_\odot$ MS star. Another class of stars are the rare progenitors of extremely low mass (ELM) WDs (see \citealt{2016PASP..128h2001H} for details), some of which show surface gravities much lower than typical WDs \citep[as low as $\log(g) \sim 4.8$,][] {2016ApJ...818..155B,2019ApJ...883...51R}. Both classes of stars are often found in binaries. \citet{2019A&A...632A...3G} find that more than 50\% of the blue stragglers in the globular cluster NGC\,3201 are close binaries, whereas \citet{2016ApJ...818..155B} found about 85\% of the ELM WDs to be short-period binaries. Typically, orbital periods in ELM WD binary systems are found to be on the order of tens of minutes to hours \citep{2016ApJ...818..155B} and should, thus, lead to observable variations in the radial velocity unless the orbital inclination is very small. Unfortunately, only one single exposure was taken in the course of the MMT survey, which is why we cannot check whether B598 is radial-velocity variable.

The slowly rotating star B576 is located mid-way in the HB band; see Fig.~\ref{fig:evol-tracks-dorman}. Anticipating results from the kinematic investigation, it is more likely a low mass BHB star than a MS star; see Sect.~\ref{sec:B576}.

\begin{sidewaystable*}
\scriptsize 
\caption{\label{tab:stellar_params}Stellar parameters of the program stars.}
\renewcommand{\arraystretch}{1.3}
\begin{tabular}{lrrrrrrrrrrrrrrrrrrrrrrrr}
\hline\hline
                            Object &  &     \multicolumn{2}{c}{$M$}              &  &   \multicolumn{2}{c}{$\tau$} &  &   \multicolumn{2}{c}{$E(B-V)$} &  &      \multicolumn{2}{c}{$\Omega$/$\Omega_\mathrm{crit}$} &  &     \multicolumn{2}{c}{$R$} &  &   \multicolumn{2}{c}{$\log (L/L_\sun$)} &  & \multicolumn{2}{c}{$\log (\theta)$} &  &     \multicolumn{2}{c}{$d$} \\
                            
                            \cline{3-4} \cline{6-7} \cline{9-10} \cline{12-13} \cline{15-16} \cline{18-19} \cline{21-22} \cline{24-25} & & \multicolumn{2}{c}{(M$_\sun$)} & & \multicolumn{2}{c}{(Myr)} & & \multicolumn{2}{c}{(mag)} & & & & &  \multicolumn{2}{c}{(R$_\sun$)} & & & & & & & & \multicolumn{2}{c}{(kpc)}                           \\
                            \hline
 HVS1 (\object{SDSSJ090744.99+024506.8}) &         & $3.59$ & $^{+0.06}_{-0.08}$ &         & $247$ &             $^{+10}_{-10}$ &         & $0.040$ & $^{+0.021}_{-0.021}$ &         & $0.95$ & $^{+0.01}_{-0.04}$ &         & $5.6$ & $^{+0.4}_{-0.4}$ &         & $2.50$ & $^{+0.05}_{-0.06}$ &         & $-11.80$ & $^{+0.02}_{-0.02}$ &         & $161$ &           $^{+13}_{-13}$ \\
 HVS4 (\object{SDSSJ091301.01+305119.8}) &         & $4.14$ & $^{+0.20}_{-0.13}$ &         & $120$ &               $^{+8}_{-9}$ &         & $0.001$ & $^{+0.004}_{-0.001}$ &         & $0.57$ & $^{+0.27}_{-0.20}$ &         & $3.5$ & $^{+0.3}_{-0.3}$ &         & $2.61$ & $^{+0.08}_{-0.08}$ &         & $-11.68$ & $^{+0.01}_{-0.01}$ &         &  $75$ &             $^{+7}_{-6}$ \\
 HVS5 (\object{SDSSJ091759.47+672238.3}) &         & $3.40$ & $^{+0.10}_{-0.10}$ &         & $149$ &             $^{+20}_{-26}$ &         & $0.106$ & $^{+0.013}_{-0.012}$ &         & $0.58$ & $^{+0.01}_{-0.18}$ &         & $2.8$ & $^{+0.3}_{-0.3}$ &         & $2.23$ & $^{+0.08}_{-0.08}$ &         & $-11.47$ & $^{+0.01}_{-0.01}$ &         &  $37$ &             $^{+4}_{-4}$ \\
 HVS6 (\object{SDSSJ110557.45+093439.4}) &         & $3.04$ & $^{+0.09}_{-0.09}$ &         &  $62$ &             $^{+41}_{-47}$ &         & $0.015$ & $^{+0.014}_{-0.014}$ &         & $0.38$ & $^{+0.06}_{-0.39}$ &         & $2.0$ & $^{+0.2}_{-0.2}$ &         & $1.95$ & $^{+0.08}_{-0.08}$ &         & $-11.75$ & $^{+0.01}_{-0.01}$ &         &  $52$ &             $^{+6}_{-5}$ \\
 HVS7 (\object{SDSSJ113312.12+010824.8}) &         & $3.74$ & $^{+0.20}_{-0.11}$ &         & $152$ &   $^{+\phantom{0}9}_{-16}$ &         & $0.018$ & $^{+0.017}_{-0.017}$ &         & $0.28$ & $^{+0.51}_{-0.16}$ &         & $3.3$ & $^{+0.3}_{-0.3}$ &         & $2.45$ & $^{+0.09}_{-0.08}$ &         & $-11.50$ & $^{+0.01}_{-0.01}$ &         &  $48$ &             $^{+5}_{-4}$ \\
 HVS8 (\object{SDSSJ094214.03+200322.0}) &         & $2.93$ & $^{+0.08}_{-0.07}$ &         & $208$ &             $^{+54}_{-33}$ &         & $0.026$ & $^{+0.012}_{-0.013}$ &         & $0.95$ & $^{+0.00}_{-0.10}$ &         & $2.7$ & $^{+0.2}_{-0.2}$ &         & $1.95$ & $^{+0.07}_{-0.07}$ &         & $-11.48$ & $^{+0.01}_{-0.01}$ &         &  $36$ &             $^{+3}_{-3}$ \\
 HVS9 (\object{SDSSJ102137.08-005234.7}) &         & $3.89$ & $^{+0.15}_{-0.22}$ &         & $167$ &             $^{+66}_{-15}$ &         & $0.053$ & $^{+0.014}_{-0.014}$ &         & $0.03$ & $^{+0.90}_{-0.03}$ &         & $6.2$ & $^{+0.5}_{-0.5}$ &         & $2.67$ & $^{+0.06}_{-0.07}$ &         & $-11.61$ & $^{+0.01}_{-0.01}$ &         & $115$ &           $^{+10}_{-10}$ \\
HVS10 (\object{SDSSJ120337.85+180250.3}) &         & $2.91$ & $^{+0.07}_{-0.12}$ &         & $308$ &             $^{+36}_{-30}$ &         & $0.039$ & $^{+0.020}_{-0.020}$ &         & $0.75$ & $^{+0.12}_{-0.55}$ &         & $3.0$ & $^{+0.3}_{-0.3}$ &         & $2.03$ & $^{+0.06}_{-0.08}$ &         & $-11.72$ & $^{+0.01}_{-0.01}$ &         &  $72$ &             $^{+7}_{-7}$ \\
HVS11 (\object{SDSSJ095906.47+000853.4}) &         & $2.33$ & $^{+0.05}_{-0.05}$ &         & $334$ & $^{+126}_{-\phantom{0}71}$ &         & $0.009$ & $^{+0.018}_{-0.010}$ &         & $0.86$ & $^{+0.01}_{-0.31}$ &         & $2.2$ & $^{+0.2}_{-0.2}$ &         & $1.55$ & $^{+0.05}_{-0.06}$ &         & $-11.75$ & $^{+0.01}_{-0.02}$ &         &  $56$ &             $^{+5}_{-5}$ \\
HVS12 (\object{SDSSJ105009.59+031550.6}) &         & $2.84$ & $^{+0.10}_{-0.09}$ &         & $241$ &             $^{+49}_{-25}$ &         & $0.056$ & $^{+0.023}_{-0.022}$ &         & $0.08$ & $^{+0.76}_{-0.08}$ &         & $2.5$ & $^{+0.3}_{-0.2}$ &         & $1.93$ & $^{+0.07}_{-0.07}$ &         & $-11.81$ & $^{+0.02}_{-0.02}$ &         &  $74$ &             $^{+8}_{-7}$ \\
HVS13 (\object{SDSSJ105248.30-000133.9}) &         & $3.02$ & $^{+0.07}_{-0.12}$ &         & $279$ &             $^{+42}_{-30}$ &         & $0.007$ & $^{+0.016}_{-0.008}$ &         & $0.81$ & $^{+0.03}_{-0.82}$ &         & $3.2$ & $^{+0.3}_{-0.3}$ &         & $2.09$ & $^{+0.06}_{-0.07}$ &         & $-11.92$ & $^{+0.01}_{-0.02}$ &         & $119$ &           $^{+10}_{-11}$ \\
HVS14 (\object{SDSSJ104401.75+061139.0}) &         & $3.10$ & $^{+0.09}_{-0.10}$ &         & $228$ &             $^{+23}_{-21}$ &         & $0.025$ & $^{+0.016}_{-0.016}$ &         & $0.70$ & $^{+0.12}_{-0.27}$ &         & $2.9$ & $^{+0.3}_{-0.3}$ &         & $2.11$ & $^{+0.07}_{-0.07}$ &         & $-11.86$ & $^{+0.01}_{-0.01}$ &         &  $95$ &             $^{+9}_{-9}$ \\
HVS15 (\object{SDSSJ113341.09-012114.2}) &         & $2.86$ & $^{+0.10}_{-0.11}$ &         & $250$ &             $^{+36}_{-48}$ &         & $0.061$ & $^{+0.024}_{-0.024}$ &         & $0.70$ & $^{+0.18}_{-0.17}$ &         & $2.6$ & $^{+0.3}_{-0.3}$ &         & $1.94$ & $^{+0.08}_{-0.08}$ &         & $-11.72$ & $^{+0.02}_{-0.02}$ &         &  $61$ &             $^{+7}_{-7}$ \\
HVS16 (\object{SDSSJ122523.40+052233.8}) &         & $2.73$ & $^{+0.07}_{-0.07}$ &         & $283$ &             $^{+35}_{-41}$ &         & $0.006$ & $^{+0.014}_{-0.007}$ &         & $0.78$ & $^{+0.10}_{-0.02}$ &         & $2.5$ & $^{+0.3}_{-0.2}$ &         & $1.86$ & $^{+0.07}_{-0.07}$ &         & $-11.76$ & $^{+0.01}_{-0.01}$ &         &  $65$ &             $^{+6}_{-6}$ \\
HVS17 (\object{SDSSJ164156.39+472346.1}) &         & $3.43$ & $^{+0.10}_{-0.10}$ &         & $143$ &             $^{+20}_{-20}$ &         & $0.005$ & $^{+0.010}_{-0.005}$ &         & $0.61$ & $^{+0.01}_{-0.10}$ &         & $2.8$ & $^{+0.2}_{-0.2}$ &         & $2.24$ & $^{+0.07}_{-0.07}$ &         & $-11.45$ & $^{+0.01}_{-0.01}$ &         &  $35$ &             $^{+3}_{-3}$ \\
HVS18 (\object{SDSSJ232904.94+330011.4}) &         & $3.34$ & $^{+0.13}_{-0.09}$ &         & $228$ &             $^{+10}_{-20}$ &         & $0.100$ & $^{+0.025}_{-0.024}$ &         & $0.64$ & $^{+0.20}_{-0.07}$ &         & $3.5$ & $^{+0.3}_{-0.3}$ &         & $2.29$ & $^{+0.08}_{-0.07}$ &         & $-11.79$ & $^{+0.02}_{-0.02}$ &         &  $96$ & $^{+11}_{-\phantom{0}9}$ \\
HVS19 (\object{SDSSJ113517.75+080201.4}) &         & $3.01$ & $^{+0.08}_{-0.09}$ &         & $117$ & $^{+104}_{-\phantom{0}55}$ &         & $0.083$ & $^{+0.037}_{-0.036}$ &         & $0.94$ & $^{+0.01}_{-0.54}$ &         & $2.4$ & $^{+0.3}_{-0.3}$ &         & $1.94$ & $^{+0.08}_{-0.09}$ &         & $-11.90$ & $^{+0.02}_{-0.02}$ &         &  $84$ &           $^{+11}_{-11}$ \\
HVS20 (\object{SDSSJ113637.13+033106.8}) &         & $3.03$ & $^{+0.03}_{-0.10}$ &         & $310$ &             $^{+54}_{-14}$ &         & $0.074$ & $^{+0.028}_{-0.026}$ &         & $0.94$ & $^{+0.01}_{-0.84}$ &         & $3.7$ & $^{+0.3}_{-0.4}$ &         & $2.11$ & $^{+0.05}_{-0.08}$ &         & $-11.79$ & $^{+0.02}_{-0.02}$ &         & $102$ & $^{+\phantom{0}9}_{-13}$ \\
HVS21 (\object{SDSSJ103418.25+481134.5}) &         & $3.43$ & $^{+0.13}_{-0.11}$ &         & $140$ &             $^{+22}_{-28}$ &         & $0.033$ & $^{+0.018}_{-0.018}$ &         & $0.24$ & $^{+0.23}_{-0.15}$ &         & $2.7$ & $^{+0.3}_{-0.3}$ &         & $2.25$ & $^{+0.09}_{-0.09}$ &         & $-11.89$ & $^{+0.01}_{-0.01}$ &         &  $96$ &           $^{+11}_{-10}$ \\
HVS22 (\object{SDSSJ114146.44+044217.2}) &         & $2.80$ & $^{+0.12}_{-0.08}$ &         & $344$ &             $^{+28}_{-30}$ &         & $0.099$ & $^{+0.048}_{-0.048}$ &         & $0.75$ & $^{+0.13}_{-0.08}$ &         & $3.0$ & $^{+0.4}_{-0.3}$ &         & $1.96$ & $^{+0.08}_{-0.08}$ &         & $-11.87$ & $^{+0.03}_{-0.03}$ &         &  $99$ &           $^{+16}_{-14}$ \\
HVS23 (\object{SDSSJ215629.01+005444.1}) &         & $3.37$ & $^{+0.24}_{-0.31}$ &         & $282$ & $^{+106}_{-\phantom{0}49}$ &         & $0.082$ & $^{+0.032}_{-0.037}$ &         & $0.73$ & $^{+0.22}_{-0.58}$ &         & $4.8$ & $^{+1.0}_{-0.6}$ &         & $2.39$ & $^{+0.14}_{-0.13}$ &         & $-11.89$ & $^{+0.03}_{-0.02}$ &         & $168$ &           $^{+37}_{-25}$ \\
HVS24 (\object{SDSSJ111136.44+005856.4}) &         & $3.03$ & $^{+0.08}_{-0.14}$ &         & $258$ &             $^{+46}_{-27}$ &         & $0.043$ & $^{+0.014}_{-0.015}$ &         & $0.86$ & $^{+0.05}_{-0.56}$ &         & $3.1$ & $^{+0.3}_{-0.3}$ &         & $2.07$ & $^{+0.07}_{-0.06}$ &         & $-11.66$ & $^{+0.01}_{-0.01}$ &         &  $63$ &             $^{+6}_{-6}$ \\
 B095 (\object{SDSSJ101359.79+563111.6}) &         & $2.34$ & $^{+0.07}_{-0.08}$ &         & $309$ &             $^{+91}_{-56}$ &         & $0.019$ & $^{+0.024}_{-0.020}$ &         & $0.66$ & $^{+0.14}_{-0.21}$ &         & $2.1$ & $^{+0.2}_{-0.2}$ &         & $1.56$ & $^{+0.06}_{-0.08}$ &         & $-11.80$ & $^{+0.02}_{-0.02}$ &         &  $59$ &             $^{+6}_{-6}$ \\
 B129 (\object{SDSSJ074950.24+243841.1}) &         & $3.63$ & $^{+0.14}_{-0.17}$ &         & $192$ &             $^{+63}_{-10}$ &         & $0.080$ & $^{+0.020}_{-0.019}$ &         & $0.03$ & $^{+0.92}_{-0.04}$ &         & $5.3$ & $^{+0.4}_{-0.5}$ &         & $2.51$ & $^{+0.07}_{-0.08}$ &         & $-11.56$ & $^{+0.01}_{-0.01}$ &         &  $85$ &             $^{+9}_{-8}$ \\
  B143 (\object{SDSSJ081828.07+570922.0}) &         & $2.97$ & $^{+0.09}_{-0.08}$ &         & $215$ &             $^{+25}_{-32}$ &         & $0.064$ & $^{+0.014}_{-0.014}$ &         & $0.95$ & $^{+0.00}_{-0.04}$ &         & $2.8$ & $^{+0.3}_{-0.2}$ &         & $1.99$ & $^{+0.07}_{-0.07}$ &         & $-11.36$ & $^{+0.01}_{-0.01}$ &         &  $29$ &             $^{+3}_{-3}$ \\
 B167 (\object{SDSSJ090710.07+365957.5}) &         & $2.76$ & $^{+0.07}_{-0.07}$ &         & $183$ &             $^{+39}_{-38}$ &         & $0.024$ & $^{+0.015}_{-0.015}$ &         & $0.00$ & $^{+0.54}_{-0.00}$ &         & $2.2$ & $^{+0.2}_{-0.2}$ &         & $1.84$ & $^{+0.06}_{-0.06}$ &         & $-11.52$ & $^{+0.01}_{-0.01}$ &         &  $33$ &             $^{+3}_{-3}$ \\
 B329 (\object{SDSSJ154806.92+093423.9}) &         & $3.06$ & $^{+0.26}_{-0.11}$ &         & $299$ &             $^{+18}_{-70}$ &         & $0.042$ & $^{+0.017}_{-0.017}$ &         & $0.36$ & $^{+0.60}_{-0.12}$ &         & $3.4$ & $^{+0.5}_{-0.4}$ &         & $2.16$ & $^{+0.10}_{-0.09}$ &         & $-11.67$ & $^{+0.01}_{-0.01}$ &         &  $73$ & $^{+10}_{-\phantom{0}8}$ \\
  B434 (\object{SDSSJ110224.37+025002.8}) &         & $2.82$ & $^{+0.18}_{-0.07}$ &         & $388$ &             $^{+16}_{-96}$ &         & $0.074$ & $^{+0.015}_{-0.014}$ &         & $0.49$ & $^{+0.43}_{-0.49}$ &         & $3.3$ & $^{+0.3}_{-0.3}$ &         & $2.02$ & $^{+0.07}_{-0.06}$ &         & $-11.44$ & $^{+0.01}_{-0.01}$ &         &  $41$ &             $^{+4}_{-3}$ \\
  B458 (\object{SDSSJ104318.29-013502.5}) &         & $2.79$ & $^{+0.21}_{-0.13}$ &         & $411$ & $^{+\phantom{0}38}_{-102}$ &         & $0.020$ & $^{+0.027}_{-0.020}$ &         & $0.76$ & $^{+0.04}_{-0.77}$ &         & $3.5$ & $^{+0.5}_{-0.4}$ &         & $2.00$ & $^{+0.10}_{-0.09}$ &         & $-11.72$ & $^{+0.02}_{-0.02}$ &         &  $83$ &           $^{+12}_{-10}$ \\
   B481 (\object{SDSSJ232229.47+043651.4}) &         & $3.35$ & $^{+0.21}_{-0.12}$ &         & $279$ &             $^{+31}_{-54}$ &         & $0.072$ & $^{+0.016}_{-0.017}$ &         & $0.84$ & $^{+0.11}_{-0.85}$ &         & $4.8$ & $^{+0.6}_{-0.5}$ &         & $2.36$ & $^{+0.10}_{-0.09}$ &         & $-11.33$ & $^{+0.01}_{-0.01}$ &         &  $46$ &             $^{+5}_{-6}$ \\
    B485 (\object{SDSSJ101018.82+302028.1}) &         & $5.05$ & $^{+0.45}_{-0.14}$ &         &  $79$ &   $^{+\phantom{0}4}_{-18}$ &         & $0.026$ & $^{+0.011}_{-0.010}$ &         & $0.38$ & $^{+0.58}_{-0.06}$ &         & $4.1$ & $^{+0.5}_{-0.3}$ &         & $2.96$ & $^{+0.09}_{-0.07}$ &         & $-11.26$ & $^{+0.01}_{-0.01}$ &         &  $33$ &             $^{+4}_{-3}$ \\
  B537 (\object{SDSSJ002810.33+215809.6}) &         & $3.73$ & $^{+0.26}_{-0.17}$ &         & $190$ &             $^{+17}_{-34}$ &         & $0.091$ & $^{+0.012}_{-0.014}$ &         & $0.88$ & $^{+0.08}_{-0.87}$ &         & $4.3$ & $^{+0.6}_{-0.5}$ &         & $2.50$ & $^{+0.12}_{-0.10}$ &         & $-11.33$ & $^{+0.01}_{-0.01}$ &         &  $41$ &             $^{+6}_{-5}$ \\
  B572 (\object{SDSSJ005956.06+313439.2}) &         & $2.65$ & $^{+0.22}_{-0.09}$ &         & $156$ &             $^{+94}_{-88}$ &         & $0.057$ & $^{+0.019}_{-0.018}$ &         & $0.67$ & $^{+0.29}_{-0.15}$ &         & $2.1$ & $^{+0.3}_{-0.2}$ &         & $1.74$ & $^{+0.13}_{-0.09}$ &         & $-11.46$ & $^{+0.01}_{-0.01}$ &         &  $27$ &             $^{+4}_{-3}$ \\
   B576 (\object{SDSSJ140432.38+352258.4}) &         & $3.65$ & $^{+0.08}_{-0.11}$ &         & $204$ &   $^{+19}_{-\phantom{0}7}$ &         & $0.022$ & $^{+0.011}_{-0.010}$ &         & $0.25$ & $^{+0.52}_{-0.06}$ &         & $4.5$ & $^{+0.3}_{-0.3}$ &         & $2.49$ & $^{+0.06}_{-0.07}$ &         & $-11.40$ & $^{+0.01}_{-0.01}$ &         &  $51$ &             $^{+4}_{-4}$ \\
 B598 (\object{SDSSJ141723.34+101245.7}) &         & $2.40$ & $^{+0.05}_{-0.05}$ &         &   $5$ &               $^{+1}_{-1}$ &         & $0.060$ & $^{+0.017}_{-0.017}$ &         & $0.00$ &           $^{+0.01}_{-0.00}$   &         & $1.6$ & $^{+0.1}_{-0.1}$ &         & $1.53$ & $^{+0.03}_{-0.04}$ &         & $-11.55$ & $^{+0.01}_{-0.01}$ &         &  $23$ &             $^{+2}_{-2}$ \\
  B711 (\object{SDSSJ142001.94+124404.8}) &         & $2.77$ & $^{+0.08}_{-0.06}$ &         & $315$ &             $^{+62}_{-21}$ &         & $0.026$ & $^{+0.013}_{-0.013}$ &         & $0.12$ & $^{+0.40}_{-0.12}$ &         & $2.8$ & $^{+0.2}_{-0.2}$ &         & $1.92$ & $^{+0.06}_{-0.06}$ &         & $-11.26$ & $^{+0.01}_{-0.01}$ &         &  $23$ &             $^{+2}_{-2}$ \\
 B733 (\object{SDSSJ144955.58+310351.4}) &         & $2.72$ & $^{+0.06}_{-0.07}$ &         & $273$ &             $^{+97}_{-36}$ &         & $0.002$ & $^{+0.010}_{-0.002}$ &         & $0.95$ & $^{+0.00}_{-0.25}$ &         & $2.6$ & $^{+0.2}_{-0.2}$ &         & $1.83$ & $^{+0.06}_{-0.06}$ &         & $-11.02$ & $^{+0.01}_{-0.01}$ &         &  $12$ &             $^{+1}_{-1}$ \\

B1080 (\object{SDSSJ103357.26-011507.3}) &         & $2.99$ & $^{+0.07}_{-0.08}$ &         & $307$ &             $^{+22}_{-41}$ &         & $0.096$ & $^{+0.015}_{-0.015}$ &         & $0.71$ & $^{+0.18}_{-0.48}$ &         & $3.2$ & $^{+0.3}_{-0.3}$ &         & $2.09$ & $^{+0.06}_{-0.06}$ &         & $-11.56$ & $^{+0.01}_{-0.01}$ &         &  $53$ &             $^{+5}_{-4}$ \\
B1085 (\object{SDSSJ112255.77-094734.9}) &         & $3.38$ & $^{+0.10}_{-0.11}$ &         & $224$ &             $^{+24}_{-13}$ &         & $0.022$ & $^{+0.012}_{-0.013}$ &         & $0.95$ & $^{+0.00}_{-0.13}$ &         & $3.9$ & $^{+0.3}_{-0.3}$ &         & $2.30$ & $^{+0.06}_{-0.07}$ &         & $-11.39$ & $^{+0.01}_{-0.01}$ &         &  $42$ &             $^{+4}_{-4}$ \\

B1139 (\object{SDSSJ180050.86+482424.6}) &         & $2.83$ & $^{+0.10}_{-0.09}$ &         & $116$ &             $^{+54}_{-65}$ &         & $0.029$ & $^{+0.017}_{-0.016}$ &         & $0.17$ & $^{+0.56}_{-0.07}$ &         & $2.1$ & $^{+0.3}_{-0.2}$ &         & $1.85$ & $^{+0.08}_{-0.09}$ &         & $-11.44$ & $^{+0.01}_{-0.01}$ &         &  $25$ &             $^{+3}_{-3}$ \\

\hline
\end{tabular}

\tablefoot{The derivations of the parameters are outlined in Sect.~\ref{subsection:atmos_stellar_params} and \ref{sec:distance_estimation}. The given uncertainties are $1\sigma$.}

\end{sidewaystable*}

\subsection{Comparison of spectroscopic results with \citetalias{2018A&A...615L...5I}}

By reprocessing the spectra (relative flux calibration) we were able to improve the atmospheric parameters of a few stars in comparison with their previous analysis in \citetalias{2018A&A...615L...5I}. For most stars, these differences are well within the given uncertainties, which is reassuring, because it demonstrates that our results are typically independent of the details of the applied analysis strategy. However, two cases exist where the revision of the parameters is significant enough to be mentioned explicitly.

Compared to \citetalias{2018A&A...615L...5I}, the surface gravity of HVS\,5 is now $\sim 0.1$\,dex lower. This yields a distance ($37\pm4$\,kpc) that is $\sim 6$\,kpc larger and hence closer to $45\pm5.2$\,kpc, the value by \cite{2015ApJ...804...49B}. As will be discussed in Sect.~\ref{sec:HVS5}, the interpretation of the origin of HVS\,5 is quite sensitive to its assumed distance.

In \citetalias{2018A&A...615L...5I}, we re-classified B711 as an A-type star ($T_\mathrm{eff} = 9170\pm250$\,K) based on the shape of the Balmer lines and the wealth of metal lines visible in the spectrum. Fitting the flux-calibrated data, however, reveals that the height of the Balmer jump is much better reproduced by a higher $T_\mathrm{eff} = 10\,410\pm120$\,K. Both temperatures can reproduce the spectral shape of the Balmer lines equally well because these lines almost behave the same when spreading out from their peak strength value between 9500--10\,000\,K, which leads to ambiguities in the parameter determination. The presence of many metal lines, which we used as argument for the cooler solution in \citetalias{2018A&A...615L...5I} due to the fact that we were lacking sufficient information about the height of the Balmer jump, now suggests that B711 could be metal rich.

\section{SEDs and spectrophotometric distances}\label{sec:spec-distances}

\begin{figure}
\centering
\includegraphics[width=\linewidth]{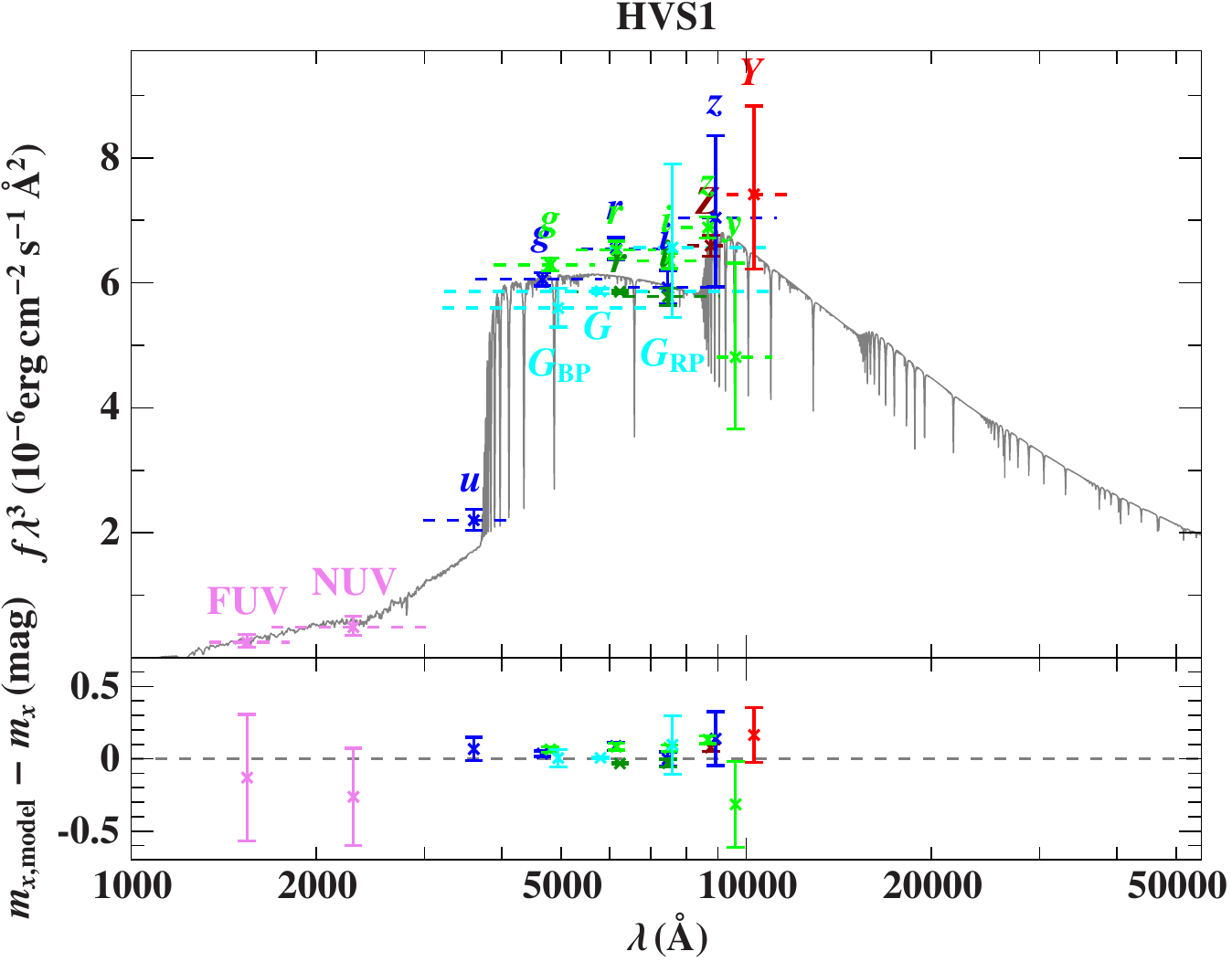}
\caption{
Comparison of synthetic and observed photometry for HVS\,1: The \textit{top panel} shows the SED. The colored data points (GALEX: violet, SDSS: blue; Pan-STARRS: green; {\it Gaia}: cyan; VST-KiDs: dark green; VISTA: dark red; UKIDSS: red) are filter-averaged fluxes which were converted from observed magnitudes (the respective filter widths are indicated by the dashed horizontal lines), while the gray solid line represents a model that is based on the spectroscopic parameters given in Table~\ref{tab:atmos-results}. Only the angular diameter and the color excess were fitted. The flux is multiplied with the wavelength to the power of three to reduce the steep slope of the SED on such a wide wavelength range. The residual panel at the \textit{bottom} shows the differences between synthetic and observed magnitudes.
\label{fig:SED-HVS1-text}}
\end{figure}

As pointed out in Sect.~\ref{sec:introduction}, the {\it Gaia} parallaxes of our program stars, which all are quite far away, are too uncertain to be of any help. Thus, we have to rely on spectrophotometric distances, which require precise photometry to construct SEDs. Angular diameters and interstellar reddening and extinction result from fits of the observed SEDs by synthetic ones. The spectrophotometric distance can then be derived from the angular diameter, the spectroscopic surface gravity, and the mass from evolutionary models.

\subsection{Photometric data}\label{sec:photometric_data}

Photometric data were compiled from a variety of surveys covering the ultraviolet, the optical, and the infrared mainly using the \texttt{VizieR}\footnote{\url{http://vizier.u-strasbg.fr/viz-bin/VizieR}} catalog access tool. The following catalogs were queried: For ultraviolet magnitudes the GALEX catalog \citep[][with corrections from \citealt{2019MNRAS.489.5046W} for the brightest targets]{2017yCat.2335....0B}. Optical photometry came from {\it Gaia} DR2 \citep[][with corrections and calibrations from \citealt{2018A&A...619A.180M}]{2018A&A...616A...4E}, SDSS DR12 \citep{2015ApJS..219...12A}, Pan-STARRS1 \citep{2016arXiv161205560C}, SkyMapper DR1.1 \citep{2019yCat.2358....0W}, BATC \citep{2005JKAS...38..203Z}, VST-ATLAS-DR3 \citep[without the $u$-band due to known zero-point calibration issues]{2015MNRAS.451.4238S}, VST-KiDS-DR3 \citep{2017A&A...604A.134D}, and APASS-DR9 \citep{2015AAS...22533616H}. Infrared magnitudes were taken from UKIDSS-DR9 \citep{2013yCat.2319....0L} and VISTA \citep{2012A&A...548A.119C}. Measurements with large or unknown uncertainties as well as measurements that turned out to be obvious outliers during the fitting with synthetic SEDs have been omitted.

\subsection{Angular diameter and interstellar extinction}\label{sec:sed}

The resulting observed SEDs were compared to synthetic ones based on {\sc Atlas12} model computations (see Sect.~\ref{sec:models}). For each star, the model parameters were fixed to their respective spectroscopic values (see Sect.~\ref{subsection:atmos_stellar_params}), which is why we were left with only two free fitting parameters, the angular diameter $\Theta$ as distance scaling factor and the color excess $E(B-V)$ as indicator for interstellar reddening and extinction. We used the extinction curve by \citet{1999PASP..111...63F} to account for wavelength-dependent reddening and keep the extinction parameter $R_V$ fixed at $3.1$, that is, its typical value for the interstellar medium. The best parameter values and their corresponding uncertainties were obtained via $\chi^2$ minimization, see \citet{2018OAst...27...35H} for details. In the fitting procedure only a small fraction of data had to be dismissed as outliers leaving us with an excellent coverage of the optical spectral range (see Figs.~\ref{fig:SED-HVS1-text}, \ref{fig:SED-HVS22-text}, and \ref{fig:SED-HVS4}-\ref{fig:SED-B1139}) and consequently high precision angular diameters. The color excesses are small ($E(B-V)\leq 0.1$\,mag) as expected for stars at high Galactic latitudes.

\subsection{Spectrophotometric distances}\label{sec:distance_estimation}

The resulting angular diameters $\Theta = 2 R / d$ were combined with the stellar radii $R = (G M / g)^{1/2}$ which are based on the surface gravities $g$ from spectroscopy and the stellar masses derived from evolutionary tracks (see Sect.~\ref{subsection:atmos_stellar_params}). This allowed us to calculate the spectrophotometric distances $d$ which are listed in Table~\ref{tab:stellar_params}. For the candidate pre-ELM WD B598 (see Sect.~\ref{subsection:atmos_stellar_params}), we calculated an additional distance assuming a typical ELM mass of $M=0.25\,M_\odot$, which results in $d=6.5^{+1.0}_{-0.9}$\,kpc. We furthermore examined the BHB case for candidate B576 and determined $d=18.7^{+1.6}_{-1.5}$\,kpc based on the assumption of a typical BHB mass of $M=0.5\,M_\odot$.

\section{Kinematic analysis}\label{sec:kinematics}

The strategy of the kinematic analysis is identical to that of \citetalias{2018A&A...620A..48I}. Complementing our revised radial velocities and spectrophotometric distances with positions and proper motions measured with {\it Gaia} or HST gave us the full six-dimensional phase space information that is needed to calculate the trajectories of the program stars back in time to unravel their (spatial) origin. Unfortunately, proper motion measurements were not available for HVS\,11, HVS\,14, and HVS\,23, which is why the kinematic analysis encompasses only 37 out of the 40 stars in the sample. Propagation of uncertainties in the spectrophotometric distance, radial velocity, and proper motions was achieved using a standard Monte Carlo approach with 1.5\,million realizations that accounts also for the correlation in proper motions as provided by {\it Gaia} DR2.

\subsection{Proper motions}

Proper motions were mainly taken from {\it Gaia} DR2 \citep{2018A&A...616A...2L}. We used the ``renormalized unit weight error'' (RUWE, see \citealt{RUWE}) as given in the ARI {\it Gaia} Data Service\footnote{\url{http://gaia.ari.uni-heidelberg.de/}} as primary indicator for the quality of the astrometric solution. Unlike other quality indicators, RUWE is by design independent of the color of the object, which makes it the best choice when studying blue stars as done here. For all objects in the sample, RUWE is below the recommended value of $1.4$, indicating that the astrometric solutions are well-behaved. For a subsample of 15 stars, pre-{\it Gaia} proper motions measured with the HST instruments WFC3 and ACS were available \citep{2015ApJ...804...49B}. As discussed in \citetalias{2018A&A...620A..48I}, proper motions from both sources are consistent with each other --~except for the outlier B711~-- suggesting that the {\it Gaia} DR2 proper motions of the other 22 program stars are reliable as well. In analogy to \citetalias{2018A&A...620A..48I}, we chose the more precise measurement when {\it Gaia} as well as HST proper motions were available, which implies that we used HST data for HVS\,1, HVS\,10, HVS\,12, and HVS\,13 and {\it Gaia} DR2 data for B711.

\subsection{Galactic gravitational potentials}\label{sec:gravitational_potentials}

The trajectories of all targets were computed in two different Milky Way mass models which primarily differ in the mass and analytic form of the dark matter halo. Model~I is a revision of the popular model by \citet{1991RMxAA..22..255A} and Model~II is based on the flat rotation curve model by \citet{1999MNRAS.310..645W}. Both models have been calibrated using the same observational constraints \citep[for details see][]{2013A&A...549A.137I} and are consistent with latest results based on {\it Gaia} DR2 data (see \citetalias{2018A&A...620A..48I}). In contrast to \citetalias{2018A&A...620A..48I}, we omit Model~III because it predicts escape velocities that are most likely too large. 

\subsection{Places of origin}\label{sec:places_of_origin}

Tracing back the trajectories of the targets also gave us information about their spatial origin within the Galactic plane. To this end, we considered only those out of the 1.5\,million Monte Carlo orbits that cross the Galactic plane within the upper 99\% confidence limit for the respective stellar age. For B598, this limit was set to 100\,Myr, which clearly exceeds its derived flight time and, thus, did not affect the outcome at all.

\subsection{Ejection velocities}

Although the current Galactic rest-frame velocity provides a good first impression of how extreme the underlying disk ejection event might have been, it may still be misleading because it does not account for the intrinsic rotation of the disk. For instance, stars ejected in direction of Galactic rotation may be boosted by more than 200\,km\,s${}^{-1}$ while those ejected against Galactic rotation may be slowed down by the same amount. Consequently, the ejection velocity $\varv_\mathrm{ej,p}$, that is, the Galactic rest-frame velocity at plane intersection corrected for Galactic rotation, is a more useful quantity to look at when studying the nature and origin of runaway and HVSs. In particular, it can help to distinguish between the various disk ejection mechanisms outlined in Sect.~\ref{sec:introduction}, see, for example, \citetalias{2018A&A...620A..48I} and \cite{2019A&A...628L...5I}.

\section{Results of the kinematic analyses}\label{sec:results}

The detailed results of the kinematic analyses are listed in Tables~\ref{tab:kinematic-AS} and \ref{tab:kinematic-TF} for both Galactic mass models considered here. The quantities shown there are based on a right-handed Cartesian Galactic coordinate system in which the Sun is located on the negative $x$-axis and the $z$-axis points to the Galactic north pole. Plane-crossing quantities are labeled by the subscript ``p'' and are based on all orbits that crossed the Galactic plane within the maximum backward integration time, which was set to 15\,Gyr. The Galactic rest-frame velocity $\varv_{\mathrm{Grf}}=(\varv_x^2+\varv_y^2+\varv_z^2)^{1/2}$, the local Galactic escape velocity $\varv_{\mathrm{esc}}$, the Galactocentric radius $r=(x^2+y^2+z^2)^{1/2}$, the ejection velocity $\varv_{\mathrm{ej}}$, and the flight time $\tau_{\mathrm{flight}}$ are listed in addition to Cartesian positions and velocities. 

\subsection{Flight time vs.\ evolutionary time}

\begin{figure}
\centering
\includegraphics[width=\linewidth]{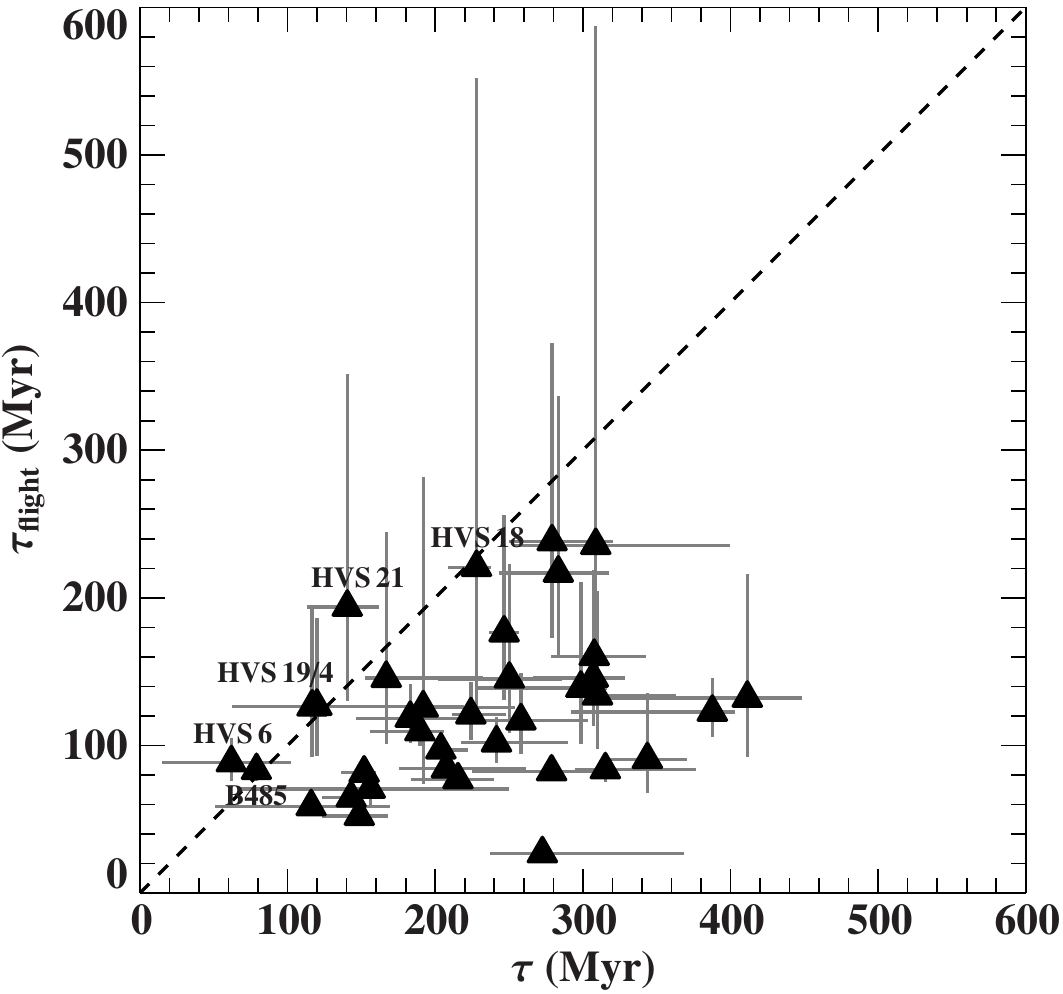}
      \caption{Time of flight from the Galactic plane to the current position in the halo vs.\ inferred evolutionary age assuming a single-star MS nature. The dashed line is the identity line. B598 is omitted in this plot because we cannot derive its MS age (see Sect.~\ref{subsection:atmos_stellar_params}). All stars are consistent with an ejection scenario from the Galactic plane.}
         \label{fig:age_vs_tof}
\end{figure}

In order to check whether the program stars are consistent with an ejection scenario, we compare their flight times, which result from tracing back their orbits to the Galactic plane, to their inferred evolutionary ages; see Fig.~\ref{fig:age_vs_tof}. All stars except B598 have ages that, within uncertainties, exceed their respective flight times, that is, they can reach their present-day position in the Galactic halo within their derived MS lifetimes. The object B598 does not pass this test owing to its location below the ZAMS in the Kiel diagram (see Sect.~\ref{subsection:atmos_stellar_params}), which does not allow for a determination of a reasonable MS age.

\subsection{Places of origin}\label{sec:places_origin}

      \begin{figure*}
   \centering
   \includegraphics[width=\linewidth]{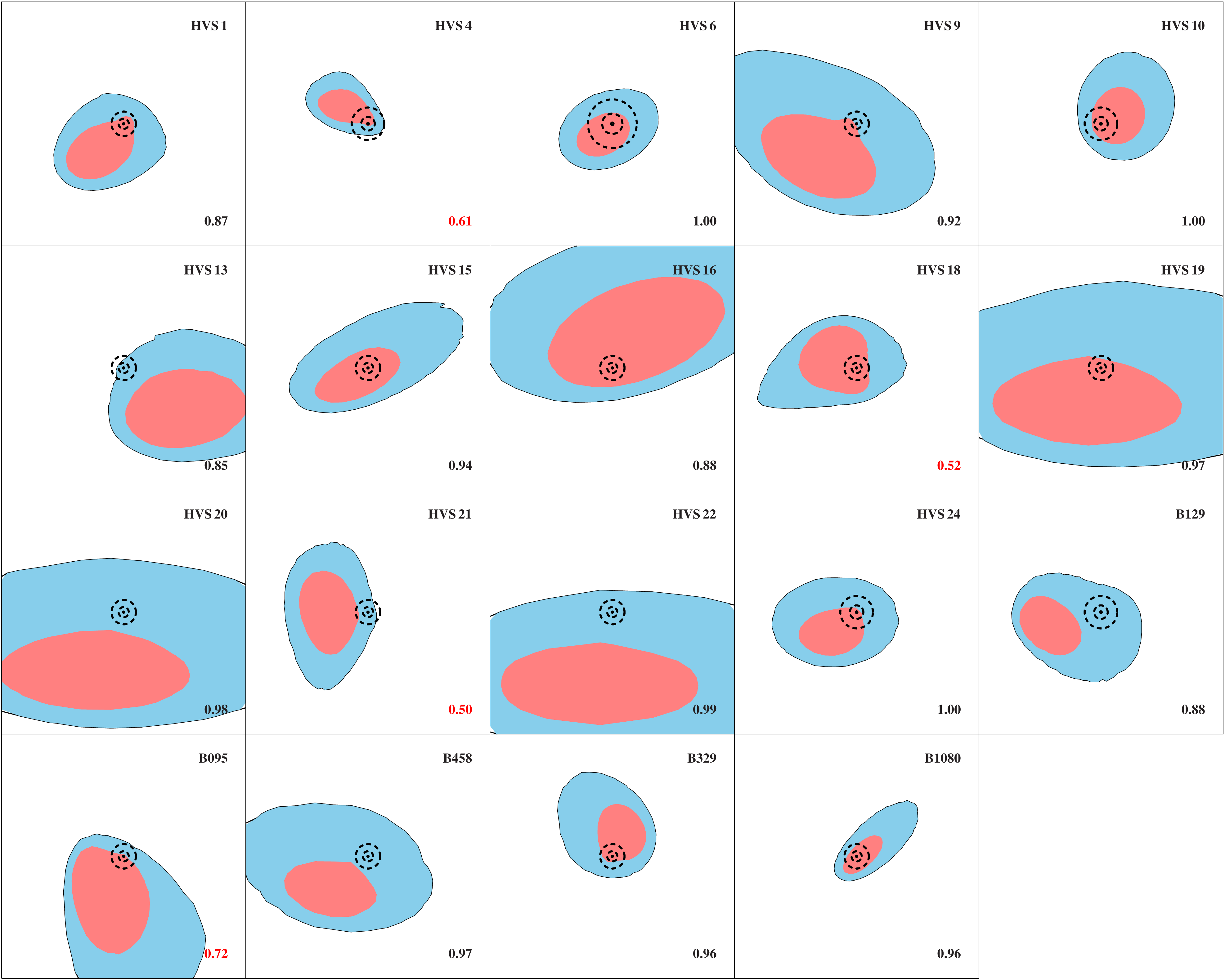}
      \caption{Galactic plane-crossing locations for those objects whose origin is not constrained. The black rimmed, red and blue shaded areas mark regions where 68\% and 95\% ($1\sigma$ and $2\sigma$) of the 1.5\,million Monte Carlo trajectories intersect the Galactic plane. Three black circles with different radii are overplotted for reference: a central circle (solid; 1\,kpc), the solar radius (dashed; 8.3\,kpc), and the Galactic disk (dashed; 20\,kpc). To account for the finite lifetimes of the stars, only orbits that cross the Galactic plane within the upper 99\% confidence limit for the respective stellar age are considered. The number in the lower right corner denotes this fraction of orbits and is displayed in red if the age restriction removes more than 20\% of the Monte Carlo trials.}
         \label{fig:cat4-contours}
   \end{figure*}

In the following, we group the stars in three categories based on their inferred spatial origin. The first consists of stars for which the available data are insufficient to constrain the place of origin because the error contours enclose the entire Galactic disk and even more (Sect.~\ref{sec:unconstrained}). The second category comprises objects whose origin lies far outside the solar circle at the rim of the Galactic disk (Sect.~\ref{sec:rim_origin}). Finally, the third group consists of the best constrained objects (Sect.~\ref{sec:disk_origin}). Particularly interesting stars are discussed in separate paragraphs. Unless stated otherwise, numbers are always taken from Table~\ref{tab:kinematic-AS}, that is, they are based on Model~I for the gravitational potential. A comparison with Model~II is presented in Sect. \ref{sec:bound}. 

\subsubsection{Unconstrained origin}\label{sec:unconstrained}

Owing to the large uncertainties for the kinematic input parameters that are mainly caused by the objects' huge distances, the origin of 19 stars in the sample is not really constrained; see Fig.~\ref{fig:cat4-contours}. All of them could possibly stem from the Galactic disk, and in particular from the GC. Because the vast majority of trajectories of most stars intersects the Galactic plane outside of the 20\,kpc circle which we use here as a rough boundary for the Galactic disk, the given ejection velocities from the plane should be considered with caution. With the outstanding exception of HVS\,22 (see below), the current Galactic rest-frame velocities in this group lie between $\varv_\mathrm{Grf}=390^{+150}_{-\phantom{0}80}$\,km\,s${}^{-1}$ (B129) and $\varv_\mathrm{Grf}=970^{+480}_{-360}$\,km\,s${}^{-1}$ (HVS\,20). Apart from HVS\,21 ($\varv\sin(i)=47^{+80}_{-47}$\,km\,s${}^{-1}$) and B329 ($\varv\sin(i)=58^{+35}_{-49}$\,km\,s${}^{-1}$), all stars in this group exhibit projected rotational velocities that are significantly larger than 50\,km\,s${}^{-1}$, which hints at a MS nature. Moreover, the boundness probability of all stars except B1080 ($P_\mathrm{b}=67$\%) is lower than 50\%, for most of them it is even equal to zero. Although the most plausible explanation for the presence of those apparently young massive stars in the far-distant Galactic halo is the ejection from the Galactic disk by a very powerful mechanism, the precision of the currently available data is just not high enough to definitely proof or discard it.

\paragraph{HVS\,22}\label{sec:hvs22}

\begin{figure}
\centering
\includegraphics[width=\linewidth]{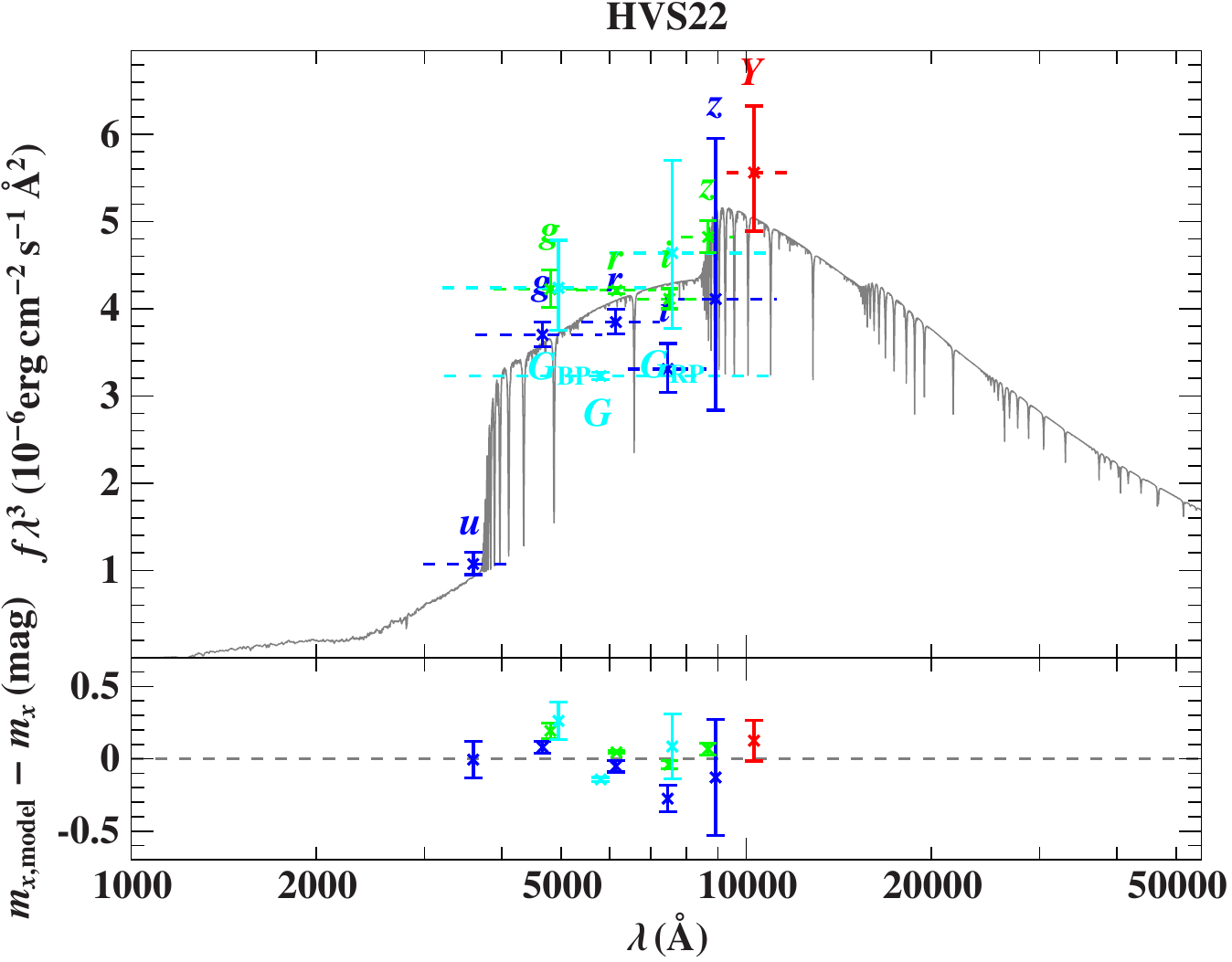}\caption{Same as Fig.~\ref{fig:SED-HVS1-text} but for HVS22. SDSS: blue; Pan-STARRS: green; UKIDSS: red; {\it Gaia}: cyan.}\label{fig:SED-HVS22-text}
\end{figure}

With a current Galactic rest-frame velocity of $\varv_\mathrm{Grf} = 1530^{+690}_{-560}$\,km\,s${}^{-1}$, HVS\,22 is the star with the most outstanding kinematic properties in the sample. However, its place of origin is completely unconstrained; see Fig.~\ref{fig:cat4-contours}. The extreme velocity of HVS\,22 is a consequence of its large inferred spectrophotometric distance of $d=99^{+16}_{-14}$\,kpc. Key ingredients for the distance determination are the spectroscopic surface gravity and the assumed MS mass. Although the S/N of the available spectra is relatively low, there is currently no indication in the spectral fit nor in the SED (see Fig.~\ref{fig:SED-HVS22-text}) that the derived surface gravity might be incorrect. Similarly, the projected rotation of $\varv\sin(i)=156^{+32}_{-31}$\,km\,s${}^{-1}$ as well as the fact that the position of the star in the Kiel diagram (see Fig.~\ref{fig:evol-tracks-dorman}) is well below the ZAHB support the idea that HVS\,22 is indeed a MS star.

\subsubsection{Possible outer rim origin}\label{sec:rim_origin}

   \begin{figure*}
   \centering
   \includegraphics[width=\linewidth]{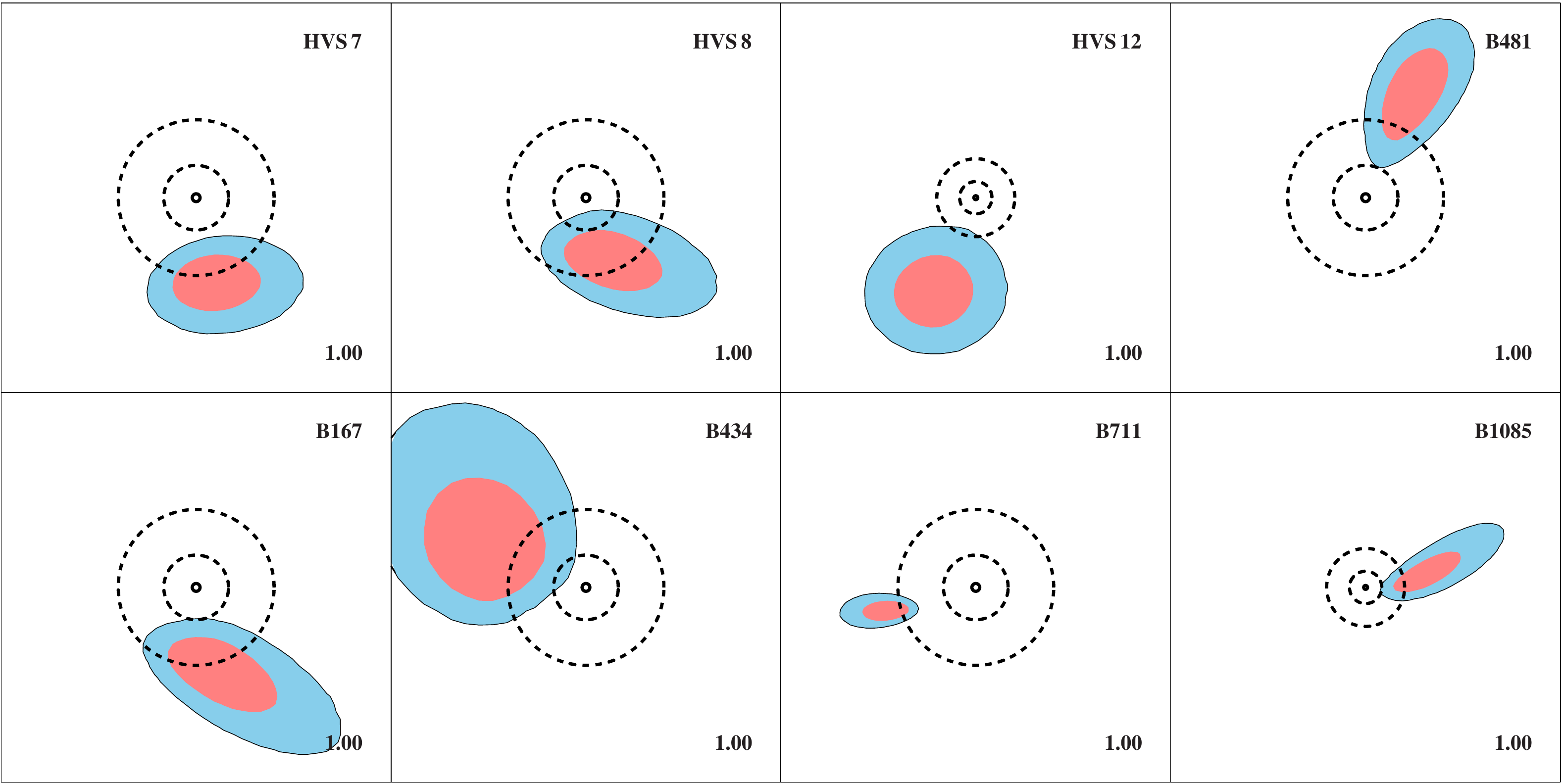}
      \caption{Same as Fig.~\ref{fig:cat4-contours} but for objects that may stem from the outer rim of the Galactic disk or not from the disk at all.}
         \label{fig:cat2-contours}
   \end{figure*}

For eight stars in the sample, the precision of the kinematic analysis is sufficient to conclude that those stars either stem from the outer rim of the Galactic disk or not from the disk at all; see Fig.~\ref{fig:cat2-contours}. The best candidate for an ejection from the disk is HVS\,8 albeit its ejection velocity of $\varv_\mathrm{ej,p}=440^{+40}_{-20}$\,km\,s${}^{-1}$ is close to $\sim 500$\,km\,s${}^{-1}$, that is, to the upper limit of what classical disk ejection mechanisms are capable of. For all other stars in this group, the majority of the Monte Carlo trajectories intersects the Galactic plane outside of the 20\,kpc circle. Consequently and similar to the previous group, it is unclear whether their derived ejection velocities from the Galactic plane are physically meaningful at all. If the stars indeed came from the very outskirts of the Galactic disk, the ejection velocities of HVS\,7 ($\varv_\mathrm{ej,p}=530\pm30$\,km\,s${}^{-1}$), HVS\,12 ($\varv_\mathrm{ej,p}=610^{+80}_{-70}$\,km\,s${}^{-1}$), B434 ($\varv_\mathrm{ej,p}=580\pm20$\,km\,s${}^{-1}$), B481 ($\varv_\mathrm{ej,p}=630^{+50}_{-40}$\,km\,s${}^{-1}$), and B576 ($\varv_\mathrm{ej,p}=740^{+60}_{-50}$\,km\,s${}^{-1}$) would more or less clearly exceed this aforementioned limit and hence hint at the existence of another powerful but yet neglected ejection channel, see, e.g., \citetalias{2018A&A...620A..48I} and \cite{2019A&A...628L...5I}. An alternative explanation would be that these objects either do not originate in the Galactic disk, for example, because they are of extragalactic origin, or that they are not MS stars, which would render our spectrophotometric distance estimation incorrect. However, all stars in this group either rotate fast or lie outside of the BHB band of the Kiel diagram (see Fig.~\ref{fig:evol-tracks-dorman}), which corroborates the idea that they are indeed MS stars. 

Assuming a MS nature, only three stars (B167, B434, B711) have boundness probabilities larger than 50\%, that is, most objects in this group are likely unbound.

\subsubsection{Disk origin}\label{sec:disk_origin}

      \begin{figure*}
   \centering
   \includegraphics[width=\linewidth]{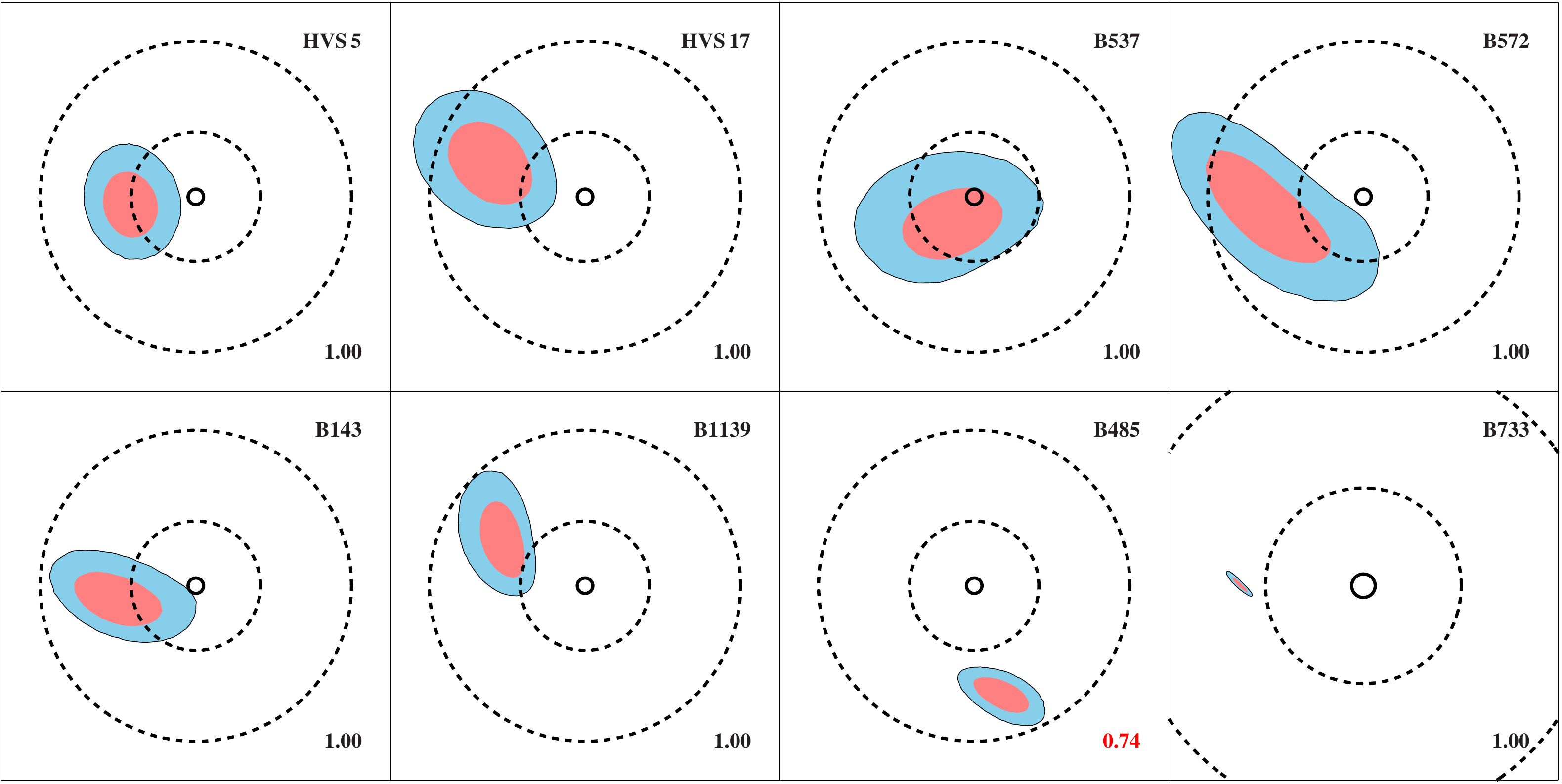}
      \caption{Same as Fig.~\ref{fig:cat4-contours} but for objects for which an origin in the Galactic disk is very likely.}
         \label{fig:cat1-contours}
   \end{figure*}

\begin{figure}
   \centering
   \includegraphics[width=\linewidth]{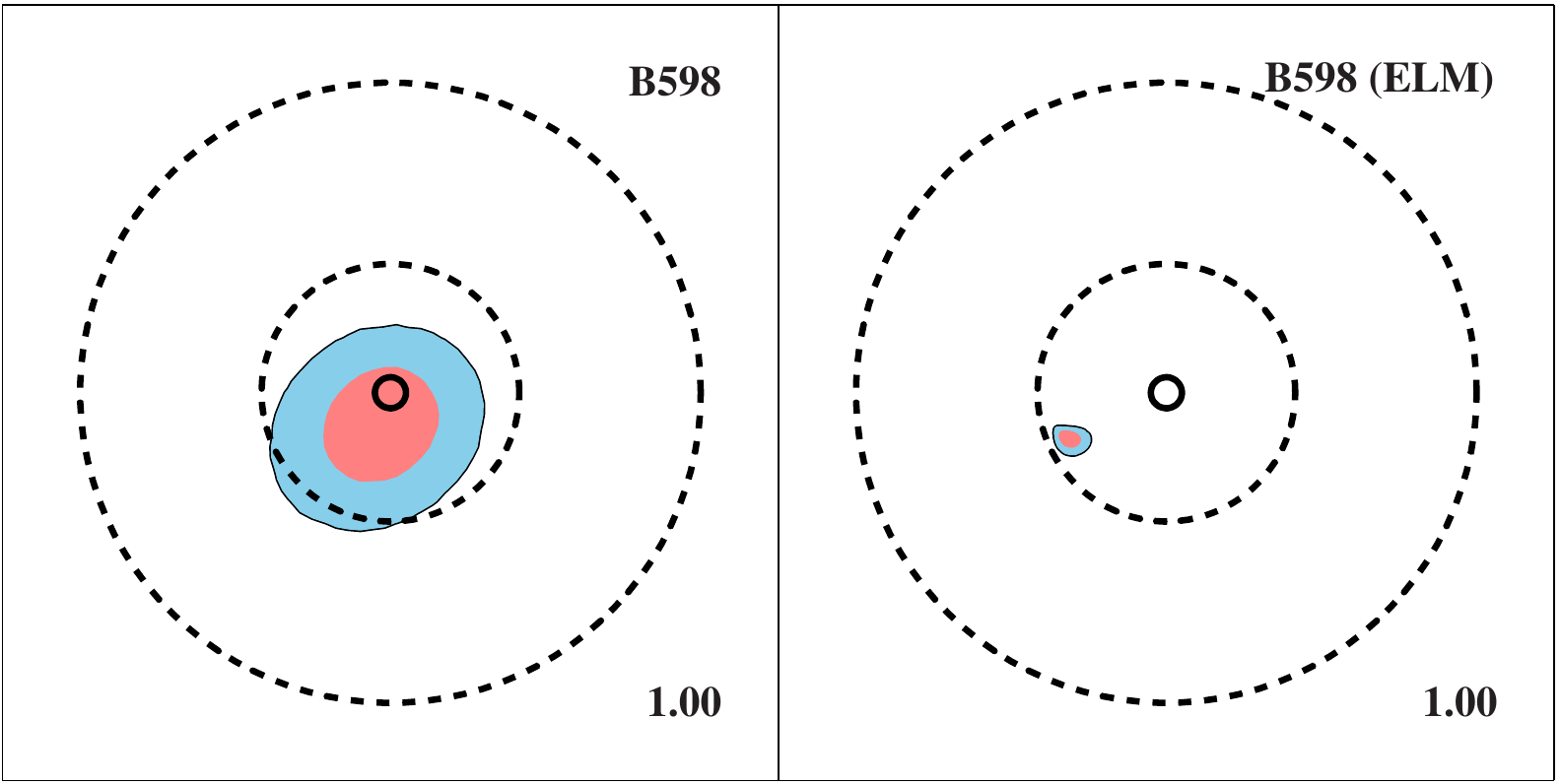}
      \caption{Same as Fig.~\ref{fig:cat4-contours} but only for B598. The sole difference in the two panels is the underlying distance, which is $23\pm2$\,kpc assuming a MS nature (\textit{left panel}) and $6.5^{+1.0}_{-0.9}$\,kpc assuming a pre-ELM WD nature (\textit{right panel}). A three-dimensional representation of the trajectory for the ELM version is shown in Fig.~\ref{fig:B598-cube}.}
         \label{fig:B598-contours}
\end{figure}

\begin{figure}
   \centering
   \includegraphics[width=\linewidth]{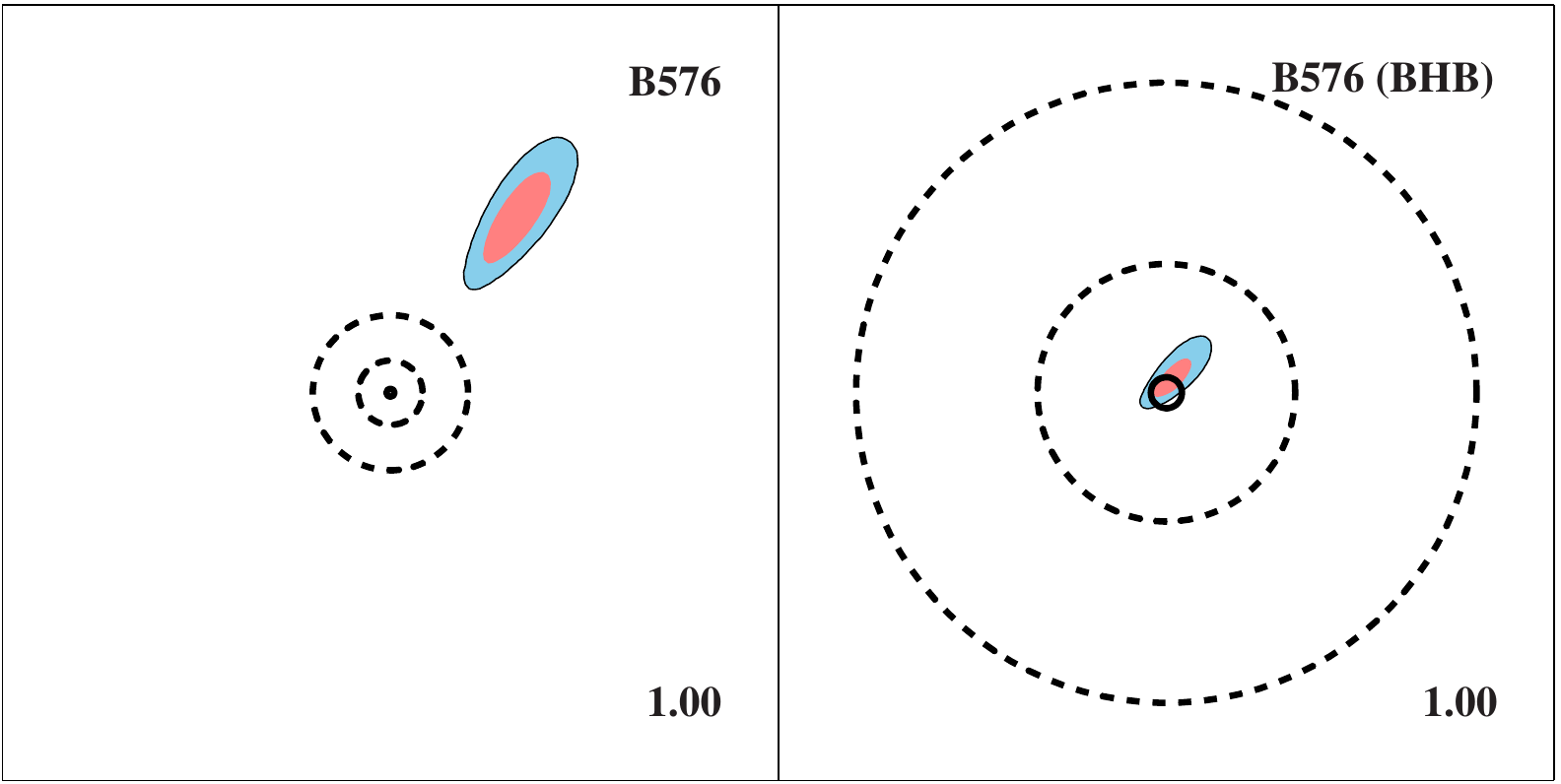}
      \caption{Same as Fig.~\ref{fig:cat4-contours} but only for B576. The sole difference in the two panels is the underlying distance, which is $51\pm4$\,kpc assuming a MS nature (\textit{left panel}) and $18.7^{+1.6}_{-1.5}$\,kpc assuming a BHB nature (\textit{right panel}). While the MS option suggests an extragalactic origin, the BHB option points to a GC origin. A three-dimensional representation of the trajectory for the BHB version is shown in Fig.~\ref{fig:B576-cube}.}
         \label{fig:B576-contours}
\end{figure}

For ten stars in the sample, the precision of the astrometric input data is high enough to conclude that their spatial origin is very likely located within the Galactic disk; see Figs.~\ref{fig:cat1-contours}, \ref{fig:B598-contours}, and \ref{fig:B576-contours}. It is worthwhile to comment on particularly interesting individual objects and we shall also discuss the case of B576 here, because, assuming a BHB nature, its place of origin is well constrained.

\paragraph{B733}\label{sec:B733}

With a derived spectrophotometric distance of $12\pm1$\,kpc, B733 is the second closest object in the sample. Combined with the unprecedented astrometric precision of {\it Gaia}, it is possible to pinpoint the star's place of origin to a narrow region close but slightly outside of the solar radius; see Fig.~\ref{fig:cat1-contours}. The derived ejection velocity $\varv_\mathrm{ej,p}=470\pm10$\,km\,s${}^{-1}$ is comparable to the fastest known disk runaway stars (see, e.g., \citealt{2011MNRAS.411.2596S} and \citealt{2019A&A...628L...5I}). Elemental abundances could help us to better understand the object. Unfortunately, the star rotates so fast ($\varv\sin(i)=278^{+12}_{-10}$\,km\,s${}^{-1}$) that a high-precision abundance analysis will be almost impossible even if spectra of much better quality were available. 

\paragraph{HVS\,17, B485, and B1139}

The places of origin for these stars are rather well constrained to lie between the solar circle and the outer rim of the Galaxy (20\,kpc). Their ejection velocities $\varv_\mathrm{ej,p}=270$--430\,km\,s${}^{-1}$ are also comparable to the high-velocity tail of the sample by \cite{2011MNRAS.411.2596S}.

\paragraph{B143 and B572}
The Galactic plane-crossing contours for both stars locate their origin somewhat beyond the solar circle. However, the 2$\sigma$ contours come close to the GC. Hence, the possibility of a GC origin should not be completely dismissed. Ejection velocities of 
$\varv_\mathrm{ej,p}=410^{+60}_{-50}$ and $310^{+60}_{-40}$\,km\,s${}^{-1}$ place them among the fastest disk runaways known.

\paragraph{B537}

This star is the only one that likely originates in the inner disk, that is, inside the solar circle. More precise proper motions are needed to confirm or rule out an origin in the GC. The ejection velocity ($\varv_\mathrm{ej,p}=460^{+210}_{-\phantom{0}90}$\,km\,s${}^{-1}$) is comparable to the most extreme disk runaways known to date.

\paragraph{HVS\,5}\label{sec:HVS5}

\begin{figure}
   \centering
   \includegraphics[width=\linewidth]{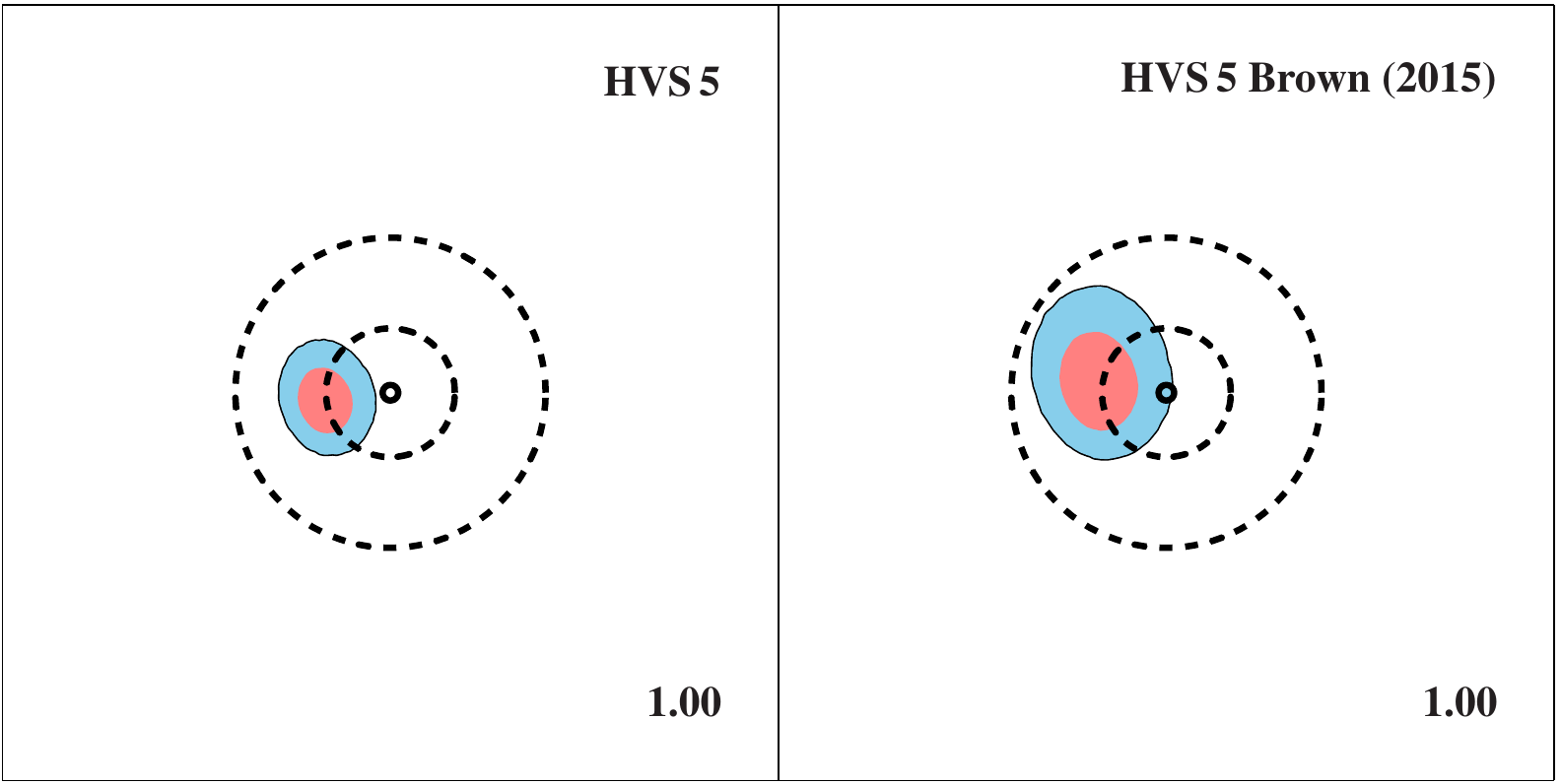}
      \caption{Same as Fig.~\ref{fig:cat4-contours} but only for HVS\,5. The sole difference in the two panels is the underlying distance, which is our value of $37\pm4$\,kpc for the \textit{left panel} and $45\pm5.2$\,kpc \citep{2015ApJ...804...49B} for the \textit{right panel}. The smaller distance renders an origin in the GC even more unlikely.}
         \label{fig:hvs5-contours}
\end{figure}

This object is one of the most interesting program stars because it is both, very extreme but at the same time relatively well constrained. Nevertheless, the conclusions about its spatial origin diverge. While \cite{2018ApJ...866...39B} argue for an ejection from the GC by the Hills mechanism, we discarded this option in \citetalias{2018A&A...620A..48I} because the GC was not within the region where 95\% of all Monte Carlo orbits intersected the Galactic plane. A major difference between the two kinematic analyses was the assumed distance. \cite{2018ApJ...866...39B} probably used $45\pm5.2$\,kpc \citep{2015ApJ...804...49B} while we derived $31.2^{+3.2}_{-2.5}$\,kpc. Our revised atmospheric parameters now yield $37\pm4$\,kpc, which is closer to the value by \cite{2015ApJ...804...49B}. In Fig.~\ref{fig:hvs5-contours}, we illustrate the impact of the different distance estimates on the outcome of the kinematic analysis. Even though the distance estimates are now in better agreement, we can still rule out the GC with more than $2\sigma$ confidence. With an ejection velocity of $\varv_\mathrm{Grf}=670^{+80}_{-60}$\,km\,s${}^{-1}$, HVS\,5 is clearly above the limit of what classical scenarios are capable of.

\paragraph{B598}\label{sec:B598}

\begin{figure}
\centering
\includegraphics[width=\linewidth]{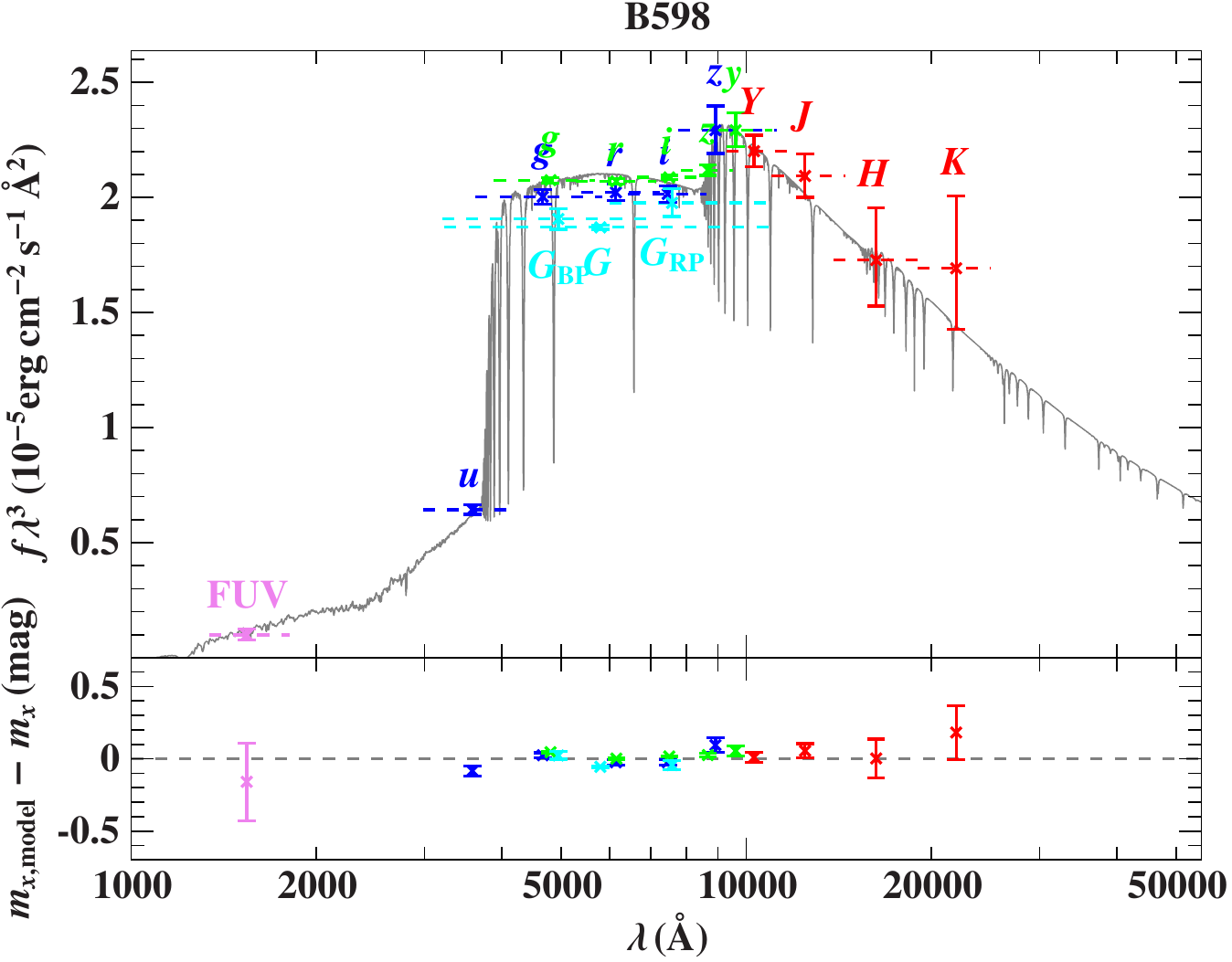}\caption{Same as Fig.~\ref{fig:SED-HVS1-text} but for B598. SDSS: blue; Pan-STARRS: green; UKIDSS: red; {\it Gaia}: cyan; GALEX: violet.}\label{fig:SED-B598-text}
\end{figure}
\begin{figure}
   \centering
   \includegraphics[width=\linewidth]{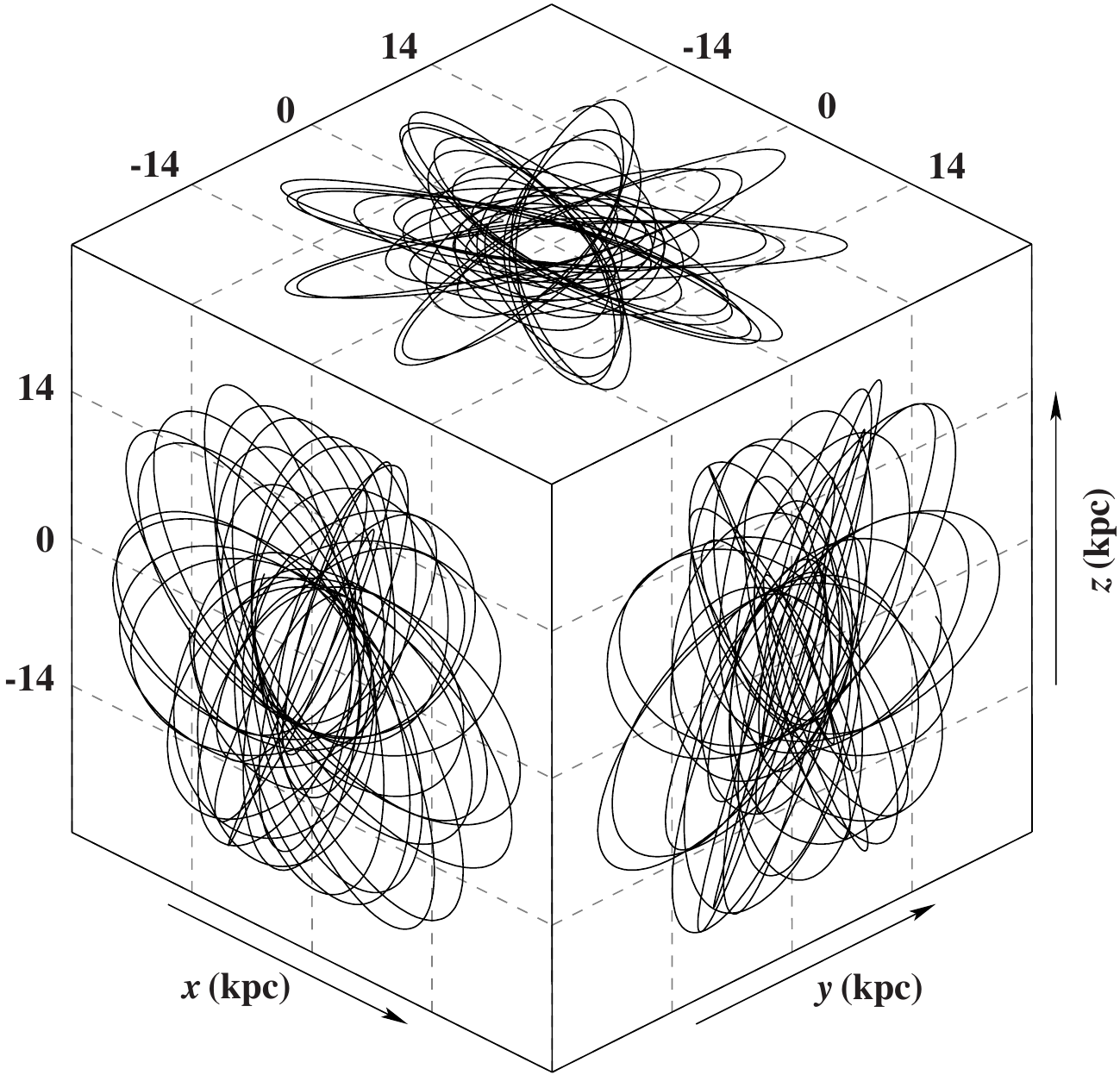}
      \caption{Three-dimensional orbit of B598 assuming a typical (pre-)ELM WD mass of $0.2\,M_\odot$ in the Galactic Cartesian coordinate system introduced in Sect.~\ref{sec:results}. The orbit is calculated 15\,Gyr back in time using Model~I and is typical of a halo star.}
         \label{fig:B598-cube}
\end{figure}

As already outlined in Sect.~\ref{subsection:atmos_stellar_params}, B598 is a puzzling object because its measured $\log(g) = 4.52\pm0.05$ seems to be too high to be compatible with a MS nature. The SED of the target (Fig.~\ref{fig:SED-B598-text}), which covers observations from the far ultraviolet to the infrared, shows that there is no reason to doubt our spectroscopically derived atmospheric parameters. Elemental abundances could be the key to unravel the nature of B598. However, the spectral smearing due to the high projected rotation ($\varv\sin(i)=192^{+18}_{-15}$\,km\,s${}^{-1}$) will make this a very challenging task and require spectra of superb quality. For the time being, we note that if B598 was a MS star, it would be one of our best candidates for the ejection by the Hills mechanism given its large ejection velocity ($\varv_\mathrm{ej,p}=610\pm50$\,km\,s${}^{-1}$) and its inferred place of origin, which encloses the GC; see Fig.~\ref{fig:B598-contours}. However, as discussed in Sect.~\ref{subsection:atmos_stellar_params}, the star is likely an evolved low mass star of $\approx0.2\,M_\odot$ in transition from the red giant branch to a low mass helium-core WD. The corresponding trajectory (see Fig.~\ref{fig:B598-cube}) would then be typical of a halo star.

\paragraph{B576}\label{sec:B576}

\begin{figure}
   \centering
   \includegraphics[width=\linewidth]{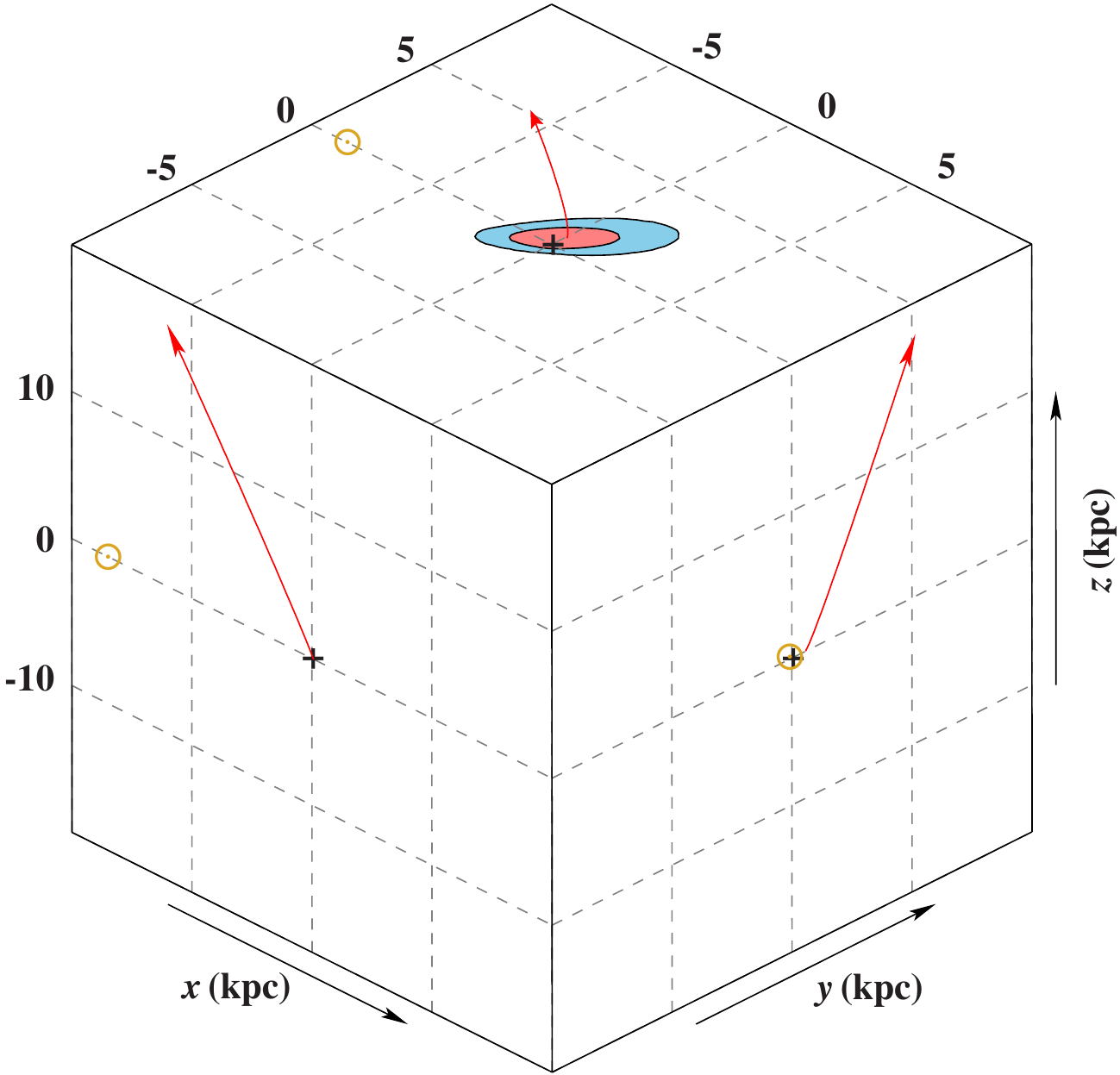}
      \caption{Three-dimensional orbit (red line; the arrow indicates the current position of the star) of B576 assuming a typical BHB mass of $0.5\,M_\odot$ in the Galactic Cartesian coordinate system introduced in Sect.~\ref{sec:results}. The meaning of the shaded areas is identical to Fig.~\ref{fig:cat4-contours}. The positions of the Sun and the GC are marked by a yellow $\sun$ and a black $+$, respectively.}
         \label{fig:B576-cube}
\end{figure}

Assuming a MS nature for B576, it would be the only star in the sample for which an origin in the Galactic disk could be ruled out with more than $2\sigma$ confidence (see Fig.~\ref{fig:B576-contours}). Its current Galactic rest-frame velocity $\varv_\mathrm{Grf}=680^{+70}_{-60}$\,km\,s${}^{-1}$ would then render the object clearly unbound, which would imply that it were just passing through our Milky Way. However, the measured rotational velocity of $\varv\sin(i)=47^{+15}_{-16}$\,km\,s${}^{-1}$ in combination with the inferred atmospheric parameters, which place the object right between the ZAHB and the TAHB in the Kiel diagram (see Fig.~\ref{fig:evol-tracks-dorman}), make it much more plausible that B576 is actually a low-mass BHB star, which would yield a smaller spectrophotometric distance ($18.7^{+1.6}_{-1.5}$\,kpc) and hence a lower current Galactic rest-frame velocity ($\varv_\mathrm{Grf}=330\pm10$\,km\,s${}^{-1}$). Interestingly, the resulting trajectory (see Figs.~\ref{fig:B576-cube}) would have almost no angular momentum and thus very closely pass the GC, which would be quite uncommon for a halo star. A possible explanation for the lack of angular momentum would be that the star stems from the central region of the Milky Way. The inferred ejection velocity ($\varv_\mathrm{ej,p}=730\pm80$\,km\,s${}^{-1}$) would strongly hint at the Hills mechanism. Combined with how precisely the location of Galactic plane-crossing is known (see Fig.~\ref{fig:B576-contours}), B576 could be the second star after S5-HVS1 \citep{2020MNRAS.491.2465K} for which an origin in the GC could be confirmed. Despite its large ejection velocity, B576 would be gravitationally bound to the Milky Way due to the strong deceleration in the bulge region. Follow-up observations with large telescopes are needed to unravel the nature of this interesting object, for example, in order to determine whether the chemical composition in the star's atmosphere is characteristic of BHB stars.

\subsubsection{Bound Probabilities in different Galactic mass models}\label{sec:bound}

The original MMT HVS sample identified 42 B-type stars whose Galactic rest-frame radial velocity exceeds $+275$\,km\,s${}^{-1}$, sixteen of which were considered bound to the Galaxy as indicated by their name starting with the letter ``B''. With new proper motions from {\it Gaia} DR2 being available, it is worthwhile to reconsider the bound probabilities of the sample. Whether a star is bound to the Galaxy or not is determined by the Galactic potential, in particular the mass of the Galactic dark matter halo. 

The (analytic) representations for the different Galactic components that we use are consistent with various {\it Gaia} DR2 based studies (see Sect.~\ref{sec:gravitational_potentials}). Except for three stars (B485, B576, B1085), all stars originally considered bound likely remain bound (probability $P_\mathrm{b} > $ 5\%) irrespective of the choice of the Galactic mass model (see Tables~\ref{tab:kinematic-AS} and \ref{tab:kinematic-TF}). With $P_\mathrm{b}$ = 2\%, B485 and B1085 would likely be unbound in the lighter mass model II. B576 would be clearly unbound ($P_\mathrm{b}$ = 0\%) in both Galactic potentials if it were a MS star. However, it is more probable that the star is actually an evolved star of low mass that is bound to the Galaxy (Sect.~\ref{sec:B576}).

It is worthwhile to also reconsider the stars originally considered unbound. We find two objects (HVS\,15, HVS\,24) that are possibly bound ($P_\mathrm{b} >$ 5\%) in both Galactic potentials and another four (HVS\,7, HVS\,8, HVS\,16, HVS\,17) for which this is the case at least in Model~I.

\subsection{Discussion}\label{sec:discussion}

The main goal of our investigation was to pin down the place of origin of the program stars, determine their ejection velocities, and identify the ejection channel. The Hills mechanism would require the stars to be ejected from the GC. As demonstrated in Sect.~\ref{sec:unconstrained}, data quality is insufficient to provide constraints for 19 stars. When more precise measurements are available (Sects.~\ref{sec:rim_origin} and \ref{sec:disk_origin}) the favored places of origin are in the Galactic disk, for some it may even be its outer part rather than the inner. The GC is actually excluded for 16 stars with a significance of 2$\sigma$ or more. There remain two objects (B537 and B576) that may have been ejected from the GC. In particular, the star B576, which is most likely a BHB star, appears to be ejected close to the GC.

Ejection velocities have been derived for all stars of the kinematic sample. Those discussed in Sects.~\ref{sec:unconstrained} and \ref{sec:rim_origin}, however, have to be taken with a grain of salt because their places of origin are quite uncertain. Therefore we restrain the discussion here to the most reliable results (see Sect.~\ref{sec:disk_origin}). 
The ejection velocities for the eight stars (HVS\,5, HVS\,17, B143, B485, B572, B598, B733, and B1139) for which we excluded an origin in the GC and favor the Galactic disk, range from 270\,km\,s$^{-1}$ to 670\,km\,s$^{-1}$ with HVS\,5 being the fastest at $670^{+80}_{-60}$\,km\,s$^{-1}$. Ejection velocities in excess of $\sim 500$\,km\,s$^{-1}$
are at variance with predictions of classical scenarios (binary supernova or dynamical cluster ejection; see \citealt{2019A&A...628L...5I} for a detailed discussion). Interaction with intermediate-mass black holes have been suggested as a viable ejection process. However, evidence for the existence of intermediate-mass black holes is lacking. 

The potential BHB star B576 is of particular interest because its ejection velocity of 730$\pm$80\,km\,s$^{-1}$ exceeds even that of HVS\,5. Combined with its probable GC origin, this suggests that B576 has been ejected by the Hills mechanism. The highest velocity in the sample is that of HVS\,22, which exceeds 1000\,km\,s$^{-1}$. However, its place of origin remains unconstrained. Nevertheless, the extraordinarily high speed favors the Hills mechanism and, hence, the GC. Its velocity is solely superseded by S5-HVS1, which is the only HVS for which a GC origin has been inferred beyond any reasonable doubt \citep{2020MNRAS.491.2465K}.

\section{Summary and conclusion}\label{sec:conclusion}
 
We carried out an extensive analysis of the sample of HVSs of \citet{2014ApJ...787...89B} using their spectra taken with the MMT, which we flux calibrated anew. State-of-the-art model atmospheres that take non-LTE effects into account were used to perform quantitative spectroscopic analyses of 40 HVS candidates. Applying a well-tested fitting technique, we derived effective temperatures, surface gravities, and projected rotational velocities. The location of the stars on the predicted MS band along with their high projected rotational velocities supports the MS nature of all but two objects. B576 turns out to be most likely a BHB star and B598 is probably a very low-mass (0.2\,$M_\sun$) stripped red giant star evolving into a helium-core WD. Comparing the atmospheric parameters to predictions from evolutionary models, masses and ages were derived. SEDs were constructed to derive the spectrophotometric distances, which are the most important ingredients for a kinematic study but cannot be measured precisely enough by {\it Gaia} because the objects are too distant. However, the second data release of the {\it Gaia} mission provided proper motions of unprecedented precision and accuracy. Following the procedure already applied to 14 HVSs in \citetalias{2018A&A...620A..48I}, we studied the trajectories of 37 HVS candidates in two different Galactic gravitational potentials to trace their place of origin in the Galaxy. While the available data are still not precise enough to constrain the place of origin for 19 program stars, a group of eight stars unexpectedly appears to come from the outer rim of the Galactic disk. Nine stars (including B576) are identified to stem from the Galactic disk while B598 shows typical kinematics of a halo star. For almost all targets with reasonably well-constrained spatial origin, the GC is discarded as possible place of origin. The most notable exception is the BHB star B576, the place of origin of which coincides very well with the GC. In addition, its very high ejection velocity of 730$\pm$80 km\,s$^{-1}$ points to the Hills' slingshot as the most likely mechanism for acceleration. HVS\,22 is by far the most extreme object in the sample. Although its origin is unconstrained, its current very high Galactic rest-frame velocity of $1530^{+690}_{-560}$\,km\,s$^{-1}$ hints at the Hills mechanism as the most plausible explanation.

\begin{acknowledgements}
We thank Warren Brown for providing the MMT spectra, Stephan Geier for organizing the X-shooter spectrum of B1085, and John E.\ Davis for the development of the {\sc slxfig} module used to prepare the figures in this paper.
Based on observations made with ESO Telescopes at the La Silla Paranal Observatory under programme ID 0102.D-0092(A). 
AI and UH acknowledge funding by the Deutsche Forschungsgemeinschaft (DFG) through grant IR190/1-1 and HE1356/71-1.
This research has made use of the Keck Observatory Archive (KOA), which is operated by the W.\ M.\ Keck Observatory and the NASA Exoplanet Science Institute (NExScI), under contract with the National Aeronautics and Space Administration. 
This research has made use of NASA’s Astrophysics Data System.
This research has made use of the VizieR catalogue access tool, CDS, Strasbourg, France.
This work has made use of data from the European Space Agency (ESA) mission {\it Gaia} (\url{https://www.cosmos.esa.int/gaia}), processed by the {\it Gaia} Data Processing and Analysis Consortium (DPAC, \url{https://www.cosmos.esa.int/web/gaia/dpac/consortium}). Funding for the DPAC has been provided by national institutions, in particular the institutions participating in the {\it Gaia} Multilateral Agreement. 
Funding for SDSS-III has been provided by the Alfred P. Sloan Foundation, the Participating Institutions, the National Science Foundation, and the U.S.\ Department of Energy Office of Science. The SDSS-III web site is \url{http://www.sdss3.org/}. SDSS-III is managed by the Astrophysical Research Consortium for the Participating Institutions of the SDSS-III Collaboration including the University of Arizona, the Brazilian Participation Group, Brookhaven National Laboratory, University of Cambridge, University of Florida, the French Participation Group, the German Participation Group, the Instituto de Astrofisica de Canarias, the Michigan State/Notre Dame/JINA Participation Group, Johns Hopkins University, Lawrence Berkeley National Laboratory, Max Planck Institute for Astrophysics, New Mexico State University, New York University, Ohio State University, Pennsylvania State University, University of Portsmouth, Princeton University, the Spanish Participation Group, University of Tokyo, University of Utah, Vanderbilt University, University of Virginia, University of Washington, and Yale University. 
The Pan-STARRS1 Surveys were made possible through contributions by the Institute for Astronomy, the University of Hawaii, the Pan-STARRS Project Office, the Max-Planck Society and its participating institutes, the Max Planck Institute for Astronomy, Heidelberg and the Max Planck Institute for Extraterrestrial Physics, Garching, The Johns Hopkins University, Durham University, the University of Edinburgh, the Queen's University Belfast, the Harvard-Smithsonian Center for Astrophysics, the Las Cumbres Observatory Global Telescope Network Incorporated, the National Central University of Taiwan, the Space Telescope Science Institute, and the National Aeronautics and Space Administration under Grant No. NNX08AR22G issued through the Planetary Science Division of the NASA Science Mission Directorate, the National Science Foundation Grant No. AST-1238877, the University of Maryland, Eotvos Lorand University (ELTE), and the Los Alamos National Laboratory.The Pan-STARRS1 Surveys are archived at the Space Telescope Science Institute (STScI) and can be accessed through MAST, the Mikulski Archive for Space Telescopes. Additional support for the Pan-STARRS1 public science archive is provided by the Gordon and Betty Moore Foundation. 
The national facility capability for SkyMapper has been funded through ARC LIEF grant LE130100104 from the Australian Research Council, awarded to the University of Sydney, the Australian National University, Swinburne University of Technology, the University of Queensland, the University of Western Australia, the University of Melbourne, Curtin University of Technology, Monash University and the Australian Astronomical Observatory. SkyMapper is owned and operated by The Australian National University's Research School of Astronomy and Astrophysics. The survey data were processed and provided by the SkyMapper Team at ANU. The SkyMapper node of the All-Sky Virtual Observatory (ASVO) is hosted at the National Computational Infrastructure (NCI). Development and support the SkyMapper node of the ASVO has been funded in part by Astronomy Australia Limited (AAL) and the Australian Government through the Commonwealth's Education Investment Fund (EIF) and National Collaborative Research Infrastructure Strategy (NCRIS), particularly the National eResearch Collaboration Tools and Resources (NeCTAR) and the Australian National Data Service Projects (ANDS). 
This publication makes use of data products from the AAVSO Photometric All Sky Survey (APASS). Funded by the Robert Martin Ayers Sciences Fund and the National Science Foundation. 
Based on data products from observations made with ESO Telescopes at the La Silla Paranal Observatory under programme IDs 177.A-3016, 177.A-3017 and 177.A-3018, and on data products produced by Target/OmegaCEN, INAF-OACN, INAF-OAPD and the KiDS production team, on behalf of the KiDS consortium. OmegaCEN and the KiDS production team acknowledge support by NOVA and NWO-M grants. Members of INAF-OAPD and INAF-OACN also acknowledge the support from the Department of Physics \& Astronomy of the University of Padova, and of the Department of Physics of Univ.\ Federico II (Naples). 
This work is based in part on data obtained as part of the UKIRT Infrared Deep Sky Survey. 
We have used data from the VISTA Data Flow System pipeline processing and science archive, which are described in Irwin et al.~(2004), Hambly et al.~(2008) and Cross et al.~(2012). 
\end{acknowledgements}

\bibliographystyle{aa}
\bibliography{bib}

\begin{appendix}
\section{Kinematic Tables}\label{app:kinematic-tables}
\newpage
\begin{table*}
\caption{Kinematic quantities for the program stars based on Model I.}\label{tab:kinematic-AS}
\tiny
\centering
\setlength{\tabcolsep}{0.075cm}
\renewcommand{\arraystretch}{1.05}
\scalebox{.84}{\begin{tabular}{lrrrrrrrrrrrrrrrrrrrrr}
\hline\hline
 Object & $x$ & $y$ & $z$ & & $\varv_x$ & $\varv_y$ & $\varv_z$ & $\varv_{\mathrm{Grf}}$ & $\varv_{\mathrm{Grf}}-\varv_{\mathrm{esc}}$ 		& $P_{\mathrm{b}}$ & $x_{\mathrm{p}}$ & $y_{\mathrm{p}}$ & $z_{\mathrm{p}}$ & $r_{\mathrm{p}}$ & & $\varv_{x\mathrm{,p}}$ & 		$\varv_{y\mathrm{,p}}$ & $\varv_{z\mathrm{,p}}$ & $\varv_{\mathrm{Grf,p}}$ &  $\varv_{\mathrm{ej,p}}$ & $\tau_{\mathrm{flight,p}}$ \\\cline{2-4} \cline{6-10} \cline{12-15} \cline{17-21}& \multicolumn{3}{c}{(kpc)} & & \multicolumn{5}{c}{$(\mathrm{km\,s^{-1}})$} & (\%) & \multicolumn{4}{c}{(kpc)} & & \multicolumn{5}{c}{$(\mathrm{km\,s^{-1}})$} & (Myr) \\
\hline \hline
HVS1 & $-101.4$ & $-100.9$ & $83.5$ &   & $-390$ & $-340$ & $450$ & $710$ & $400$ & $0$ & $-26.5$ & $-34.5$ & $0.0$ & $66.1$ &   & $-450$ & $-400$ & $480$ & $780$ & $780$ & $176$ \\ Stat. & $^{+7.6}_{-6.9}$ & $^{+8.2}_{-7.5}$ & $^{+6.2}_{-6.8}$ &   & $^{+160}_{-160}$ & $^{+130}_{-140}$ & $^{+170}_{-170}$ & $^{+70}_{-40}$ & $^{+70}_{-40}$ & \ldots & $^{+59.8}_{-37.6}$ & $^{+50.8}_{-31.7}$ & $^{+0.0}_{-0.0}$ & $^{+33.0}_{-32.0}$ &   & $^{+170}_{-110}$ & $^{+150}_{-100}$ & $^{+150}_{-100}$ & $^{+40}_{-30}$ & $^{+130}_{-110}$ & $^{+80}_{-46}$ \\
HVS4 & $-61.9$ & $-14.1$ & $50.8$ &   & $-360$ & $-300$ & $360$ & $630$ & $230$ & $0$ & $-9.8$ & $23.0$ & $0.0$ & $43.1$ &   & $-450$ & $-250$ & $440$ & $680$ & $830$ & $127$ \\ Stat. & $^{+4.1}_{-4.8}$ & $^{+1.1}_{-1.3}$ & $^{+4.5}_{-3.9}$ &   & $^{+180}_{-180}$ & $^{+170}_{-180}$ & $^{+160}_{-160}$ & $^{+110}_{-\phantom{0}60}$ & $^{+110}_{-\phantom{0}60}$ & \ldots & $^{+51.6}_{-32.0}$ & $^{+20.9}_{-17.4}$ & $^{+0.0}_{-0.0}$ & $^{+27.6}_{-21.4}$ &   & $^{+210}_{-\phantom{0}90}$ & $^{+140}_{-200}$ & $^{+110}_{-120}$ & $^{+70}_{-30}$ & $^{+\phantom{0}70}_{-140}$ & $^{+60}_{-34}$ \\
HVS5 & $-32.1$ & $15.9$ & $22.9$ &   & $-400$ & $310$ & $410$ & $650$ & $190$ & $0$ & $-8.6$ & $-1.1$ & $0.0$ & $9.1$ &   & $-510$ & $330$ & $460$ & $760$ & $670$ & $52$ \\ Stat. & $^{+2.2}_{-2.4}$ & $^{+1.7}_{-1.5}$ & $^{+2.4}_{-2.1}$ &   & $^{+30}_{-30}$ & $^{+50}_{-50}$ & $^{+30}_{-30}$ & $^{+10}_{-10}$ & $^{+20}_{-20}$ & \ldots & $^{+2.4}_{-2.3}$ & $^{+2.9}_{-2.7}$ & $^{+0.0}_{-0.0}$ & $^{+2.3}_{-2.3}$ &   & $^{+40}_{-50}$ & $^{+40}_{-30}$ & $^{+20}_{-20}$ & $^{+30}_{-20}$ & $^{+80}_{-60}$ & $^{+6}_{-5}$ \\
HVS5 (Brown 2015) & $-37.6$ & $19.5$ & $28.1$ &   & $-420$ & $270$ & $420$ & $650$ & $210$ & $0$ & $-8.8$ & $1.6$ & $0.0$ & $9.8$ &   & $-530$ & $310$ & $480$ & $770$ & $740$ & $62$ \\ Stat. & $^{+3.4}_{-3.4}$ & $^{+2.3}_{-2.3}$ & $^{+3.3}_{-3.3}$ &   & $^{+40}_{-40}$ & $^{+60}_{-70}$ & $^{+40}_{-30}$ & $^{+10}_{-10}$ & $^{+20}_{-20}$ & \ldots & $^{+3.5}_{-3.3}$ & $^{+4.6}_{-4.0}$ & $^{+0.0}_{-0.0}$ & $^{+3.5}_{-3.3}$ &   & $^{+50}_{-50}$ & $^{+30}_{-50}$ & $^{+20}_{-20}$ & $^{+40}_{-20}$ & $^{+110}_{-\phantom{0}90}$ & $^{+8}_{-7}$ \\
HVS6 & $-20.4$ & $-23.6$ & $45.0$ &   & $-150$ & $-150$ & $450$ & $530$ & $90$ & $0$ & $-4.4$ & $-7.0$ & $0.0$ & $17.2$ &   & $-220$ & $-240$ & $560$ & $640$ & $670$ & $88$ \\ Stat. & $^{+1.1}_{-1.2}$ & $^{+2.2}_{-2.4}$ & $^{+4.5}_{-4.1}$ &   & $^{+160}_{-160}$ & $^{+120}_{-120}$ & $^{+80}_{-80}$ & $^{+60}_{-30}$ & $^{+60}_{-30}$ & \ldots & $^{+15.0}_{-13.9}$ & $^{+12.7}_{-11.2}$ & $^{+0.0}_{-0.0}$ & $^{+11.4}_{-\phantom{0}9.1}$ &   & $^{+150}_{-\phantom{0}80}$ & $^{+130}_{-\phantom{0}60}$ & $^{+40}_{-50}$ & $^{+50}_{-40}$ & $^{+90}_{-80}$ & $^{+17}_{-13}$ \\
HVS7 & $-11.1$ & $-25.1$ & $40.3$ &   & $-200$ & $0$ & $450$ & $500$ & $50$ & $7$ & $6.0$ & $-21.8$ & $0.0$ & $23.8$ &   & $-200$ & $-100$ & $510$ & $560$ & $530$ & $81$ \\ Stat. & $^{+0.3}_{-0.3}$ & $^{+2.1}_{-2.5}$ & $^{+3.9}_{-3.3}$ &   & $^{+90}_{-90}$ & $^{+50}_{-50}$ & $^{+40}_{-40}$ & $^{+50}_{-40}$ & $^{+50}_{-40}$ & \ldots & $^{+8.2}_{-7.0}$ & $^{+4.7}_{-4.9}$ & $^{+0.0}_{-0.0}$ & $^{+5.4}_{-4.7}$ &   & $^{+70}_{-80}$ & $^{+60}_{-60}$ & $^{+30}_{-20}$ & $^{+30}_{-20}$ & $^{+30}_{-30}$ & $^{+10}_{-\phantom{0}8}$ \\
HVS8 & $-29.7$ & $-13.2$ & $26.3$ &   & $-410$ & $80$ & $260$ & $500$ & $40$ & $18$ & $8.5$ & $-16.1$ & $0.0$ & $19.1$ &   & $-450$ & $-40$ & $350$ & $570$ & $440$ & $84$ \\ Stat. & $^{+1.6}_{-1.8}$ & $^{+1.0}_{-1.1}$ & $^{+2.2}_{-2.0}$ &   & $^{+60}_{-60}$ & $^{+60}_{-60}$ & $^{+40}_{-40}$ & $^{+50}_{-40}$ & $^{+50}_{-40}$ & \ldots & $^{+9.8}_{-7.5}$ & $^{+4.9}_{-5.4}$ & $^{+0.0}_{-0.0}$ & $^{+8.2}_{-5.7}$ &   & $^{+30}_{-40}$ & $^{+70}_{-70}$ & $^{+30}_{-40}$ & $^{+20}_{-10}$ & $^{+40}_{-20}$ & $^{+14}_{-11}$ \\
HVS9 & $-44.8$ & $-70.6$ & $80.9$ &   & $50$ & $-170$ & $520$ & $710$ & $350$ & $0$ & $-47.8$ & $-40.5$ & $0.0$ & $94.7$ &   & $0$ & $-230$ & $550$ & $710$ & $770$ & $146$ \\ Stat. & $^{+3.7}_{-3.9}$ & $^{+7.1}_{-7.4}$ & $^{+8.5}_{-8.1}$ &   & $^{+420}_{-420}$ & $^{+300}_{-300}$ & $^{+250}_{-250}$ & $^{+280}_{-180}$ & $^{+280}_{-180}$ & \ldots & $^{+70.8}_{-61.0}$ & $^{+78.1}_{-41.9}$ & $^{+0.0}_{-0.0}$ & $^{+49.2}_{-37.4}$ &   & $^{+440}_{-350}$ & $^{+310}_{-240}$ & $^{+230}_{-190}$ & $^{+270}_{-120}$ & $^{+260}_{-160}$ & $^{+99}_{-45}$ \\
HVS10 & $-14.5$ & $-16.8$ & $70.2$ &   & $-250$ & $-190$ & $360$ & $500$ & $90$ & $1$ & $24.8$ & $14.1$ & $0.0$ & $35.9$ &   & $-190$ & $-160$ & $490$ & $570$ & $600$ & $160$ \\ Stat. & $^{+0.6}_{-0.6}$ & $^{+1.6}_{-1.6}$ & $^{+6.6}_{-6.4}$ &   & $^{+130}_{-140}$ & $^{+140}_{-150}$ & $^{+40}_{-40}$ & $^{+90}_{-60}$ & $^{+90}_{-60}$ & \ldots & $^{+24.7}_{-19.0}$ & $^{+27.0}_{-19.5}$ & $^{+0.0}_{-0.0}$ & $^{+27.7}_{-19.1}$ &   & $^{+\phantom{0}80}_{-130}$ & $^{+\phantom{0}70}_{-120}$ & $^{+60}_{-60}$ & $^{+50}_{-20}$ & $^{+50}_{-40}$ & $^{+26}_{-21}$ \\
HVS12 & $-26.2$ & $-42.1$ & $59.5$ &   & $-50$ & $70$ & $550$ & $570$ & $170$ & $0$ & $-19.4$ & $-46.5$ & $0.0$ & $52.1$ &   & $-80$ & $10$ & $580$ & $600$ & $610$ & $102$ \\ Stat. & $^{+1.7}_{-1.9}$ & $^{+4.0}_{-4.5}$ & $^{+6.4}_{-5.6}$ &   & $^{+130}_{-130}$ & $^{+110}_{-110}$ & $^{+80}_{-80}$ & $^{+100}_{-\phantom{0}80}$ & $^{+100}_{-\phantom{0}80}$ & \ldots & $^{+13.6}_{-13.0}$ & $^{+12.2}_{-12.1}$ & $^{+0.0}_{-0.0}$ & $^{+12.3}_{-12.3}$ &   & $^{+130}_{-120}$ & $^{+120}_{-120}$ & $^{+80}_{-70}$ & $^{+80}_{-50}$ & $^{+80}_{-70}$ & $^{+18}_{-14}$ \\
HVS13 & $-32.4$ & $-72.1$ & $92.7$ &   & $-660$ & $-40$ & $340$ & $780$ & $430$ & $0$ & $123.6$ & $-53.3$ & $0.0$ & $146.7$ &   & $-610$ & $-100$ & $400$ & $760$ & $710$ & $238$ \\ Stat. & $^{+2.3}_{-2.0}$ & $^{+6.7}_{-5.9}$ & $^{+7.6}_{-8.6}$ &   & $^{+220}_{-230}$ & $^{+200}_{-200}$ & $^{+150}_{-150}$ & $^{+210}_{-180}$ & $^{+220}_{-180}$ & \ldots & $^{+122.7}_{-\phantom{0}65.5}$ & $^{+77.0}_{-43.2}$ & $^{+0.0}_{-0.0}$ & $^{+111.9}_{-\phantom{0}55.5}$ &   & $^{+220}_{-240}$ & $^{+210}_{-190}$ & $^{+130}_{-140}$ & $^{+210}_{-160}$ & $^{+210}_{-160}$ & $^{+135}_{-\phantom{0}66}$ \\
HVS15 & $-10.5$ & $-34.2$ & $50.7$ &   & $-60$ & $-160$ & $280$ & $450$ & $30$ & $41$ & $-0.1$ & $-3.7$ & $0.0$ & $44.6$ &   & $-80$ & $-240$ & $410$ & $510$ & $570$ & $145$ \\ Stat. & $^{+0.3}_{-0.3}$ & $^{+3.6}_{-4.1}$ & $^{+6.1}_{-5.3}$ &   & $^{+340}_{-350}$ & $^{+190}_{-190}$ & $^{+140}_{-140}$ & $^{+210}_{-100}$ & $^{+210}_{-100}$ & \ldots & $^{+63.4}_{-42.6}$ & $^{+48.6}_{-28.7}$ & $^{+0.0}_{-0.0}$ & $^{+50.2}_{-25.1}$ &   & $^{+300}_{-240}$ & $^{+190}_{-\phantom{0}90}$ & $^{+\phantom{0}80}_{-140}$ & $^{+130}_{-\phantom{0}50}$ & $^{+120}_{-100}$ & $^{+78}_{-37}$ \\
HVS16 & $-1.5$ & $-24.2$ & $60.4$ &   & $-270$ & $-470$ & $210$ & $680$ & $260$ & $10$ & $51.7$ & $76.7$ & $0.0$ & $126.2$ &   & $-210$ & $-420$ & $300$ & $640$ & $660$ & $217$ \\ Stat. & $^{+0.7}_{-0.6}$ & $^{+2.2}_{-2.2}$ & $^{+5.5}_{-5.3}$ &   & $^{+480}_{-480}$ & $^{+240}_{-250}$ & $^{+\phantom{0}90}_{-100}$ & $^{+350}_{-230}$ & $^{+360}_{-230}$ & \ldots & $^{+123.7}_{-\phantom{0}93.3}$ & $^{+128.2}_{-\phantom{0}59.7}$ & $^{+0.0}_{-0.0}$ & $^{+159.0}_{-\phantom{0}77.1}$ &   & $^{+340}_{-490}$ & $^{+210}_{-240}$ & $^{+110}_{-110}$ & $^{+340}_{-150}$ & $^{+320}_{-130}$ & $^{+121}_{-\phantom{0}57}$ \\
HVS17 & $-0.8$ & $25.6$ & $23.3$ &   & $190$ & $280$ & $310$ & $460$ & $-20$ & $83$ & $-12.3$ & $4.2$ & $0.0$ & $13.4$ &   & $110$ & $380$ & $400$ & $560$ & $430$ & $65$ \\ Stat. & $^{+0.7}_{-0.6}$ & $^{+2.1}_{-2.0}$ & $^{+1.9}_{-1.8}$ &   & $^{+60}_{-60}$ & $^{+30}_{-40}$ & $^{+30}_{-30}$ & $^{+20}_{-20}$ & $^{+30}_{-20}$ & \ldots & $^{+3.4}_{-3.8}$ & $^{+3.7}_{-3.4}$ & $^{+0.0}_{-0.0}$ & $^{+4.1}_{-3.7}$ &   & $^{+70}_{-70}$ & $^{+40}_{-40}$ & $^{+30}_{-20}$ & $^{+30}_{-10}$ & $^{+30}_{-20}$ & $^{+6}_{-5}$ \\
HVS18 & $-28.7$ & $83.5$ & $-43.3$ &   & $30$ & $470$ & $-130$ & $560$ & $190$ & $0$ & $-24.2$ & $-27.6$ & $0.0$ & $77.3$ &   & $-70$ & $470$ & $-240$ & $570$ & $540$ & $221$ \\ Stat. & $^{+1.9}_{-2.1}$ & $^{+8.7}_{-7.5}$ & $^{+3.9}_{-4.5}$ &   & $^{+300}_{-300}$ & $^{+130}_{-130}$ & $^{+200}_{-200}$ & $^{+160}_{-\phantom{0}90}$ & $^{+160}_{-\phantom{0}90}$ & \ldots & $^{+64.9}_{-70.4}$ & $^{+\phantom{0}63.2}_{-191.7}$ & $^{+0.0}_{-0.0}$ & $^{+185.5}_{-\phantom{0}40.4}$ &   & $^{+310}_{-160}$ & $^{+80}_{-60}$ & $^{+140}_{-130}$ & $^{+100}_{-\phantom{0}70}$ & $^{+190}_{-\phantom{0}80}$ & $^{+332}_{-\phantom{0}99}$ \\
HVS19 & $-17.6$ & $-36.9$ & $77.0$ &   & $-230$ & $90$ & $560$ & $920$ & $540$ & $0$ & $11.9$ & $-44.5$ & $0.0$ & $98.1$ &   & $-210$ & $50$ & $610$ & $910$ & $920$ & $126$ \\ Stat. & $^{+1.1}_{-1.2}$ & $^{+4.5}_{-4.5}$ & $^{+9.3}_{-9.2}$ &   & $^{+750}_{-760}$ & $^{+410}_{-400}$ & $^{+220}_{-220}$ & $^{+470}_{-300}$ & $^{+470}_{-300}$ & \ldots & $^{+130.9}_{-\phantom{0}86.3}$ & $^{+67.2}_{-40.8}$ & $^{+0.0}_{-0.0}$ & $^{+86.5}_{-44.4}$ &   & $^{+700}_{-730}$ & $^{+410}_{-360}$ & $^{+200}_{-200}$ & $^{+450}_{-260}$ & $^{+450}_{-230}$ & $^{+69}_{-34}$ \\
HVS20 & $-15.0$ & $-50.8$ & $90.2$ &   & $40$ & $300$ & $620$ & $970$ & $600$ & $0$ & $-18.8$ & $-86.8$ & $0.0$ & $127.3$ &   & $10$ & $260$ & $650$ & $960$ & $970$ & $134$ \\ Stat. & $^{+0.9}_{-0.7}$ & $^{+6.3}_{-4.6}$ & $^{+\phantom{0}8.2}_{-11.1}$ &   & $^{+720}_{-720}$ & $^{+410}_{-400}$ & $^{+240}_{-240}$ & $^{+480}_{-360}$ & $^{+480}_{-360}$ & \ldots & $^{+111.3}_{-\phantom{0}92.6}$ & $^{+61.5}_{-37.5}$ & $^{+0.0}_{-0.0}$ & $^{+59.9}_{-45.2}$ &   & $^{+720}_{-670}$ & $^{+420}_{-410}$ & $^{+230}_{-200}$ & $^{+470}_{-340}$ & $^{+470}_{-320}$ & $^{+72}_{-36}$ \\
HVS21 & $-60.2$ & $13.6$ & $79.8$ &   & $-130$ & $200$ & $360$ & $590$ & $220$ & $0$ & $-28.4$ & $-22.3$ & $0.0$ & $87.3$ &   & $-200$ & $150$ & $450$ & $590$ & $620$ & $194$ \\ Stat. & $^{+5.3}_{-6.2}$ & $^{+1.7}_{-1.4}$ & $^{+9.5}_{-8.1}$ &   & $^{+290}_{-290}$ & $^{+370}_{-370}$ & $^{+220}_{-220}$ & $^{+250}_{-150}$ & $^{+250}_{-150}$ & \ldots & $^{+108.8}_{-\phantom{0}50.9}$ & $^{+\phantom{0}60.7}_{-120.6}$ & $^{+0.0}_{-0.0}$ & $^{+95.1}_{-41.7}$ &   & $^{+310}_{-190}$ & $^{+330}_{-290}$ & $^{+160}_{-200}$ & $^{+200}_{-\phantom{0}70}$ & $^{+180}_{-130}$ & $^{+158}_{-\phantom{0}64}$ \\
HVS22 & $-13.7$ & $-45.4$ & $86.2$ &   & $-380$ & $730$ & $910$ & $1530$ & $1150$ & $0$ & $21.1$ & $-108.5$ & $0.0$ & $139.1$ &   & $-370$ & $700$ & $920$ & $1520$ & $1510$ & $91$ \\ Stat. & $^{+0.8}_{-0.9}$ & $^{+6.5}_{-7.6}$ & $^{+14.4}_{-12.3}$ &   & $^{+950}_{-970}$ & $^{+590}_{-560}$ & $^{+320}_{-310}$ & $^{+690}_{-560}$ & $^{+690}_{-560}$ & \ldots & $^{+114.2}_{-\phantom{0}83.0}$ & $^{+46.5}_{-39.2}$ & $^{+0.0}_{-0.0}$ & $^{+62.9}_{-43.6}$ &   & $^{+930}_{-960}$ & $^{+600}_{-560}$ & $^{+320}_{-290}$ & $^{+680}_{-550}$ & $^{+680}_{-550}$ & $^{+45}_{-23}$ \\
HVS24 & $-17.1$ & $-35.4$ & $51.2$ &   & $80$ & $-90$ & $390$ & $460$ & $40$ & $33$ & $-23.1$ & $-20.6$ & $0.0$ & $40.1$ &   & $10$ & $-170$ & $460$ & $520$ & $610$ & $117$ \\ Stat. & $^{+0.8}_{-0.8}$ & $^{+3.2}_{-3.0}$ & $^{+4.4}_{-4.6}$ &   & $^{+230}_{-230}$ & $^{+140}_{-150}$ & $^{+110}_{-110}$ & $^{+140}_{-\phantom{0}80}$ & $^{+140}_{-\phantom{0}80}$ & \ldots & $^{+26.9}_{-26.0}$ & $^{+23.7}_{-17.2}$ & $^{+0.0}_{-0.0}$ & $^{+20.5}_{-19.2}$ &   & $^{+250}_{-170}$ & $^{+160}_{-120}$ & $^{+80}_{-60}$ & $^{+90}_{-40}$ & $^{+\phantom{0}90}_{-100}$ & $^{+33}_{-23}$ \\
B095 & $-43.6$ & $16.1$ & $45.3$ &   & $-120$ & $470$ & $100$ & $550$ & $130$ & $22$ & $-6.0$ & $-82.2$ & $0.0$ & $95.5$ &   & $-180$ & $350$ & $220$ & $480$ & $450$ & $236$ \\ Stat. & $^{+3.5}_{-3.5}$ & $^{+1.6}_{-1.6}$ & $^{+4.5}_{-4.5}$ &   & $^{+200}_{-200}$ & $^{+230}_{-230}$ & $^{+180}_{-180}$ & $^{+210}_{-170}$ & $^{+210}_{-170}$ & \ldots & $^{+115.3}_{-\phantom{0}44.3}$ & $^{+\phantom{0}61.3}_{-197.1}$ & $^{+0.0}_{-0.0}$ & $^{+216.9}_{-\phantom{0}49.3}$ &   & $^{+190}_{-100}$ & $^{+210}_{-190}$ & $^{+130}_{-130}$ & $^{+150}_{-\phantom{0}70}$ & $^{+160}_{-\phantom{0}90}$ & $^{+352}_{-\phantom{0}98}$ \\
B129 & $-84.1$ & $-21.8$ & $33.8$ &   & $-190$ & $-60$ & $250$ & $390$ & $10$ & $47$ & $-54.8$ & $-10.7$ & $0.0$ & $67.9$ &   & $-260$ & $-90$ & $270$ & $440$ & $490$ & $126$ \\ Stat. & $^{+6.8}_{-6.9}$ & $^{+2.0}_{-2.0}$ & $^{+3.1}_{-3.1}$ &   & $^{+120}_{-120}$ & $^{+200}_{-200}$ & $^{+190}_{-190}$ & $^{+150}_{-\phantom{0}80}$ & $^{+150}_{-\phantom{0}80}$ & \ldots & $^{+73.2}_{-23.3}$ & $^{+21.8}_{-24.4}$ & $^{+0.0}_{-0.0}$ & $^{+19.8}_{-26.5}$ &   & $^{+150}_{-120}$ & $^{+150}_{-170}$ & $^{+180}_{-110}$ & $^{+110}_{-\phantom{0}50}$ & $^{+170}_{-160}$ & $^{+157}_{-\phantom{0}52}$ \\
B143 & $-30.8$ & $7.9$ & $16.1$ &   & $-200$ & $120$ & $180$ & $290$ & $-190$ & $100$ & $-10.2$ & $-1.9$ & $0.0$ & $10.6$ &   & $-370$ & $120$ & $240$ & $460$ & $410$ & $77$ \\ Stat. & $^{+1.8}_{-2.0}$ & $^{+0.7}_{-0.7}$ & $^{+1.5}_{-1.3}$ &   & $^{+20}_{-20}$ & $^{+30}_{-30}$ & $^{+30}_{-30}$ & $^{+10}_{-10}$ & $^{+10}_{-10}$ & \ldots & $^{+4.0}_{-3.6}$ & $^{+2.4}_{-2.2}$ & $^{+0.0}_{-0.0}$ & $^{+3.3}_{-3.5}$ &   & $^{+50}_{-60}$ & $^{+20}_{-20}$ & $^{+20}_{-10}$ & $^{+50}_{-40}$ & $^{+60}_{-50}$ & $^{+9}_{-8}$ \\
B167 & $-32.5$ & $-2.7$ & $22.0$ &   & $-280$ & $210$ & $130$ & $370$ & $-100$ & $99$ & $8.3$ & $-22.1$ & $0.0$ & $24.2$ &   & $-360$ & $70$ & $230$ & $430$ & $310$ & $119$ \\ Stat. & $^{+1.8}_{-2.0}$ & $^{+0.2}_{-0.3}$ & $^{+1.8}_{-1.7}$ &   & $^{+40}_{-40}$ & $^{+50}_{-50}$ & $^{+40}_{-40}$ & $^{+40}_{-40}$ & $^{+40}_{-40}$ & \ldots & $^{+11.0}_{-\phantom{0}8.3}$ & $^{+5.8}_{-7.4}$ & $^{+0.0}_{-0.0}$ & $^{+10.5}_{-\phantom{0}6.7}$ &   & $^{+10}_{-10}$ & $^{+60}_{-60}$ & $^{+30}_{-30}$ & $^{+10}_{-20}$ & $^{+30}_{-30}$ & $^{+24}_{-18}$ \\
B329 & $40.7$ & $16.6$ & $51.5$ &   & $160$ & $-240$ & $330$ & $480$ & $70$ & $33$ & $12.6$ & $45.2$ & $0.0$ & $57.2$ &   & $220$ & $-160$ & $390$ & $500$ & $480$ & $139$ \\ Stat. & $^{+6.3}_{-5.3}$ & $^{+2.1}_{-1.8}$ & $^{+6.6}_{-5.6}$ &   & $^{+160}_{-160}$ & $^{+210}_{-220}$ & $^{+160}_{-150}$ & $^{+190}_{-140}$ & $^{+190}_{-140}$ & \ldots & $^{+24.9}_{-41.7}$ & $^{+36.1}_{-28.6}$ & $^{+0.0}_{-0.0}$ & $^{+34.5}_{-25.2}$ &   & $^{+100}_{-160}$ & $^{+220}_{-240}$ & $^{+130}_{-110}$ & $^{+160}_{-\phantom{0}70}$ & $^{+150}_{-110}$ & $^{+72}_{-39}$ \\
B434 & $-16.2$ & $-22.8$ & $33.7$ &   & $130$ & $-280$ & $210$ & $380$ & $-80$ & $92$ & $-25.4$ & $14.7$ & $0.0$ & $30.8$ &   & $0$ & $-280$ & $310$ & $430$ & $580$ & $123$ \\ Stat. & $^{+0.6}_{-0.7}$ & $^{+1.7}_{-2.1}$ & $^{+3.1}_{-2.4}$ &   & $^{+80}_{-80}$ & $^{+60}_{-60}$ & $^{+40}_{-40}$ & $^{+50}_{-40}$ & $^{+60}_{-40}$ & \ldots & $^{+\phantom{0}9.4}_{-10.5}$ & $^{+12.6}_{-\phantom{0}9.0}$ & $^{+0.0}_{-0.0}$ & $^{+12.5}_{-\phantom{0}9.5}$ &   & $^{+90}_{-90}$ & $^{+30}_{-40}$ & $^{+30}_{-40}$ & $^{+20}_{-10}$ & $^{+20}_{-20}$ & $^{+23}_{-17}$ \\
B458 & $-26.8$ & $-52.5$ & $61.5$ &   & $170$ & $-20$ & $430$ & $570$ & $180$ & $17$ & $-45.1$ & $-44.8$ & $0.0$ & $79.8$ &   & $110$ & $-80$ & $460$ & $570$ & $640$ & $132$ \\ Stat. & $^{+2.0}_{-2.3}$ & $^{+5.7}_{-6.4}$ & $^{+7.5}_{-6.7}$ &   & $^{+360}_{-350}$ & $^{+220}_{-220}$ & $^{+210}_{-200}$ & $^{+270}_{-190}$ & $^{+280}_{-190}$ & \ldots & $^{+49.4}_{-46.2}$ & $^{+48.8}_{-26.5}$ & $^{+0.0}_{-0.0}$ & $^{+33.6}_{-29.8}$ &   & $^{+370}_{-300}$ & $^{+240}_{-200}$ & $^{+190}_{-140}$ & $^{+260}_{-130}$ & $^{+260}_{-150}$ & $^{+85}_{-40}$ \\
B481 & $-6.1$ & $28.7$ & $-36.4$ &   & $-220$ & $-30$ & $-400$ & $460$ & $10$ & $45$ & $12.5$ & $27.5$ & $0.0$ & $30.6$ &   & $-210$ & $60$ & $-450$ & $500$ & $630$ & $82$ \\ Stat. & $^{+0.3}_{-0.3}$ & $^{+3.0}_{-3.1}$ & $^{+3.9}_{-3.7}$ &   & $^{+70}_{-80}$ & $^{+60}_{-60}$ & $^{+50}_{-50}$ & $^{+80}_{-70}$ & $^{+90}_{-70}$ & \ldots & $^{+6.0}_{-5.4}$ & $^{+7.7}_{-7.6}$ & $^{+0.0}_{-0.0}$ & $^{+8.5}_{-8.4}$ &   & $^{+70}_{-70}$ & $^{+80}_{-80}$ & $^{+40}_{-40}$ & $^{+60}_{-40}$ & $^{+50}_{-40}$ & $^{+8}_{-7}$ \\
B485 & $-26.8$ & $-6.0$ & $27.4$ &   & $-330$ & $140$ & $270$ & $450$ & $-20$ & $91$ & $4.2$ & $-14.7$ & $0.0$ & $15.4$ &   & $-390$ & $20$ & $380$ & $540$ & $420$ & $83$ \\ Stat. & $^{+1.4}_{-1.6}$ & $^{+0.5}_{-0.6}$ & $^{+2.4}_{-2.1}$ &   & $^{+30}_{-30}$ & $^{+20}_{-20}$ & $^{+20}_{-20}$ & $^{+20}_{-20}$ & $^{+20}_{-20}$ & \ldots & $^{+3.4}_{-2.8}$ & $^{+1.7}_{-2.0}$ & $^{+0.0}_{-0.0}$ & $^{+2.8}_{-2.1}$ &   & $^{+10}_{-10}$ & $^{+30}_{-30}$ & $^{+10}_{-20}$ & $^{+10}_{-10}$ & $^{+10}_{-10}$ & $^{+9}_{-8}$ \\
B537 & $-22.0$ & $28.2$ & $-26.8$ &   & $-120$ & $240$ & $-180$ & $320$ & $-130$ & $100$ & $-2.5$ & $-3.1$ & $0.0$ & $5.5$ &   & $-300$ & $300$ & $-370$ & $570$ & $460$ & $109$ \\ Stat. & $^{+1.6}_{-1.9}$ & $^{+3.8}_{-3.3}$ & $^{+3.1}_{-3.7}$ &   & $^{+50}_{-50}$ & $^{+30}_{-30}$ & $^{+20}_{-30}$ & $^{+20}_{-20}$ & $^{+20}_{-20}$ & \ldots & $^{+3.8}_{-4.7}$ & $^{+3.1}_{-2.9}$ & $^{+0.0}_{-0.0}$ & $^{+3.8}_{-2.8}$ &   & $^{+80}_{-50}$ & $^{+60}_{-50}$ & $^{+60}_{-90}$ & $^{+70}_{-60}$ & $^{+210}_{-\phantom{0}90}$ & $^{+12}_{-10}$ \\
B572 & $-21.6$ & $18.9$ & $-14.0$ &   & $-70$ & $250$ & $-170$ & $320$ & $-170$ & $100$ & $-12.5$ & $-1.6$ & $0.0$ & $13.2$ &   & $-220$ & $300$ & $-230$ & $430$ & $310$ & $70$ \\ Stat. & $^{+1.4}_{-2.0}$ & $^{+2.9}_{-2.0}$ & $^{+1.5}_{-2.1}$ &   & $^{+50}_{-50}$ & $^{+30}_{-30}$ & $^{+50}_{-50}$ & $^{+20}_{-20}$ & $^{+30}_{-20}$ & \ldots & $^{+5.8}_{-4.8}$ & $^{+4.8}_{-4.7}$ & $^{+0.0}_{-0.0}$ & $^{+4.5}_{-4.1}$ &   & $^{+80}_{-90}$ & $^{+20}_{-30}$ & $^{+20}_{-30}$ & $^{+50}_{-30}$ & $^{+60}_{-40}$ & $^{+14}_{-11}$ \\
B576 & $-1.9$ & $14.1$ & $48.9$ &   & $-360$ & $-350$ & $460$ & $680$ & $240$ & $0$ & $32.9$ & $46.0$ & $0.0$ & $56.8$ &   & $-330$ & $-290$ & $510$ & $670$ & $740$ & $97$ \\ Stat. & $^{+0.5}_{-0.5}$ & $^{+1.1}_{-1.1}$ & $^{+3.8}_{-3.6}$ &   & $^{+50}_{-50}$ & $^{+60}_{-60}$ & $^{+30}_{-20}$ & $^{+70}_{-60}$ & $^{+70}_{-70}$ & \ldots & $^{+6.1}_{-5.6}$ & $^{+8.2}_{-7.5}$ & $^{+0.0}_{-0.0}$ & $^{+9.6}_{-8.9}$ &   & $^{+50}_{-50}$ & $^{+60}_{-70}$ & $^{+20}_{-20}$ & $^{+60}_{-60}$ & $^{+60}_{-50}$ & $^{+5}_{-5}$ \\
B576 (BHB) & $-6.0$ & $5.2$ & $17.8$ &   & $-110$ & $70$ & $300$ & $330$ & $-200$ & $100$ & $0.1$ & $0.6$ & $0.0$ & $1.0$ &   & $-170$ & $200$ & $590$ & $650$ & $730$ & $45$ \\ Stat. & $^{+0.2}_{-0.2}$ & $^{+0.5}_{-0.4}$ & $^{+1.6}_{-1.4}$ &   & $^{+20}_{-20}$ & $^{+30}_{-30}$ & $^{+10}_{-10}$ & $^{+10}_{-10}$ & $^{+20}_{-20}$ & \ldots & $^{+1.0}_{-0.8}$ & $^{+1.1}_{-0.8}$ & $^{+0.0}_{-0.0}$ & $^{+1.0}_{-0.6}$ &   & $^{+\phantom{0}60}_{-110}$ & $^{+40}_{-70}$ & $^{+60}_{-40}$ & $^{+80}_{-50}$ & $^{+80}_{-80}$ & $^{+4}_{-3}$ \\
B598 & $1.7$ & $-0.5$ & $20.4$ &   & $40$ & $50$ & $300$ & $310$ & $-210$ & $100$ & $-0.7$ & $-2.1$ & $0.0$ & $3.4$ &   & $30$ & $-20$ & $550$ & $550$ & $610$ & $50$ \\ Stat. & $^{+0.7}_{-0.7}$ & $^{+0.1}_{-0.1}$ & $^{+1.4}_{-1.3}$ &   & $^{+60}_{-60}$ & $^{+60}_{-60}$ & $^{+30}_{-30}$ & $^{+30}_{-20}$ & $^{+30}_{-20}$ & \ldots & $^{+2.4}_{-2.6}$ & $^{+2.3}_{-2.6}$ & $^{+0.0}_{-0.0}$ & $^{+2.5}_{-1.9}$ &   & $^{+40}_{-40}$ & $^{+40}_{-40}$ & $^{+60}_{-60}$ & $^{+60}_{-60}$ & $^{+50}_{-50}$ & $^{+5}_{-4}$ \\
B598 (ELM) & $-5.5$ & $-0.1$ & $5.8$ &   & $110$ & $190$ & $270$ & $350$ & $-260$ & $100$ & $-6.7$ & $-3.4$ & $0.0$ & $7.5$ &   & $0$ & $160$ & $340$ & $380$ & $370$ & $18$ \\ Stat. & $^{+0.5}_{-0.4}$ & $^{+0.1}_{-0.1}$ & $^{+0.9}_{-0.8}$ &   & $^{+20}_{-20}$ & $^{+20}_{-20}$ & $^{+10}_{-10}$ & $^{+20}_{-20}$ & $^{+20}_{-10}$ & \ldots & $^{+0.6}_{-0.5}$ & $^{+0.4}_{-0.4}$ & $^{+0.0}_{-0.0}$ & $^{+0.4}_{-0.5}$ &   & $^{+20}_{-20}$ & $^{+20}_{-30}$ & $^{+20}_{-20}$ & $^{+20}_{-20}$ & $^{+20}_{-20}$ & $^{+3}_{-2}$ \\
B711 & $1.4$ & $0.4$ & $20.8$ &   & $340$ & $90$ & $150$ & $380$ & $-150$ & $100$ & $-23.4$ & $-6.3$ & $0.0$ & $24.3$ &   & $210$ & $60$ & $290$ & $370$ & $430$ & $84$ \\ Stat. & $^{+0.8}_{-0.7}$ & $^{+0.1}_{-0.1}$ & $^{+1.6}_{-1.5}$ &   & $^{+30}_{-30}$ & $^{+30}_{-30}$ & $^{+20}_{-20}$ & $^{+20}_{-20}$ & $^{+30}_{-20}$ & \ldots & $^{+3.5}_{-4.4}$ & $^{+1.7}_{-1.7}$ & $^{+0.0}_{-0.0}$ & $^{+4.4}_{-3.5}$ &   & $^{+30}_{-30}$ & $^{+20}_{-20}$ & $^{+20}_{-20}$ & $^{+10}_{-10}$ & $^{+10}_{-10}$ & $^{+10}_{-\phantom{0}9}$ \\
B733 & $-4.9$ & $4.0$ & $11.0$ &   & $260$ & $150$ & $360$ & $470$ & $-100$ & $100$ & $-11.0$ & $-0.4$ & $0.0$ & $11.0$ &   & $160$ & $170$ & $430$ & $490$ & $470$ & $27$ \\ Stat. & $^{+0.3}_{-0.3}$ & $^{+0.3}_{-0.3}$ & $^{+0.8}_{-0.8}$ &   & $^{+20}_{-20}$ & $^{+20}_{-20}$ & $^{+10}_{-10}$ & $^{+10}_{-10}$ & $^{+10}_{-10}$ & \ldots & $^{+0.5}_{-0.5}$ & $^{+0.5}_{-0.5}$ & $^{+0.0}_{-0.0}$ & $^{+0.5}_{-0.5}$ &   & $^{+20}_{-20}$ & $^{+20}_{-20}$ & $^{+10}_{-10}$ & $^{+10}_{-10}$ & $^{+10}_{-10}$ & $^{+2}_{-2}$ \\
B1080 & $-22.1$ & $-33.7$ & $38.2$ &   & $-250$ & $-230$ & $180$ & $400$ & $-30$ & $67$ & $15.5$ & $7.5$ & $0.0$ & $24.4$ &   & $-210$ & $-290$ & $350$ & $500$ & $480$ & $146$ \\ Stat. & $^{+1.1}_{-1.2}$ & $^{+2.5}_{-2.8}$ & $^{+3.2}_{-2.9}$ &   & $^{+130}_{-130}$ & $^{+100}_{-100}$ & $^{+110}_{-110}$ & $^{+90}_{-50}$ & $^{+90}_{-50}$ & \ldots & $^{+35.0}_{-17.6}$ & $^{+32.8}_{-18.2}$ & $^{+0.0}_{-0.0}$ & $^{+40.9}_{-14.2}$ &   & $^{+60}_{-80}$ & $^{+50}_{-40}$ & $^{+\phantom{0}70}_{-120}$ & $^{+60}_{-40}$ & $^{+100}_{-\phantom{0}50}$ & $^{+74}_{-33}$ \\
B1085 & $-8.5$ & $-28.7$ & $31.1$ &   & $-380$ & $-260$ & $190$ & $500$ & $40$ & $21$ & $34.5$ & $8.8$ & $0.0$ & $35.9$ &   & $-280$ & $-320$ & $300$ & $520$ & $460$ & $121$ \\ Stat. & $^{+0.1}_{-0.1}$ & $^{+2.3}_{-2.5}$ & $^{+2.7}_{-2.5}$ &   & $^{+60}_{-60}$ & $^{+30}_{-40}$ & $^{+30}_{-30}$ & $^{+50}_{-50}$ & $^{+60}_{-50}$ & \ldots & $^{+14.3}_{-10.5}$ & $^{+8.9}_{-6.2}$ & $^{+0.0}_{-0.0}$ & $^{+15.8}_{-11.3}$ &   & $^{+60}_{-70}$ & $^{+20}_{-20}$ & $^{+40}_{-40}$ & $^{+30}_{-20}$ & $^{+40}_{-30}$ & $^{+23}_{-18}$ \\
B1139 & $-2.8$ & $21.9$ & $12.0$ &   & $160$ & $200$ & $170$ & $310$ & $-200$ & $100$ & $-10.9$ & $5.8$ & $0.0$ & $12.6$ &   & $70$ & $330$ & $230$ & $410$ & $270$ & $59$ \\ Stat. & $^{+0.7}_{-0.6}$ & $^{+2.6}_{-2.3}$ & $^{+1.4}_{-1.3}$ &   & $^{+40}_{-30}$ & $^{+20}_{-20}$ & $^{+30}_{-30}$ & $^{+20}_{-20}$ & $^{+30}_{-20}$ & \ldots & $^{+1.8}_{-2.2}$ & $^{+3.5}_{-3.1}$ & $^{+0.0}_{-0.0}$ & $^{+3.0}_{-2.5}$ &   & $^{+50}_{-50}$ & $^{+30}_{-30}$ & $^{+20}_{-20}$ & $^{+20}_{-20}$ & $^{+30}_{-20}$ & $^{+7}_{-6}$ \\
\hline\end{tabular}}

\tablefoot{Results and statistical uncertainties (\textit{``Stat.'' row}) are given as median values and $1\sigma$ confidence limits. Quantities are described in Sect.~\ref{sec:results}.}
\end{table*}

\begin{table*}
\caption{Kinematic quantities for the program stars based on Model II.}\label{tab:kinematic-TF}
\tiny
\centering
\setlength{\tabcolsep}{0.075cm}
\renewcommand{\arraystretch}{1.05}\scalebox{.84}{\begin{tabular}{lrrrrrrrrrrrrrrrrrrrrr}
\hline\hline
Object & $x$ & $y$ & $z$ & & $\varv_x$ & $\varv_y$ & $\varv_z$ & $\varv_{\mathrm{Grf}}$ & $\varv_{\mathrm{Grf}}-\varv_{\mathrm{esc}}$ 		& $P_{\mathrm{b}}$ & $x_{\mathrm{p}}$ & $y_{\mathrm{p}}$ & $z_{\mathrm{p}}$ & $r_{\mathrm{p}}$ & & $\varv_{x\mathrm{,p}}$ & 		$\varv_{y\mathrm{,p}}$ & $\varv_{z\mathrm{,p}}$ & $\varv_{\mathrm{Grf,p}}$ &  $\varv_{\mathrm{ej,p}}$ & $\tau_{\mathrm{flight,p}}$ \\\cline{2-4} \cline{6-10} \cline{12-15} \cline{17-21}& \multicolumn{3}{c}{(kpc)} & & \multicolumn{5}{c}{$(\mathrm{km\,s^{-1}})$} & (\%) & \multicolumn{4}{c}{(kpc)} & & \multicolumn{5}{c}{$(\mathrm{km\,s^{-1}})$} & (Myr) \\
\hline \hline
HVS1 & $-101.3$ & $-100.9$ & $83.5$ &   & $-390$ & $-340$ & $450$ & $710$ & $440$ & $0$ & $-26.8$ & $-34.6$ & $0.0$ & $66.7$ &   & $-440$ & $-390$ & $470$ & $770$ & $770$ & $177$ \\ Stat. & $^{+7.6}_{-6.9}$ & $^{+8.2}_{-7.5}$ & $^{+6.2}_{-6.8}$ &   & $^{+160}_{-160}$ & $^{+130}_{-140}$ & $^{+170}_{-170}$ & $^{+70}_{-40}$ & $^{+70}_{-40}$ & \ldots & $^{+60.9}_{-37.7}$ & $^{+51.6}_{-31.7}$ & $^{+0.0}_{-0.0}$ & $^{+33.2}_{-32.1}$ &   & $^{+170}_{-110}$ & $^{+150}_{-100}$ & $^{+150}_{-110}$ & $^{+50}_{-30}$ & $^{+120}_{-100}$ & $^{+83}_{-47}$ \\
HVS4 & $-61.9$ & $-14.1$ & $50.8$ &   & $-360$ & $-300$ & $360$ & $630$ & $280$ & $0$ & $-10.0$ & $23.5$ & $0.0$ & $43.8$ &   & $-440$ & $-260$ & $440$ & $670$ & $820$ & $128$ \\ Stat. & $^{+4.1}_{-4.8}$ & $^{+1.1}_{-1.3}$ & $^{+4.5}_{-3.9}$ &   & $^{+180}_{-180}$ & $^{+170}_{-180}$ & $^{+160}_{-160}$ & $^{+110}_{-\phantom{0}60}$ & $^{+110}_{-\phantom{0}60}$ & \ldots & $^{+53.0}_{-32.1}$ & $^{+21.4}_{-17.6}$ & $^{+0.0}_{-0.0}$ & $^{+28.2}_{-21.6}$ &   & $^{+210}_{-\phantom{0}90}$ & $^{+150}_{-200}$ & $^{+110}_{-120}$ & $^{+80}_{-30}$ & $^{+\phantom{0}70}_{-130}$ & $^{+62}_{-35}$ \\
HVS5 & $-32.1$ & $15.9$ & $22.9$ &   & $-400$ & $300$ & $410$ & $650$ & $240$ & $0$ & $-8.6$ & $-1.0$ & $0.0$ & $9.1$ &   & $-510$ & $320$ & $460$ & $760$ & $670$ & $52$ \\ Stat. & $^{+2.2}_{-2.4}$ & $^{+1.7}_{-1.5}$ & $^{+2.4}_{-2.1}$ &   & $^{+30}_{-30}$ & $^{+50}_{-50}$ & $^{+30}_{-30}$ & $^{+10}_{-10}$ & $^{+20}_{-20}$ & \ldots & $^{+2.4}_{-2.3}$ & $^{+3.0}_{-2.7}$ & $^{+0.0}_{-0.0}$ & $^{+2.3}_{-2.3}$ &   & $^{+40}_{-50}$ & $^{+40}_{-30}$ & $^{+20}_{-20}$ & $^{+30}_{-20}$ & $^{+80}_{-60}$ & $^{+6}_{-5}$ \\
HVS5 (Brown 2015) & $-37.5$ & $19.5$ & $28.1$ &   & $-420$ & $260$ & $420$ & $650$ & $260$ & $0$ & $-8.8$ & $1.8$ & $0.0$ & $9.9$ &   & $-520$ & $310$ & $470$ & $760$ & $740$ & $62$ \\ Stat. & $^{+3.4}_{-3.4}$ & $^{+2.3}_{-2.3}$ & $^{+3.3}_{-3.3}$ &   & $^{+40}_{-40}$ & $^{+60}_{-70}$ & $^{+40}_{-30}$ & $^{+10}_{-10}$ & $^{+20}_{-20}$ & \ldots & $^{+3.5}_{-3.3}$ & $^{+4.6}_{-4.0}$ & $^{+0.0}_{-0.0}$ & $^{+3.5}_{-3.3}$ &   & $^{+50}_{-50}$ & $^{+40}_{-50}$ & $^{+20}_{-20}$ & $^{+30}_{-20}$ & $^{+110}_{-\phantom{0}90}$ & $^{+8}_{-7}$ \\
HVS6 & $-20.3$ & $-23.6$ & $45.0$ &   & $-150$ & $-150$ & $450$ & $530$ & $140$ & $0$ & $-4.4$ & $-7.0$ & $0.0$ & $17.3$ &   & $-220$ & $-230$ & $560$ & $640$ & $660$ & $89$ \\ Stat. & $^{+1.1}_{-1.2}$ & $^{+2.2}_{-2.4}$ & $^{+4.5}_{-4.1}$ &   & $^{+160}_{-160}$ & $^{+120}_{-120}$ & $^{+80}_{-80}$ & $^{+60}_{-30}$ & $^{+60}_{-30}$ & \ldots & $^{+15.2}_{-14.0}$ & $^{+12.9}_{-11.3}$ & $^{+0.0}_{-0.0}$ & $^{+11.5}_{-\phantom{0}9.1}$ &   & $^{+150}_{-\phantom{0}80}$ & $^{+130}_{-\phantom{0}60}$ & $^{+40}_{-50}$ & $^{+50}_{-40}$ & $^{+90}_{-80}$ & $^{+17}_{-13}$ \\
HVS7 & $-11.1$ & $-25.1$ & $40.3$ &   & $-200$ & $0$ & $450$ & $500$ & $100$ & $0$ & $6.1$ & $-21.9$ & $0.0$ & $23.9$ &   & $-200$ & $-90$ & $510$ & $560$ & $520$ & $82$ \\ Stat. & $^{+0.3}_{-0.3}$ & $^{+2.1}_{-2.5}$ & $^{+3.9}_{-3.3}$ &   & $^{+90}_{-90}$ & $^{+50}_{-50}$ & $^{+40}_{-40}$ & $^{+50}_{-40}$ & $^{+50}_{-40}$ & \ldots & $^{+8.3}_{-7.1}$ & $^{+4.7}_{-5.0}$ & $^{+0.0}_{-0.0}$ & $^{+5.5}_{-4.8}$ &   & $^{+70}_{-80}$ & $^{+60}_{-60}$ & $^{+30}_{-20}$ & $^{+30}_{-20}$ & $^{+30}_{-30}$ & $^{+10}_{-\phantom{0}8}$ \\
HVS8 & $-29.7$ & $-13.2$ & $26.3$ &   & $-410$ & $80$ & $260$ & $500$ & $80$ & $0$ & $8.6$ & $-16.2$ & $0.0$ & $19.2$ &   & $-440$ & $-40$ & $350$ & $570$ & $440$ & $85$ \\ Stat. & $^{+1.6}_{-1.8}$ & $^{+1.0}_{-1.1}$ & $^{+2.2}_{-2.0}$ &   & $^{+60}_{-60}$ & $^{+60}_{-60}$ & $^{+40}_{-40}$ & $^{+50}_{-40}$ & $^{+50}_{-40}$ & \ldots & $^{+9.9}_{-7.6}$ & $^{+4.9}_{-5.5}$ & $^{+0.0}_{-0.0}$ & $^{+8.4}_{-5.8}$ &   & $^{+30}_{-40}$ & $^{+70}_{-70}$ & $^{+30}_{-40}$ & $^{+20}_{-10}$ & $^{+40}_{-20}$ & $^{+14}_{-11}$ \\
HVS9 & $-44.7$ & $-70.6$ & $80.9$ &   & $50$ & $-180$ & $520$ & $710$ & $400$ & $0$ & $-48.3$ & $-40.8$ & $0.0$ & $95.3$ &   & $0$ & $-220$ & $550$ & $710$ & $760$ & $146$ \\ Stat. & $^{+3.7}_{-3.9}$ & $^{+7.1}_{-7.4}$ & $^{+8.5}_{-8.1}$ &   & $^{+420}_{-420}$ & $^{+300}_{-300}$ & $^{+250}_{-250}$ & $^{+280}_{-180}$ & $^{+280}_{-180}$ & \ldots & $^{+71.5}_{-61.6}$ & $^{+79.2}_{-42.0}$ & $^{+0.0}_{-0.0}$ & $^{+50.2}_{-37.3}$ &   & $^{+440}_{-360}$ & $^{+310}_{-250}$ & $^{+240}_{-190}$ & $^{+270}_{-130}$ & $^{+270}_{-160}$ & $^{+102}_{-\phantom{0}46}$ \\
HVS10 & $-14.5$ & $-16.8$ & $70.2$ &   & $-250$ & $-190$ & $360$ & $500$ & $140$ & $0$ & $25.4$ & $14.6$ & $0.0$ & $36.9$ &   & $-190$ & $-160$ & $480$ & $570$ & $590$ & $163$ \\ Stat. & $^{+0.6}_{-0.6}$ & $^{+1.6}_{-1.6}$ & $^{+6.6}_{-6.4}$ &   & $^{+130}_{-140}$ & $^{+140}_{-150}$ & $^{+40}_{-40}$ & $^{+90}_{-60}$ & $^{+90}_{-60}$ & \ldots & $^{+25.3}_{-19.5}$ & $^{+27.9}_{-20.0}$ & $^{+0.0}_{-0.0}$ & $^{+28.5}_{-19.6}$ &   & $^{+\phantom{0}90}_{-130}$ & $^{+\phantom{0}70}_{-120}$ & $^{+60}_{-60}$ & $^{+50}_{-20}$ & $^{+50}_{-40}$ & $^{+27}_{-21}$ \\
HVS12 & $-26.1$ & $-42.1$ & $59.5$ &   & $-50$ & $70$ & $550$ & $570$ & $220$ & $0$ & $-19.5$ & $-46.7$ & $0.0$ & $52.3$ &   & $-80$ & $20$ & $580$ & $590$ & $610$ & $102$ \\ Stat. & $^{+1.7}_{-1.9}$ & $^{+4.0}_{-4.5}$ & $^{+6.4}_{-5.6}$ &   & $^{+130}_{-130}$ & $^{+110}_{-110}$ & $^{+80}_{-80}$ & $^{+100}_{-\phantom{0}80}$ & $^{+100}_{-\phantom{0}80}$ & \ldots & $^{+13.7}_{-13.1}$ & $^{+12.2}_{-12.1}$ & $^{+0.0}_{-0.0}$ & $^{+12.3}_{-12.3}$ &   & $^{+130}_{-120}$ & $^{+120}_{-120}$ & $^{+80}_{-70}$ & $^{+90}_{-60}$ & $^{+80}_{-70}$ & $^{+18}_{-14}$ \\
HVS13 & $-32.3$ & $-72.1$ & $92.7$ &   & $-660$ & $-40$ & $340$ & $780$ & $470$ & $0$ & $126.5$ & $-54.2$ & $0.0$ & $149.6$ &   & $-620$ & $-90$ & $390$ & $770$ & $720$ & $242$ \\ Stat. & $^{+2.3}_{-2.0}$ & $^{+6.7}_{-5.9}$ & $^{+7.6}_{-8.6}$ &   & $^{+220}_{-230}$ & $^{+200}_{-200}$ & $^{+150}_{-150}$ & $^{+210}_{-180}$ & $^{+210}_{-180}$ & \ldots & $^{+130.0}_{-\phantom{0}67.3}$ & $^{+79.1}_{-43.2}$ & $^{+0.0}_{-0.0}$ & $^{+119.1}_{-\phantom{0}56.9}$ &   & $^{+230}_{-240}$ & $^{+210}_{-190}$ & $^{+130}_{-140}$ & $^{+210}_{-160}$ & $^{+210}_{-170}$ & $^{+143}_{-\phantom{0}68}$ \\
HVS15 & $-10.4$ & $-34.2$ & $50.7$ &   & $-60$ & $-170$ & $280$ & $460$ & $80$ & $22$ & $-0.1$ & $-3.7$ & $0.0$ & $45.4$ &   & $-80$ & $-240$ & $400$ & $500$ & $560$ & $147$ \\ Stat. & $^{+0.3}_{-0.3}$ & $^{+3.6}_{-4.1}$ & $^{+6.1}_{-5.3}$ &   & $^{+340}_{-350}$ & $^{+190}_{-190}$ & $^{+140}_{-140}$ & $^{+210}_{-100}$ & $^{+210}_{-100}$ & \ldots & $^{+65.5}_{-43.1}$ & $^{+50.5}_{-29.0}$ & $^{+0.0}_{-0.0}$ & $^{+52.1}_{-25.5}$ &   & $^{+300}_{-250}$ & $^{+200}_{-100}$ & $^{+\phantom{0}80}_{-140}$ & $^{+140}_{-\phantom{0}50}$ & $^{+130}_{-100}$ & $^{+82}_{-39}$ \\
HVS16 & $-1.5$ & $-24.2$ & $60.4$ &   & $-270$ & $-470$ & $210$ & $680$ & $310$ & $4$ & $53.6$ & $79.6$ & $0.0$ & $130.9$ &   & $-210$ & $-430$ & $290$ & $650$ & $660$ & $222$ \\ Stat. & $^{+0.7}_{-0.6}$ & $^{+2.2}_{-2.2}$ & $^{+5.5}_{-5.3}$ &   & $^{+480}_{-480}$ & $^{+240}_{-250}$ & $^{+\phantom{0}90}_{-100}$ & $^{+350}_{-230}$ & $^{+360}_{-230}$ & \ldots & $^{+128.2}_{-\phantom{0}96.9}$ & $^{+136.7}_{-\phantom{0}62.0}$ & $^{+0.0}_{-0.0}$ & $^{+168.7}_{-\phantom{0}80.2}$ &   & $^{+350}_{-490}$ & $^{+210}_{-250}$ & $^{+110}_{-110}$ & $^{+350}_{-160}$ & $^{+330}_{-140}$ & $^{+130}_{-\phantom{0}60}$ \\
HVS17 & $-0.8$ & $25.6$ & $23.3$ &   & $190$ & $270$ & $310$ & $460$ & $30$ & $2$ & $-12.3$ & $4.4$ & $0.0$ & $13.5$ &   & $120$ & $370$ & $390$ & $550$ & $430$ & $65$ \\ Stat. & $^{+0.7}_{-0.6}$ & $^{+2.1}_{-2.0}$ & $^{+1.9}_{-1.8}$ &   & $^{+60}_{-60}$ & $^{+30}_{-40}$ & $^{+30}_{-30}$ & $^{+20}_{-20}$ & $^{+30}_{-20}$ & \ldots & $^{+3.4}_{-3.8}$ & $^{+3.7}_{-3.5}$ & $^{+0.0}_{-0.0}$ & $^{+4.1}_{-3.7}$ &   & $^{+70}_{-70}$ & $^{+40}_{-40}$ & $^{+30}_{-30}$ & $^{+30}_{-10}$ & $^{+30}_{-20}$ & $^{+6}_{-5}$ \\
HVS18 & $-28.6$ & $83.5$ & $-43.3$ &   & $30$ & $460$ & $-130$ & $560$ & $240$ & $0$ & $-25.2$ & $-25.7$ & $0.0$ & $77.4$ &   & $-60$ & $470$ & $-230$ & $570$ & $540$ & $221$ \\ Stat. & $^{+1.9}_{-2.1}$ & $^{+8.7}_{-7.5}$ & $^{+3.9}_{-4.5}$ &   & $^{+300}_{-300}$ & $^{+130}_{-130}$ & $^{+200}_{-200}$ & $^{+160}_{-\phantom{0}90}$ & $^{+160}_{-\phantom{0}90}$ & \ldots & $^{+65.6}_{-72.1}$ & $^{+\phantom{0}62.4}_{-193.2}$ & $^{+0.0}_{-0.0}$ & $^{+188.1}_{-\phantom{0}40.1}$ &   & $^{+310}_{-170}$ & $^{+80}_{-70}$ & $^{+140}_{-130}$ & $^{+110}_{-\phantom{0}60}$ & $^{+190}_{-\phantom{0}80}$ & $^{+336}_{-\phantom{0}99}$ \\
HVS19 & $-17.5$ & $-36.9$ & $77.0$ &   & $-230$ & $90$ & $560$ & $920$ & $580$ & $0$ & $12.0$ & $-44.7$ & $0.0$ & $98.8$ &   & $-220$ & $50$ & $600$ & $920$ & $920$ & $127$ \\ Stat. & $^{+1.1}_{-1.2}$ & $^{+4.5}_{-4.5}$ & $^{+9.3}_{-9.2}$ &   & $^{+750}_{-760}$ & $^{+410}_{-400}$ & $^{+220}_{-220}$ & $^{+470}_{-300}$ & $^{+470}_{-300}$ & \ldots & $^{+132.9}_{-\phantom{0}87.0}$ & $^{+68.0}_{-40.8}$ & $^{+0.0}_{-0.0}$ & $^{+88.2}_{-44.6}$ &   & $^{+700}_{-740}$ & $^{+410}_{-360}$ & $^{+200}_{-200}$ & $^{+450}_{-260}$ & $^{+450}_{-240}$ & $^{+71}_{-35}$ \\
HVS20 & $-15.0$ & $-50.8$ & $90.2$ &   & $40$ & $300$ & $620$ & $970$ & $650$ & $0$ & $-19.0$ & $-87.3$ & $0.0$ & $128.1$ &   & $20$ & $270$ & $650$ & $960$ & $970$ & $134$ \\ Stat. & $^{+0.9}_{-0.7}$ & $^{+6.3}_{-4.6}$ & $^{+\phantom{0}8.2}_{-11.1}$ &   & $^{+720}_{-720}$ & $^{+410}_{-400}$ & $^{+240}_{-240}$ & $^{+480}_{-360}$ & $^{+480}_{-360}$ & \ldots & $^{+112.7}_{-\phantom{0}93.3}$ & $^{+61.7}_{-37.4}$ & $^{+0.0}_{-0.0}$ & $^{+60.7}_{-45.2}$ &   & $^{+720}_{-670}$ & $^{+420}_{-410}$ & $^{+230}_{-200}$ & $^{+470}_{-340}$ & $^{+470}_{-330}$ & $^{+73}_{-37}$ \\
HVS21 & $-60.2$ & $13.6$ & $79.8$ &   & $-130$ & $200$ & $360$ & $590$ & $270$ & $0$ & $-29.4$ & $-22.2$ & $0.0$ & $88.4$ &   & $-190$ & $150$ & $440$ & $590$ & $620$ & $196$ \\ Stat. & $^{+5.3}_{-6.2}$ & $^{+1.7}_{-1.4}$ & $^{+9.5}_{-8.1}$ &   & $^{+290}_{-290}$ & $^{+370}_{-370}$ & $^{+220}_{-220}$ & $^{+250}_{-150}$ & $^{+250}_{-150}$ & \ldots & $^{+111.5}_{-\phantom{0}50.8}$ & $^{+\phantom{0}61.5}_{-124.5}$ & $^{+0.0}_{-0.0}$ & $^{+99.4}_{-41.9}$ &   & $^{+310}_{-200}$ & $^{+340}_{-300}$ & $^{+170}_{-200}$ & $^{+200}_{-\phantom{0}80}$ & $^{+190}_{-130}$ & $^{+166}_{-\phantom{0}65}$ \\
HVS22 & $-13.7$ & $-45.4$ & $86.2$ &   & $-380$ & $720$ & $910$ & $1530$ & $1200$ & $0$ & $21.2$ & $-108.8$ & $0.0$ & $139.5$ &   & $-370$ & $700$ & $920$ & $1520$ & $1510$ & $91$ \\ Stat. & $^{+0.8}_{-0.9}$ & $^{+6.5}_{-7.6}$ & $^{+14.4}_{-12.3}$ &   & $^{+950}_{-970}$ & $^{+590}_{-560}$ & $^{+320}_{-310}$ & $^{+690}_{-560}$ & $^{+690}_{-560}$ & \ldots & $^{+114.9}_{-\phantom{0}83.4}$ & $^{+46.5}_{-39.1}$ & $^{+0.0}_{-0.0}$ & $^{+63.3}_{-43.6}$ &   & $^{+930}_{-970}$ & $^{+600}_{-560}$ & $^{+320}_{-290}$ & $^{+690}_{-560}$ & $^{+690}_{-550}$ & $^{+46}_{-23}$ \\
HVS24 & $-17.0$ & $-35.4$ & $51.2$ &   & $80$ & $-90$ & $390$ & $460$ & $90$ & $8$ & $-23.4$ & $-20.7$ & $0.0$ & $40.5$ &   & $10$ & $-160$ & $450$ & $520$ & $600$ & $118$ \\ Stat. & $^{+0.8}_{-0.8}$ & $^{+3.2}_{-3.0}$ & $^{+4.4}_{-4.6}$ &   & $^{+230}_{-230}$ & $^{+140}_{-150}$ & $^{+110}_{-110}$ & $^{+140}_{-\phantom{0}80}$ & $^{+140}_{-\phantom{0}80}$ & \ldots & $^{+27.2}_{-26.2}$ & $^{+24.0}_{-17.3}$ & $^{+0.0}_{-0.0}$ & $^{+20.6}_{-19.3}$ &   & $^{+250}_{-180}$ & $^{+160}_{-120}$ & $^{+90}_{-70}$ & $^{+100}_{-\phantom{0}40}$ & $^{+\phantom{0}90}_{-100}$ & $^{+34}_{-24}$ \\
B095 & $-43.6$ & $16.1$ & $45.3$ &   & $-120$ & $470$ & $100$ & $550$ & $180$ & $13$ & $-8.0$ & $-82.8$ & $0.0$ & $96.3$ &   & $-170$ & $360$ & $220$ & $480$ & $460$ & $238$ \\ Stat. & $^{+3.5}_{-3.5}$ & $^{+1.6}_{-1.6}$ & $^{+4.5}_{-4.5}$ &   & $^{+200}_{-200}$ & $^{+230}_{-230}$ & $^{+180}_{-180}$ & $^{+210}_{-170}$ & $^{+210}_{-170}$ & \ldots & $^{+115.6}_{-\phantom{0}43.5}$ & $^{+\phantom{0}62.3}_{-205.8}$ & $^{+0.0}_{-0.0}$ & $^{+225.2}_{-\phantom{0}49.6}$ &   & $^{+190}_{-110}$ & $^{+210}_{-200}$ & $^{+130}_{-130}$ & $^{+160}_{-\phantom{0}80}$ & $^{+160}_{-100}$ & $^{+364}_{-100}$ \\
B129 & $-84.1$ & $-21.8$ & $33.8$ &   & $-190$ & $-70$ & $250$ & $390$ & $50$ & $27$ & $-55.4$ & $-10.6$ & $0.0$ & $68.3$ &   & $-250$ & $-90$ & $270$ & $430$ & $480$ & $126$ \\ Stat. & $^{+6.8}_{-6.9}$ & $^{+2.0}_{-2.0}$ & $^{+3.1}_{-3.1}$ &   & $^{+120}_{-120}$ & $^{+200}_{-200}$ & $^{+190}_{-190}$ & $^{+150}_{-\phantom{0}80}$ & $^{+150}_{-\phantom{0}80}$ & \ldots & $^{+72.5}_{-22.9}$ & $^{+22.1}_{-24.7}$ & $^{+0.0}_{-0.0}$ & $^{+19.7}_{-26.1}$ &   & $^{+150}_{-120}$ & $^{+150}_{-170}$ & $^{+180}_{-110}$ & $^{+120}_{-\phantom{0}50}$ & $^{+160}_{-150}$ & $^{+160}_{-\phantom{0}52}$ \\
B143 & $-30.7$ & $7.9$ & $16.1$ &   & $-200$ & $110$ & $180$ & $290$ & $-140$ & $100$ & $-10.3$ & $-1.8$ & $0.0$ & $10.7$ &   & $-360$ & $120$ & $240$ & $450$ & $410$ & $77$ \\ Stat. & $^{+1.8}_{-2.0}$ & $^{+0.7}_{-0.7}$ & $^{+1.5}_{-1.3}$ &   & $^{+20}_{-20}$ & $^{+30}_{-30}$ & $^{+30}_{-30}$ & $^{+10}_{-10}$ & $^{+10}_{-10}$ & \ldots & $^{+4.0}_{-3.6}$ & $^{+2.5}_{-2.3}$ & $^{+0.0}_{-0.0}$ & $^{+3.3}_{-3.5}$ &   & $^{+50}_{-60}$ & $^{+20}_{-20}$ & $^{+20}_{-10}$ & $^{+50}_{-40}$ & $^{+60}_{-50}$ & $^{+9}_{-8}$ \\
B167 & $-32.5$ & $-2.7$ & $22.0$ &   & $-280$ & $210$ & $130$ & $370$ & $-50$ & $90$ & $8.5$ & $-22.4$ & $0.0$ & $24.6$ &   & $-350$ & $70$ & $220$ & $430$ & $310$ & $120$ \\ Stat. & $^{+1.8}_{-2.0}$ & $^{+0.2}_{-0.3}$ & $^{+1.8}_{-1.7}$ &   & $^{+40}_{-40}$ & $^{+50}_{-50}$ & $^{+40}_{-40}$ & $^{+40}_{-40}$ & $^{+40}_{-40}$ & \ldots & $^{+11.3}_{-\phantom{0}8.5}$ & $^{+5.9}_{-7.7}$ & $^{+0.0}_{-0.0}$ & $^{+10.9}_{-\phantom{0}6.9}$ &   & $^{+10}_{-10}$ & $^{+60}_{-60}$ & $^{+30}_{-30}$ & $^{+10}_{-20}$ & $^{+40}_{-30}$ & $^{+25}_{-18}$ \\
B329 & $40.7$ & $16.6$ & $51.5$ &   & $160$ & $-240$ & $330$ & $480$ & $120$ & $20$ & $12.8$ & $46.2$ & $0.0$ & $58.2$ &   & $220$ & $-170$ & $380$ & $500$ & $480$ & $140$ \\ Stat. & $^{+6.3}_{-5.3}$ & $^{+2.1}_{-1.8}$ & $^{+6.6}_{-5.6}$ &   & $^{+160}_{-160}$ & $^{+210}_{-220}$ & $^{+160}_{-150}$ & $^{+190}_{-140}$ & $^{+190}_{-140}$ & \ldots & $^{+25.0}_{-42.8}$ & $^{+37.0}_{-29.0}$ & $^{+0.0}_{-0.0}$ & $^{+35.5}_{-25.4}$ &   & $^{+100}_{-160}$ & $^{+220}_{-240}$ & $^{+130}_{-110}$ & $^{+170}_{-\phantom{0}70}$ & $^{+160}_{-110}$ & $^{+76}_{-39}$ \\
B434 & $-16.1$ & $-22.8$ & $33.7$ &   & $130$ & $-280$ & $210$ & $380$ & $-30$ & $71$ & $-25.9$ & $15.2$ & $0.0$ & $31.5$ &   & $0$ & $-280$ & $310$ & $420$ & $570$ & $124$ \\ Stat. & $^{+0.6}_{-0.7}$ & $^{+1.7}_{-2.1}$ & $^{+3.1}_{-2.4}$ &   & $^{+80}_{-80}$ & $^{+60}_{-60}$ & $^{+40}_{-40}$ & $^{+50}_{-40}$ & $^{+60}_{-40}$ & \ldots & $^{+\phantom{0}9.6}_{-10.7}$ & $^{+13.1}_{-\phantom{0}9.2}$ & $^{+0.0}_{-0.0}$ & $^{+13.0}_{-\phantom{0}9.8}$ &   & $^{+90}_{-90}$ & $^{+30}_{-40}$ & $^{+30}_{-40}$ & $^{+20}_{-10}$ & $^{+20}_{-20}$ & $^{+24}_{-18}$ \\
B458 & $-26.7$ & $-52.5$ & $61.5$ &   & $170$ & $-20$ & $430$ & $570$ & $230$ & $8$ & $-45.5$ & $-45.1$ & $0.0$ & $80.4$ &   & $120$ & $-70$ & $460$ & $570$ & $630$ & $133$ \\ Stat. & $^{+2.0}_{-2.3}$ & $^{+5.7}_{-6.4}$ & $^{+7.5}_{-6.7}$ &   & $^{+360}_{-350}$ & $^{+220}_{-220}$ & $^{+210}_{-200}$ & $^{+270}_{-190}$ & $^{+280}_{-190}$ & \ldots & $^{+49.8}_{-46.8}$ & $^{+49.2}_{-26.5}$ & $^{+0.0}_{-0.0}$ & $^{+34.1}_{-29.6}$ &   & $^{+370}_{-300}$ & $^{+240}_{-210}$ & $^{+190}_{-150}$ & $^{+260}_{-140}$ & $^{+260}_{-150}$ & $^{+87}_{-41}$ \\
B481 & $-6.1$ & $28.7$ & $-36.4$ &   & $-220$ & $-30$ & $-400$ & $460$ & $60$ & $21$ & $12.7$ & $27.9$ & $0.0$ & $31.1$ &   & $-210$ & $60$ & $-440$ & $500$ & $620$ & $83$ \\ Stat. & $^{+0.3}_{-0.3}$ & $^{+3.0}_{-3.1}$ & $^{+3.9}_{-3.7}$ &   & $^{+70}_{-80}$ & $^{+60}_{-60}$ & $^{+50}_{-50}$ & $^{+80}_{-70}$ & $^{+90}_{-80}$ & \ldots & $^{+6.1}_{-5.5}$ & $^{+7.7}_{-7.6}$ & $^{+0.0}_{-0.0}$ & $^{+8.5}_{-8.4}$ &   & $^{+70}_{-70}$ & $^{+80}_{-80}$ & $^{+40}_{-40}$ & $^{+60}_{-40}$ & $^{+50}_{-40}$ & $^{+8}_{-7}$ \\
B485 & $-26.8$ & $-6.0$ & $27.4$ &   & $-330$ & $140$ & $270$ & $450$ & $30$ & $2$ & $4.3$ & $-14.8$ & $0.0$ & $15.5$ &   & $-380$ & $20$ & $370$ & $530$ & $420$ & $84$ \\ Stat. & $^{+1.4}_{-1.6}$ & $^{+0.5}_{-0.6}$ & $^{+2.4}_{-2.1}$ &   & $^{+30}_{-30}$ & $^{+20}_{-20}$ & $^{+20}_{-20}$ & $^{+20}_{-20}$ & $^{+20}_{-20}$ & \ldots & $^{+3.5}_{-2.9}$ & $^{+1.7}_{-2.0}$ & $^{+0.0}_{-0.0}$ & $^{+2.8}_{-2.1}$ &   & $^{+10}_{-10}$ & $^{+30}_{-30}$ & $^{+10}_{-20}$ & $^{+10}_{-10}$ & $^{+10}_{-10}$ & $^{+9}_{-8}$ \\
B537 & $-22.0$ & $28.2$ & $-26.8$ &   & $-120$ & $230$ & $-180$ & $320$ & $-90$ & $100$ & $-2.5$ & $-3.0$ & $0.0$ & $5.5$ &   & $-290$ & $300$ & $-370$ & $560$ & $450$ & $111$ \\ Stat. & $^{+1.6}_{-1.9}$ & $^{+3.8}_{-3.3}$ & $^{+3.1}_{-3.7}$ &   & $^{+50}_{-50}$ & $^{+30}_{-30}$ & $^{+20}_{-30}$ & $^{+20}_{-20}$ & $^{+20}_{-20}$ & \ldots & $^{+3.9}_{-4.8}$ & $^{+3.1}_{-3.0}$ & $^{+0.0}_{-0.0}$ & $^{+3.9}_{-2.9}$ &   & $^{+80}_{-50}$ & $^{+60}_{-50}$ & $^{+\phantom{0}60}_{-100}$ & $^{+70}_{-60}$ & $^{+220}_{-\phantom{0}90}$ & $^{+12}_{-11}$ \\
B572 & $-21.6$ & $18.9$ & $-14.0$ &   & $-70$ & $250$ & $-170$ & $320$ & $-130$ & $100$ & $-12.6$ & $-1.5$ & $0.0$ & $13.3$ &   & $-210$ & $290$ & $-230$ & $430$ & $300$ & $71$ \\ Stat. & $^{+1.4}_{-2.0}$ & $^{+2.9}_{-2.0}$ & $^{+1.5}_{-2.1}$ &   & $^{+50}_{-50}$ & $^{+30}_{-30}$ & $^{+50}_{-50}$ & $^{+20}_{-20}$ & $^{+30}_{-20}$ & \ldots & $^{+5.8}_{-4.8}$ & $^{+4.8}_{-4.7}$ & $^{+0.0}_{-0.0}$ & $^{+4.6}_{-4.2}$ &   & $^{+\phantom{0}80}_{-100}$ & $^{+20}_{-30}$ & $^{+20}_{-30}$ & $^{+50}_{-30}$ & $^{+60}_{-40}$ & $^{+14}_{-12}$ \\
B576 & $-1.9$ & $14.1$ & $48.9$ &   & $-360$ & $-350$ & $460$ & $680$ & $290$ & $0$ & $33.3$ & $46.7$ & $0.0$ & $57.5$ &   & $-330$ & $-290$ & $510$ & $670$ & $730$ & $98$ \\ Stat. & $^{+0.5}_{-0.5}$ & $^{+1.1}_{-1.1}$ & $^{+3.8}_{-3.6}$ &   & $^{+50}_{-50}$ & $^{+60}_{-60}$ & $^{+30}_{-20}$ & $^{+70}_{-60}$ & $^{+70}_{-70}$ & \ldots & $^{+6.1}_{-5.6}$ & $^{+8.2}_{-7.6}$ & $^{+0.0}_{-0.0}$ & $^{+9.7}_{-8.9}$ &   & $^{+50}_{-50}$ & $^{+60}_{-70}$ & $^{+20}_{-20}$ & $^{+60}_{-60}$ & $^{+60}_{-50}$ & $^{+5}_{-5}$ \\
B576 (BHB) & $-6.0$ & $5.2$ & $17.8$ &   & $-110$ & $70$ & $300$ & $330$ & $-160$ & $100$ & $0.1$ & $0.6$ & $0.0$ & $1.0$ &   & $-170$ & $210$ & $590$ & $660$ & $730$ & $45$ \\ Stat. & $^{+0.2}_{-0.2}$ & $^{+0.5}_{-0.4}$ & $^{+1.6}_{-1.4}$ &   & $^{+20}_{-20}$ & $^{+30}_{-30}$ & $^{+10}_{-10}$ & $^{+10}_{-10}$ & $^{+20}_{-20}$ & \ldots  & $^{+1.0}_{-0.8}$ & $^{+1.1}_{-0.8}$ & $^{+0.0}_{-0.0}$ & $^{+1.0}_{-0.6}$ &   & $^{+\phantom{0}60}_{-120}$ & $^{+40}_{-80}$ & $^{+60}_{-40}$ & $^{+70}_{-50}$ & $^{+80}_{-70}$ & $^{+4}_{-3}$ \\
B598 & $1.7$ & $-0.5$ & $20.4$ &   & $40$ & $50$ & $300$ & $310$ & $-170$ & $100$ & $-0.7$ & $-2.1$ & $0.0$ & $3.4$ &   & $30$ & $-30$ & $550$ & $550$ & $610$ & $51$ \\ Stat. & $^{+0.7}_{-0.7}$ & $^{+0.1}_{-0.1}$ & $^{+1.4}_{-1.3}$ &   & $^{+60}_{-60}$ & $^{+60}_{-60}$ & $^{+30}_{-30}$ & $^{+30}_{-20}$ & $^{+30}_{-20}$ & \ldots & $^{+2.4}_{-2.6}$ & $^{+2.3}_{-2.6}$ & $^{+0.0}_{-0.0}$ & $^{+2.5}_{-1.9}$ &   & $^{+50}_{-50}$ & $^{+50}_{-50}$ & $^{+60}_{-60}$ & $^{+70}_{-60}$ & $^{+60}_{-50}$ & $^{+5}_{-4}$ \\
B598 (ELM) & $-5.5$ & $-0.1$ & $5.8$ &   & $110$ & $190$ & $270$ & $350$ & $-220$ & $100$ & $-6.6$ & $-3.4$ & $0.0$ & $7.5$ &   & $0$ & $150$ & $340$ & $380$ & $370$ & $18$ \\ Stat. & $^{+0.5}_{-0.4}$ & $^{+0.1}_{-0.1}$ & $^{+0.9}_{-0.8}$ &   & $^{+20}_{-20}$ & $^{+20}_{-20}$ & $^{+10}_{-10}$ & $^{+20}_{-20}$ & $^{+20}_{-10}$ & \ldots & $^{+0.6}_{-0.5}$ & $^{+0.4}_{-0.4}$ & $^{+0.0}_{-0.0}$ & $^{+0.4}_{-0.5}$ &   & $^{+20}_{-20}$ & $^{+20}_{-30}$ & $^{+20}_{-20}$ & $^{+20}_{-20}$ & $^{+20}_{-20}$ & $^{+3}_{-2}$ \\
B711 & $1.4$ & $0.4$ & $20.8$ &   & $340$ & $90$ & $150$ & $380$ & $-100$ & $100$ & $-23.8$ & $-6.3$ & $0.0$ & $24.7$ &   & $210$ & $60$ & $290$ & $370$ & $420$ & $85$ \\ Stat. & $^{+0.8}_{-0.7}$ & $^{+0.1}_{-0.1}$ & $^{+1.6}_{-1.5}$ &   & $^{+30}_{-30}$ & $^{+30}_{-30}$ & $^{+20}_{-20}$ & $^{+20}_{-20}$ & $^{+30}_{-20}$ & \ldots & $^{+3.7}_{-4.6}$ & $^{+1.7}_{-1.7}$ & $^{+0.0}_{-0.0}$ & $^{+4.5}_{-3.6}$ &   & $^{+30}_{-30}$ & $^{+20}_{-20}$ & $^{+20}_{-20}$ & $^{+10}_{-10}$ & $^{+10}_{-10}$ & $^{+10}_{-\phantom{0}9}$ \\
B733 & $-4.8$ & $4.0$ & $11.0$ &   & $260$ & $150$ & $360$ & $470$ & $-60$ & $100$ & $-10.9$ & $-0.3$ & $0.0$ & $10.9$ &   & $160$ & $170$ & $430$ & $490$ & $470$ & $27$ \\ Stat. & $^{+0.3}_{-0.3}$ & $^{+0.3}_{-0.3}$ & $^{+0.8}_{-0.8}$ &   & $^{+20}_{-20}$ & $^{+20}_{-20}$ & $^{+10}_{-10}$ & $^{+10}_{-10}$ & $^{+10}_{-10}$ & \ldots & $^{+0.5}_{-0.5}$ & $^{+0.5}_{-0.5}$ & $^{+0.0}_{-0.0}$ & $^{+0.5}_{-0.5}$ &   & $^{+20}_{-20}$ & $^{+20}_{-20}$ & $^{+10}_{-10}$ & $^{+10}_{-10}$ & $^{+10}_{-10}$ & $^{+2}_{-2}$ \\
B1080 & $-22.0$ & $-33.7$ & $38.2$ &   & $-250$ & $-230$ & $180$ & $400$ & $20$ & $38$ & $15.9$ & $7.8$ & $0.0$ & $24.9$ &   & $-210$ & $-290$ & $350$ & $490$ & $480$ & $148$ \\ Stat. & $^{+1.1}_{-1.2}$ & $^{+2.5}_{-2.8}$ & $^{+3.2}_{-2.9}$ &   & $^{+130}_{-130}$ & $^{+100}_{-100}$ & $^{+110}_{-110}$ & $^{+90}_{-50}$ & $^{+90}_{-50}$ & \ldots & $^{+37.0}_{-18.0}$ & $^{+34.8}_{-18.5}$ & $^{+0.0}_{-0.0}$ & $^{+43.7}_{-14.6}$ &   & $^{+60}_{-90}$ & $^{+50}_{-40}$ & $^{+\phantom{0}70}_{-130}$ & $^{+60}_{-40}$ & $^{+100}_{-\phantom{0}50}$ & $^{+79}_{-34}$ \\
B1085 & $-8.5$ & $-28.7$ & $31.1$ &   & $-380$ & $-260$ & $190$ & $500$ & $90$ & $2$ & $35.3$ & $9.2$ & $0.0$ & $36.8$ &   & $-280$ & $-320$ & $290$ & $520$ & $460$ & $123$ \\ Stat. & $^{+0.1}_{-0.1}$ & $^{+2.3}_{-2.5}$ & $^{+2.7}_{-2.5}$ &   & $^{+60}_{-60}$ & $^{+30}_{-40}$ & $^{+30}_{-30}$ & $^{+50}_{-50}$ & $^{+60}_{-50}$ & \ldots & $^{+14.8}_{-10.8}$ & $^{+9.2}_{-6.4}$ & $^{+0.0}_{-0.0}$ & $^{+16.4}_{-11.6}$ &   & $^{+60}_{-70}$ & $^{+20}_{-20}$ & $^{+40}_{-40}$ & $^{+30}_{-20}$ & $^{+50}_{-30}$ & $^{+24}_{-18}$ \\
B1139 & $-2.8$ & $21.9$ & $12.0$ &   & $160$ & $200$ & $170$ & $310$ & $-150$ & $100$ & $-10.9$ & $6.0$ & $0.0$ & $12.8$ &   & $80$ & $330$ & $230$ & $410$ & $260$ & $59$ \\ Stat. & $^{+0.7}_{-0.6}$ & $^{+2.6}_{-2.3}$ & $^{+1.4}_{-1.3}$ &   & $^{+40}_{-30}$ & $^{+20}_{-20}$ & $^{+30}_{-30}$ & $^{+20}_{-20}$ & $^{+30}_{-30}$ & \ldots & $^{+1.8}_{-2.2}$ & $^{+3.5}_{-3.1}$ & $^{+0.0}_{-0.0}$ & $^{+3.0}_{-2.5}$ &   & $^{+50}_{-50}$ & $^{+30}_{-40}$ & $^{+20}_{-20}$ & $^{+20}_{-20}$ & $^{+30}_{-20}$ & $^{+7}_{-6}$ \\

\hline\end{tabular}}

\tablefoot{Same as Table~\ref{tab:kinematic-AS}.}
\end{table*}

\clearpage

\section{Photometry and SED fitting}\label{app:SEDs}
\begin{figure}[h]
\centering
\includegraphics[width=\hsize]{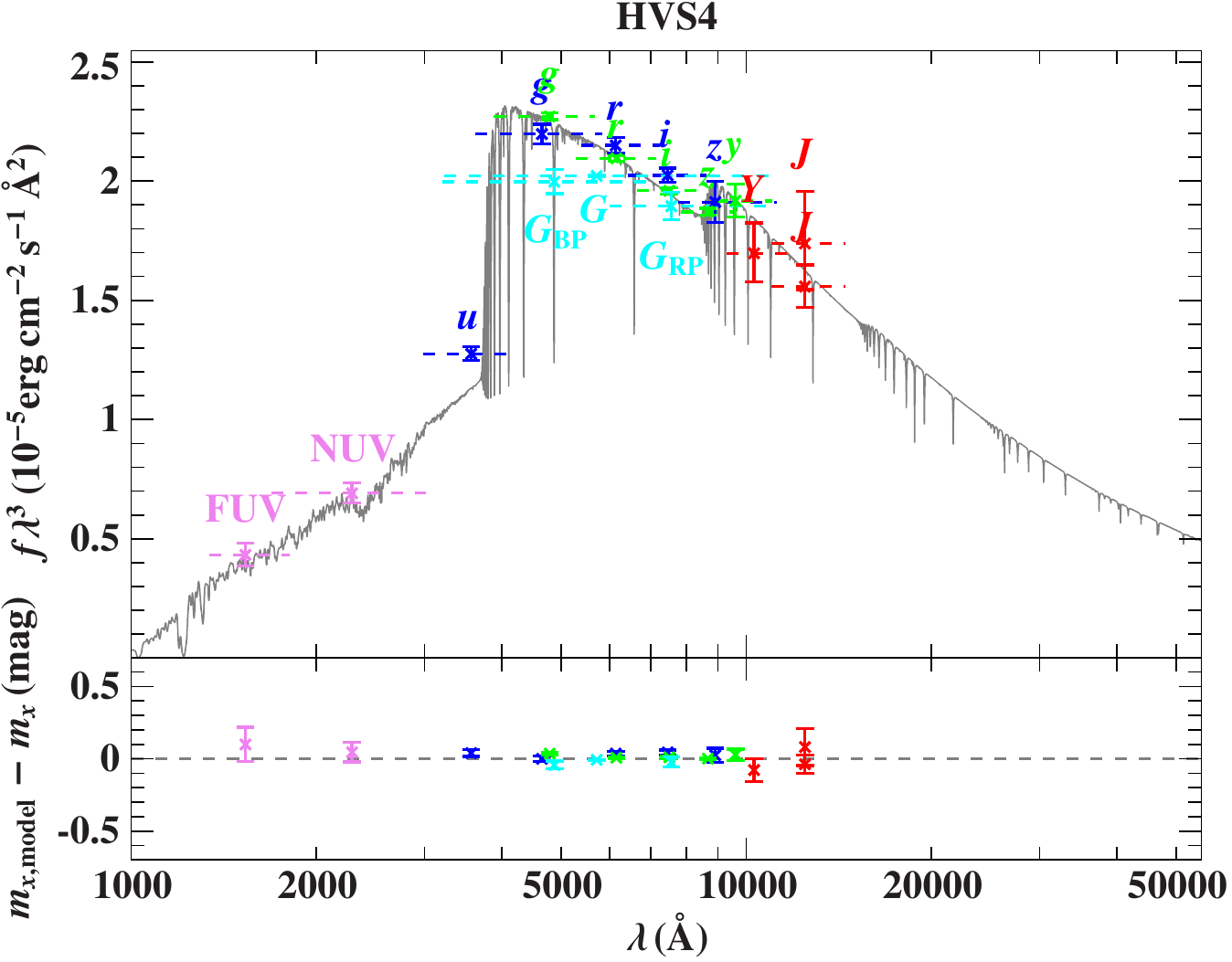}\caption{Same as Fig.~\ref{fig:SED-HVS1-text}. SDSS: blue; Pan-STARRS: green; UKIDSS: red; {\it Gaia}: cyan; GALEX: violet.}\label{fig:SED-HVS4}
\end{figure}\begin{figure}
\centering
\includegraphics[width=\hsize]{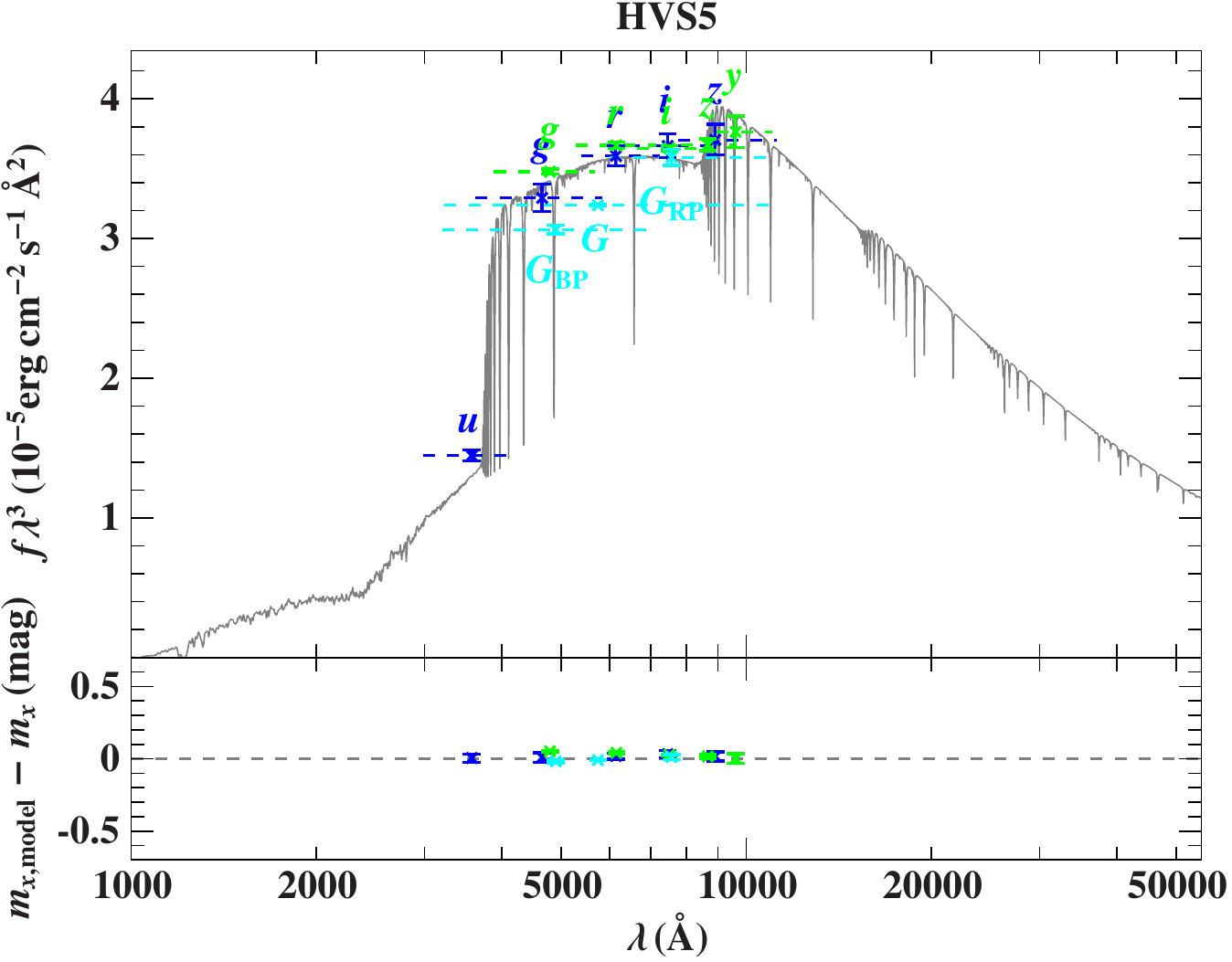}\caption{Same as Fig.~\ref{fig:SED-HVS1-text}. SDSS: blue; Pan-STARRS: green; {\it Gaia}: cyan.}\label{fig:SED-HVS5}
\end{figure}\begin{figure}
\centering
\includegraphics[width=\hsize]{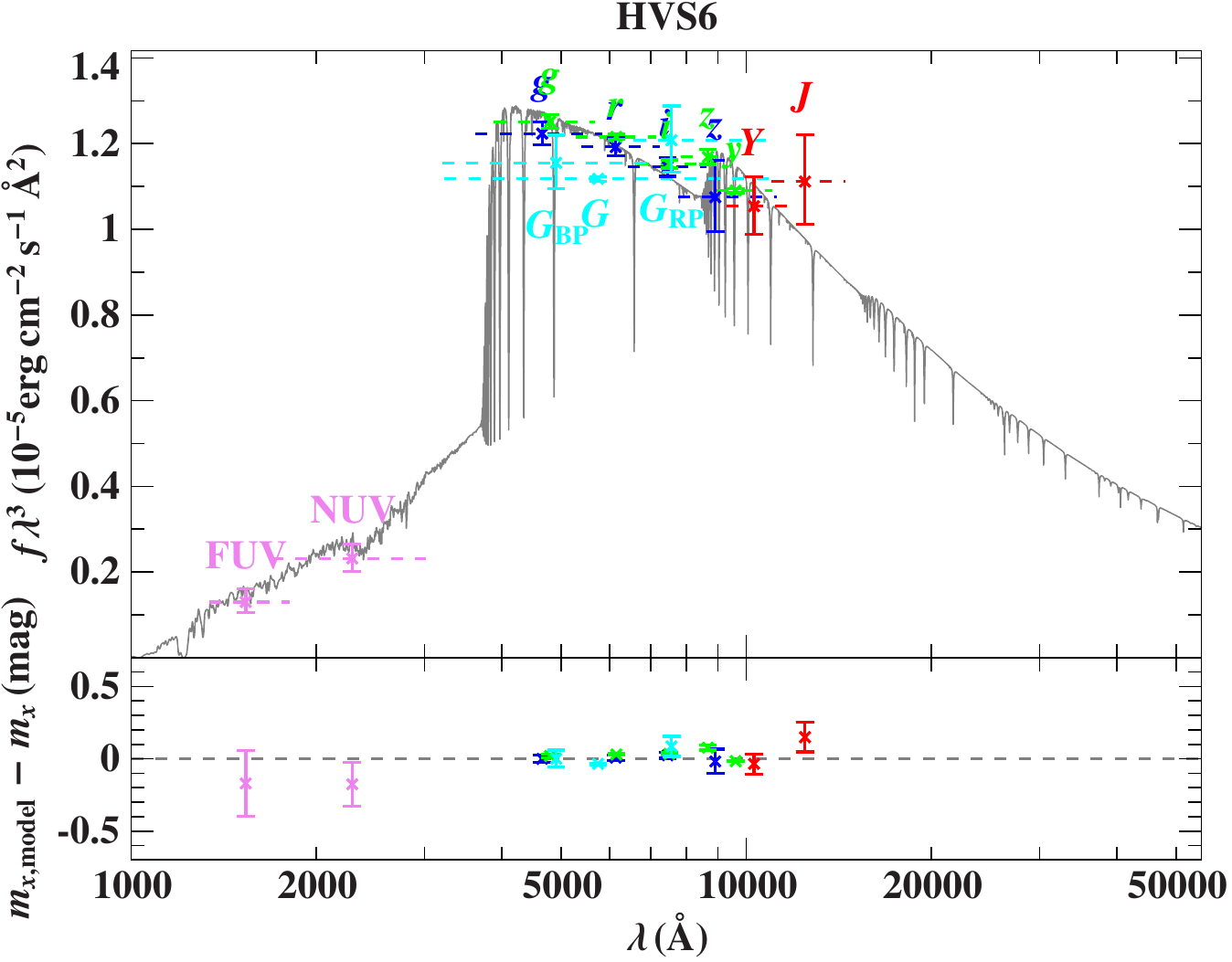}\caption{Same as Fig.~\ref{fig:SED-HVS1-text}. SDSS: blue; Pan-STARRS: green; UKIDSS: red; {\it Gaia}: cyan; GALEX: violet.}\label{fig:SED-HVS6}
\end{figure}\begin{figure}
\centering
\includegraphics[width=\hsize]{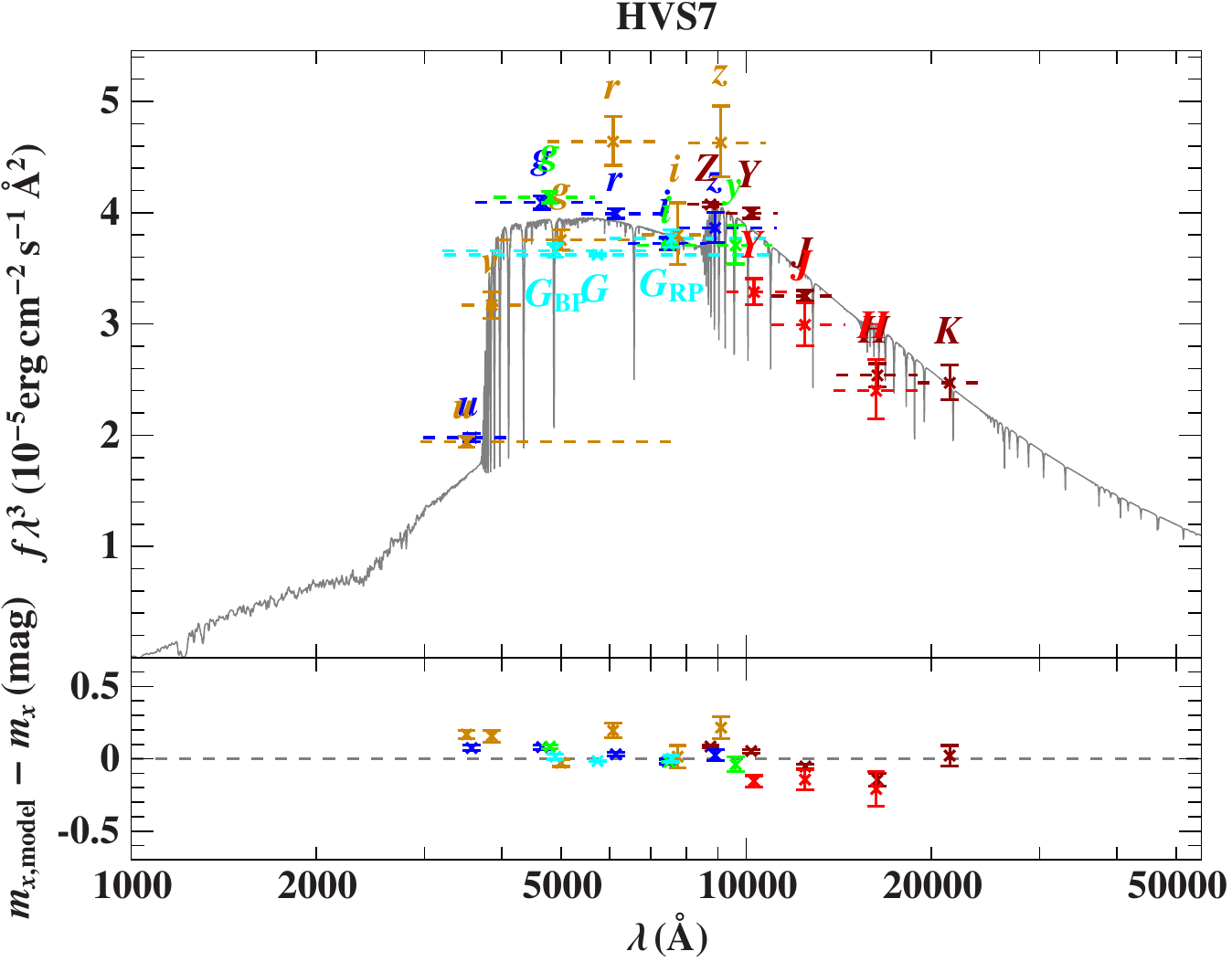}\caption{Same as Fig.~\ref{fig:SED-HVS1-text}. SDSS: blue; Pan-STARRS: green; UKIDSS: red; {\it Gaia}: cyan; VISTA: dark red; SkyMapper: orange.}\label{fig:SED-HVS7}
\end{figure}\begin{figure}
\centering
\includegraphics[width=\hsize]{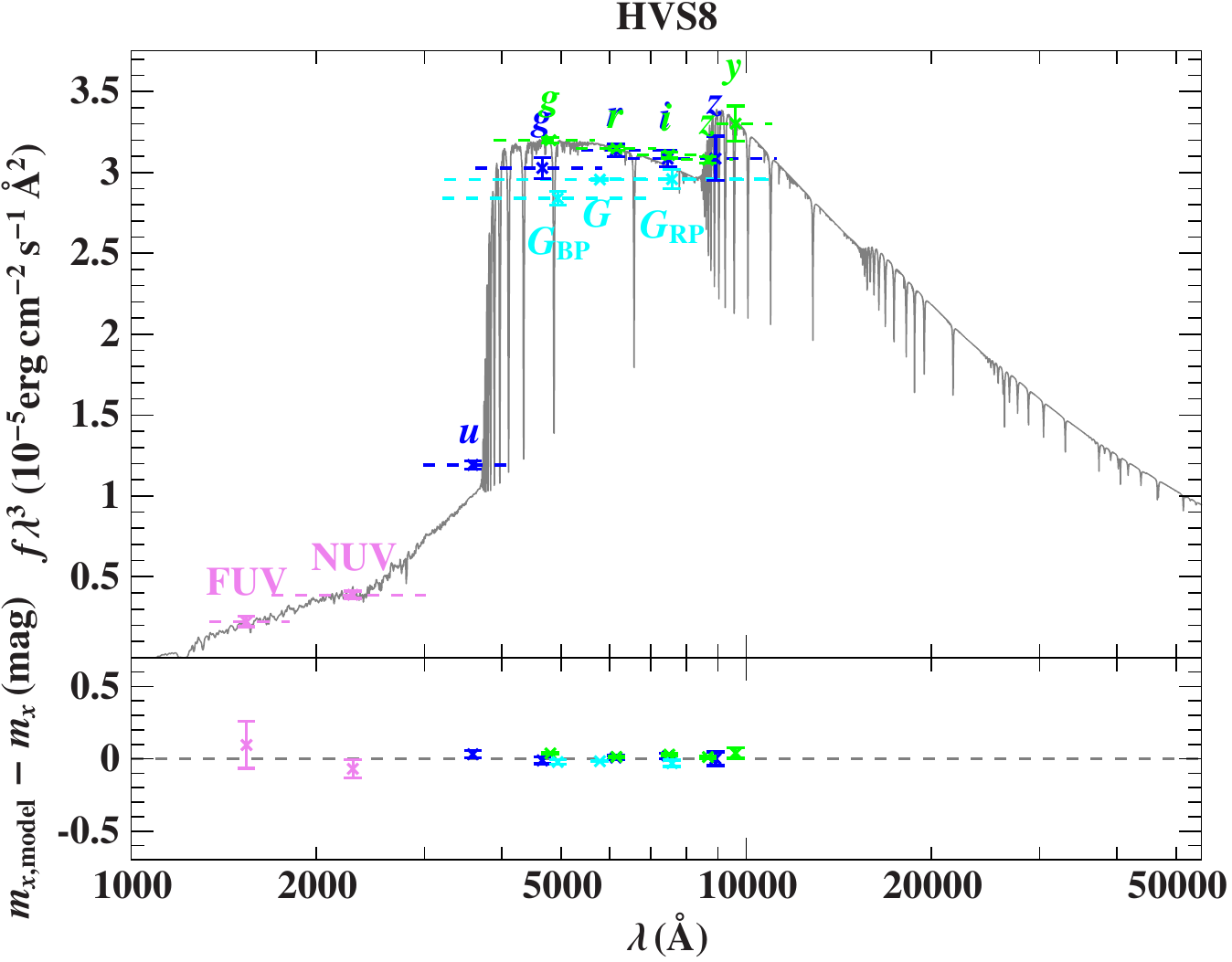}\caption{Same as Fig.~\ref{fig:SED-HVS1-text}. SDSS: blue; Pan-STARRS: green; {\it Gaia}: cyan; GALEX: violet.}\label{fig:SED-HVS8}
\end{figure}\begin{figure}
\centering
\includegraphics[width=\hsize]{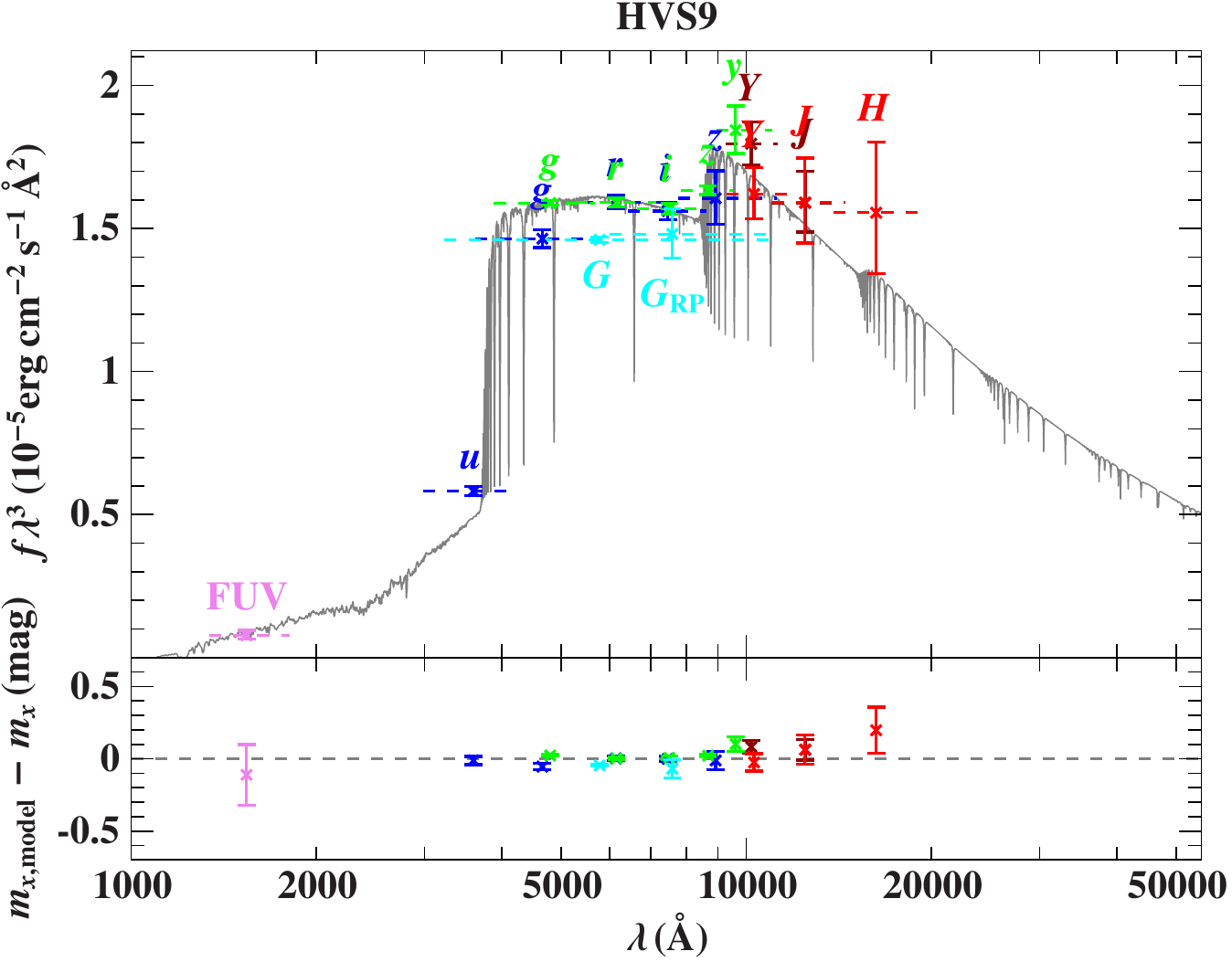}\caption{Same as Fig.~\ref{fig:SED-HVS1-text}. SDSS: blue; Pan-STARRS: green; UKIDSS: red; {\it Gaia}: cyan; GALEX: violet; VISTA: dark red.}\label{fig:SED-HVS9}
\end{figure}\begin{figure}
\centering
\includegraphics[width=\hsize]{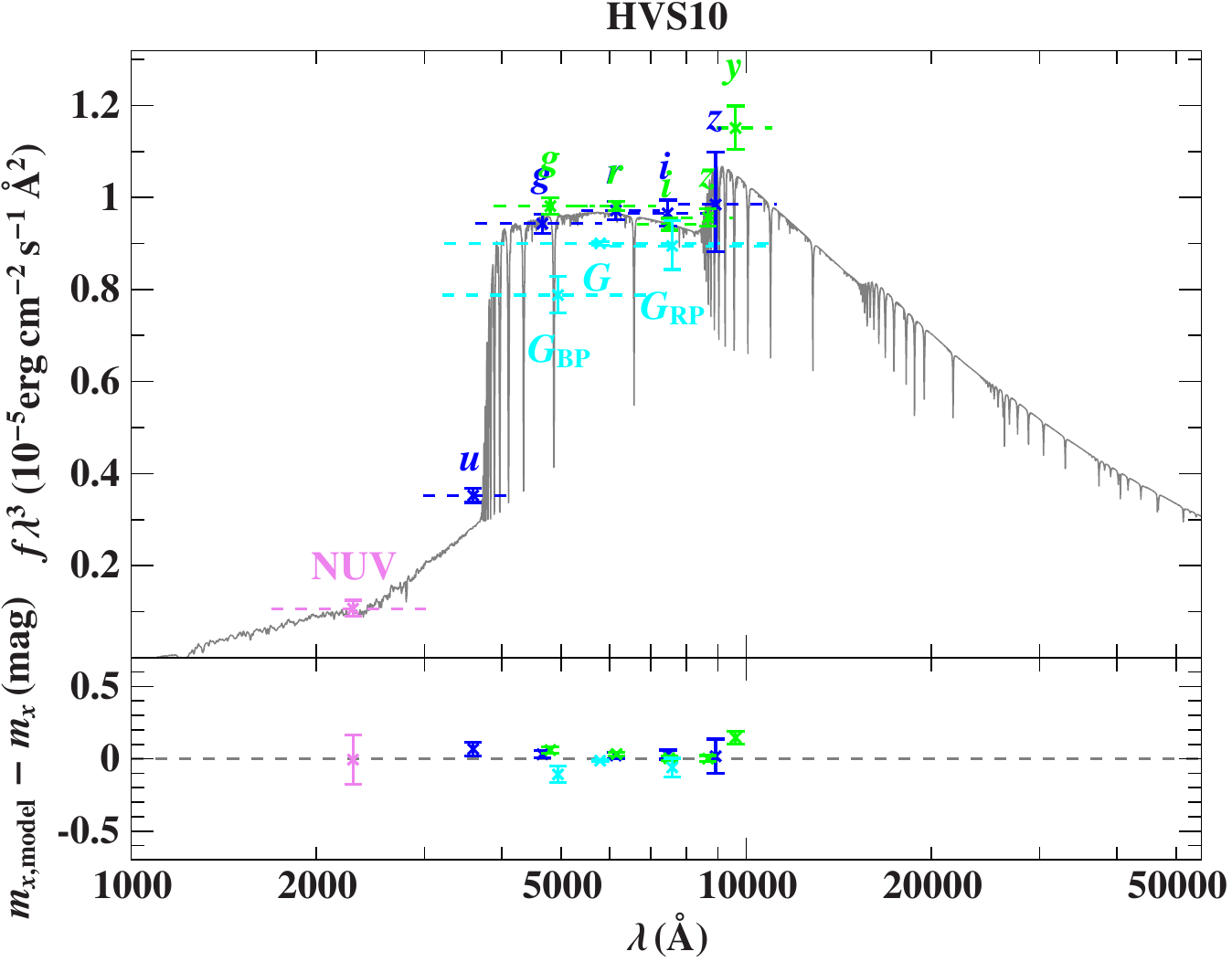}\caption{Same as Fig.~\ref{fig:SED-HVS1-text}. SDSS: blue; Pan-STARRS: green; {\it Gaia}: cyan; GALEX: violet.}\label{fig:SED-HVS10}
\end{figure}\begin{figure}
\centering
\includegraphics[width=\hsize]{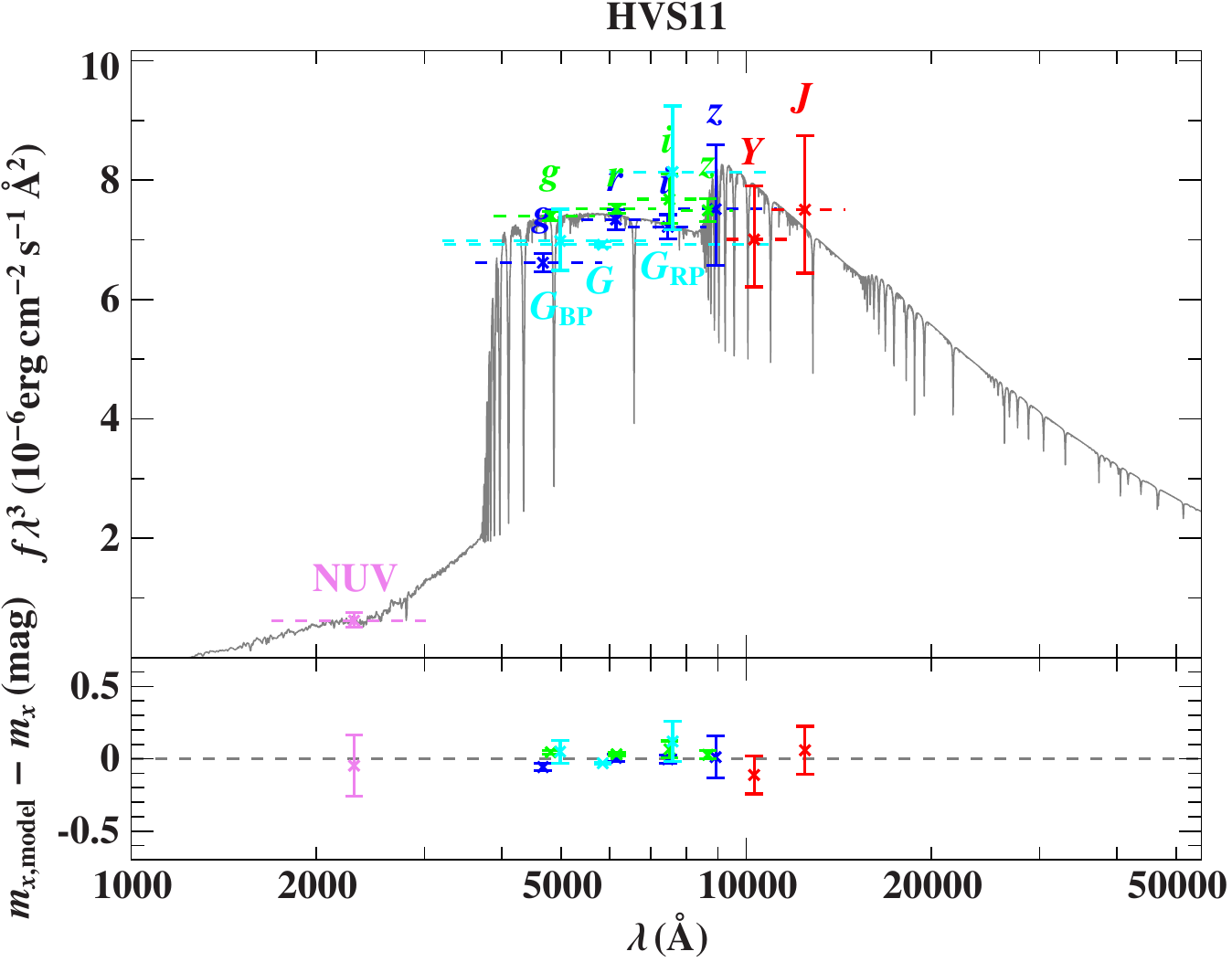}\caption{Same as Fig.~\ref{fig:SED-HVS1-text}. SDSS: blue; Pan-STARRS: green; UKIDSS: red; {\it Gaia}: cyan; GALEX: violet.}\label{fig:SED-HVS11}
\end{figure}\begin{figure}
\centering
\includegraphics[width=\hsize]{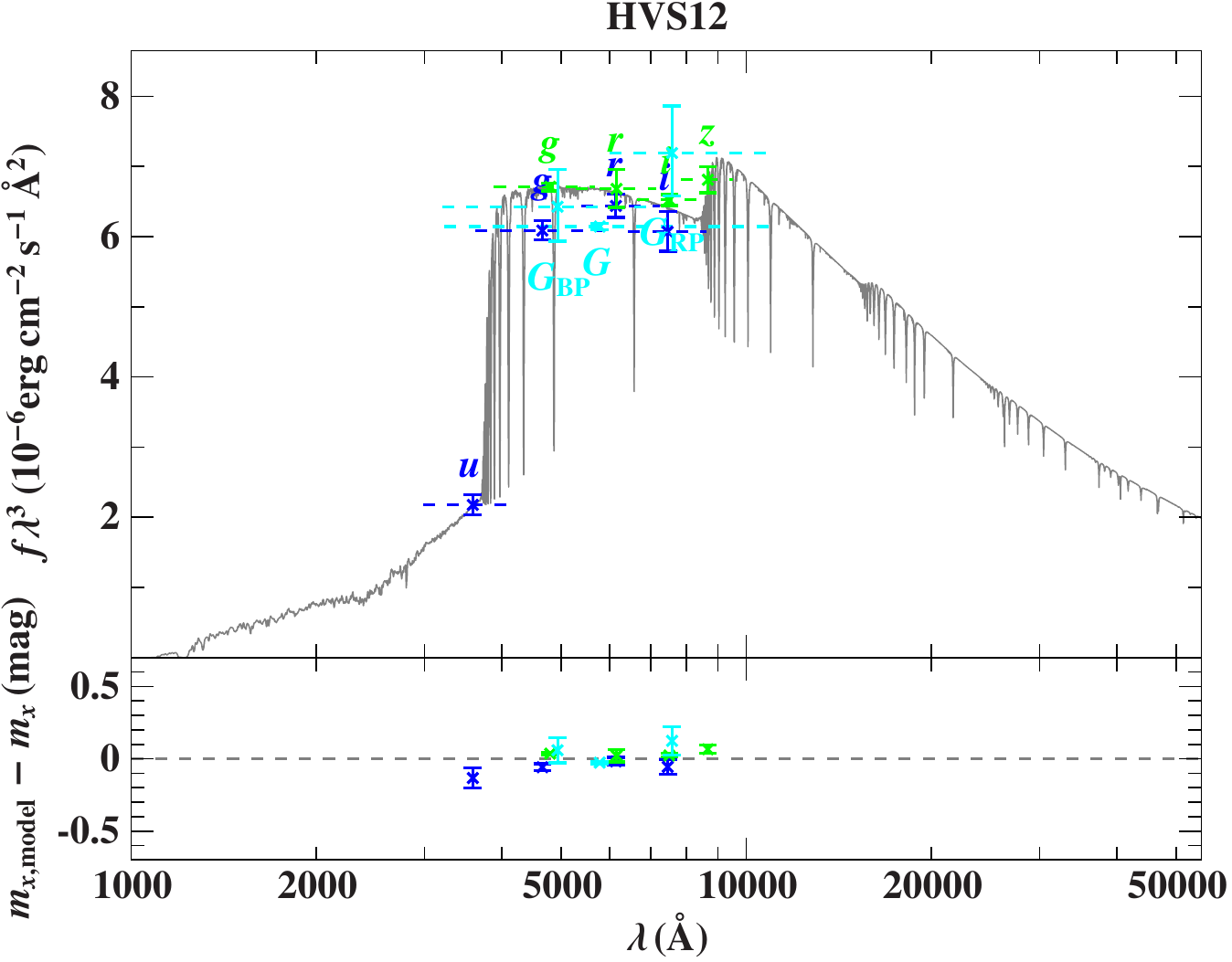}\caption{Same as Fig.~\ref{fig:SED-HVS1-text}. SDSS: blue; Pan-STARRS: green; {\it Gaia}: cyan.}\label{fig:SED-HVS12}
\end{figure}\begin{figure}
\centering
\includegraphics[width=\hsize]{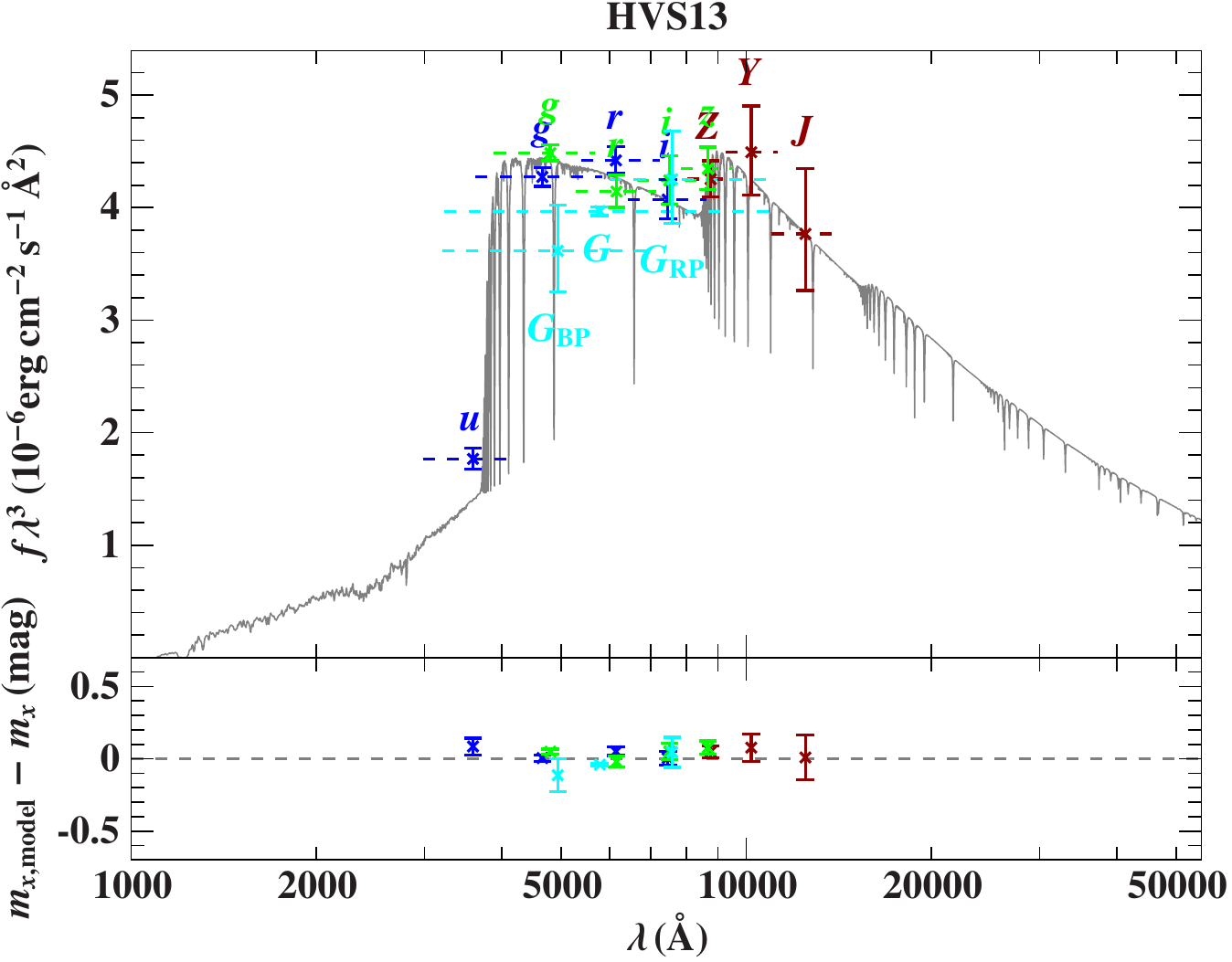}\caption{Same as Fig.~\ref{fig:SED-HVS1-text}. SDSS: blue; Pan-STARRS: green; {\it Gaia}: cyan; VISTA: dark red.}\label{fig:SED-HVS13}
\end{figure}\begin{figure}
\centering
\includegraphics[width=\hsize]{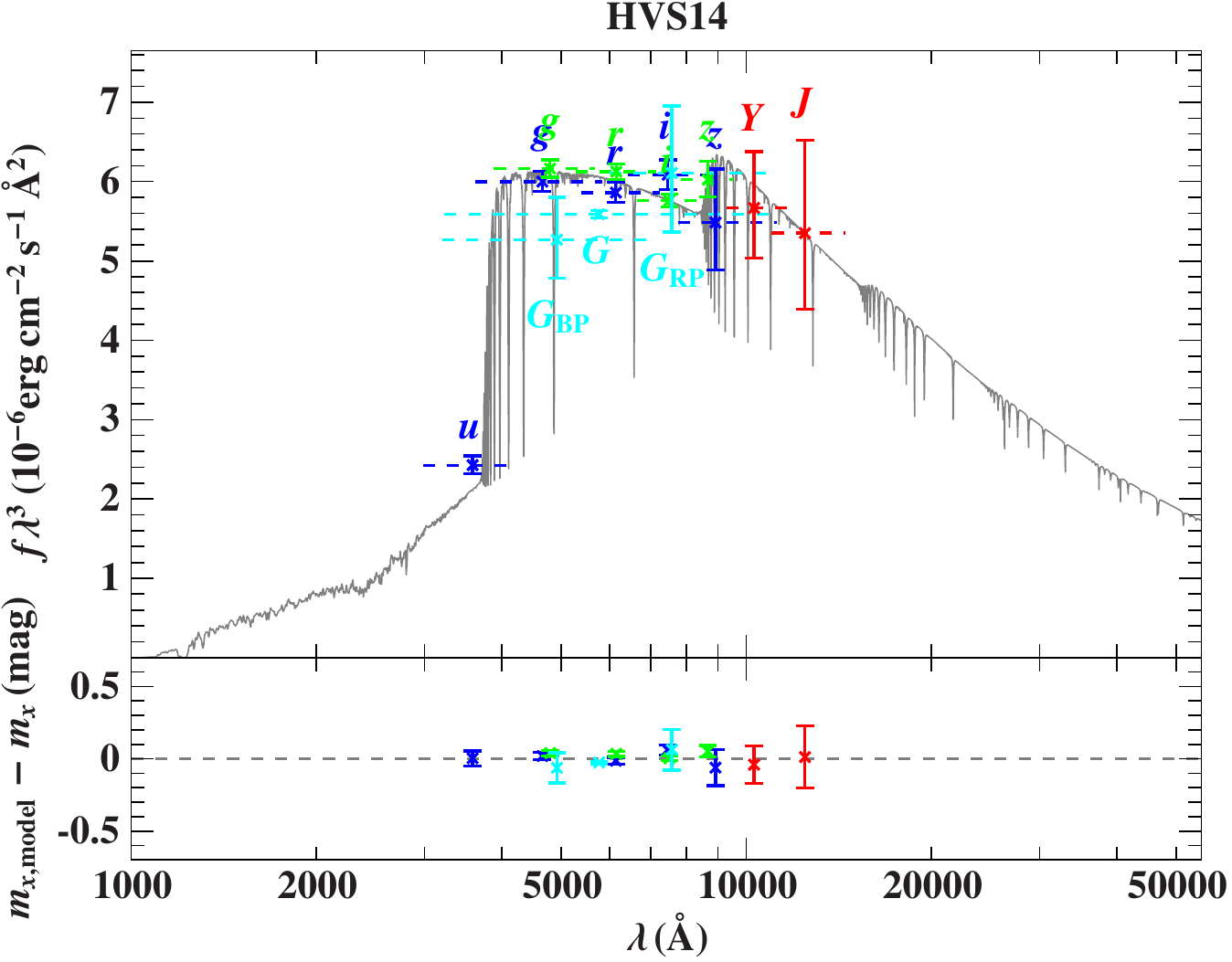}\caption{Same as Fig.~\ref{fig:SED-HVS1-text}. SDSS: blue; Pan-STARRS: green; UKIDSS: red; {\it Gaia}: cyan.}\label{fig:SED-HVS14}
\end{figure}\begin{figure}
\centering
\includegraphics[width=\hsize]{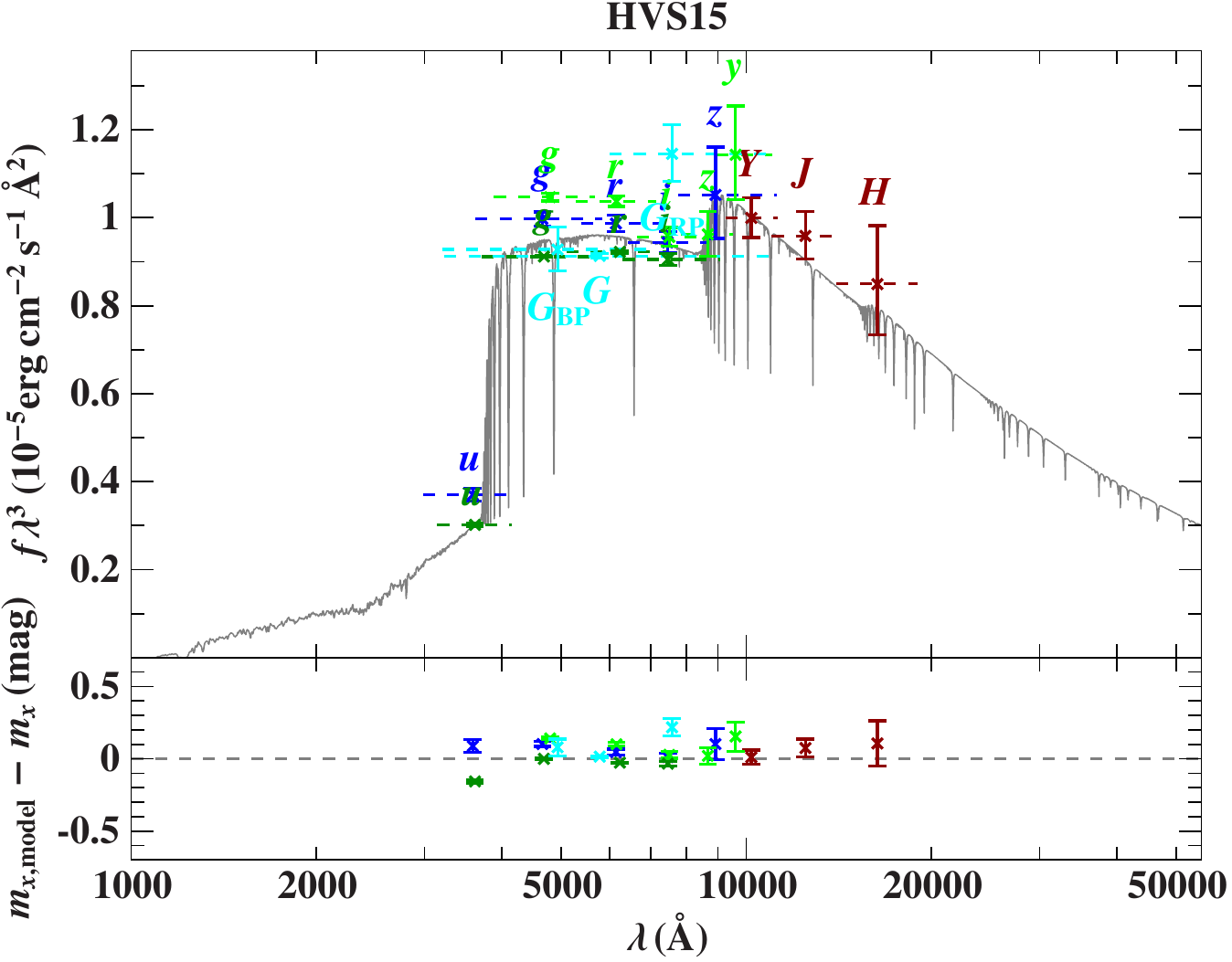}\caption{Same as Fig.~\ref{fig:SED-HVS1-text}. SDSS: blue; Pan-STARRS: green; {\it Gaia}: cyan; VISTA: dark red; VST-KiDs: dark green.}\label{fig:SED-HVS15}
\end{figure}\begin{figure}
\centering
\includegraphics[width=\hsize]{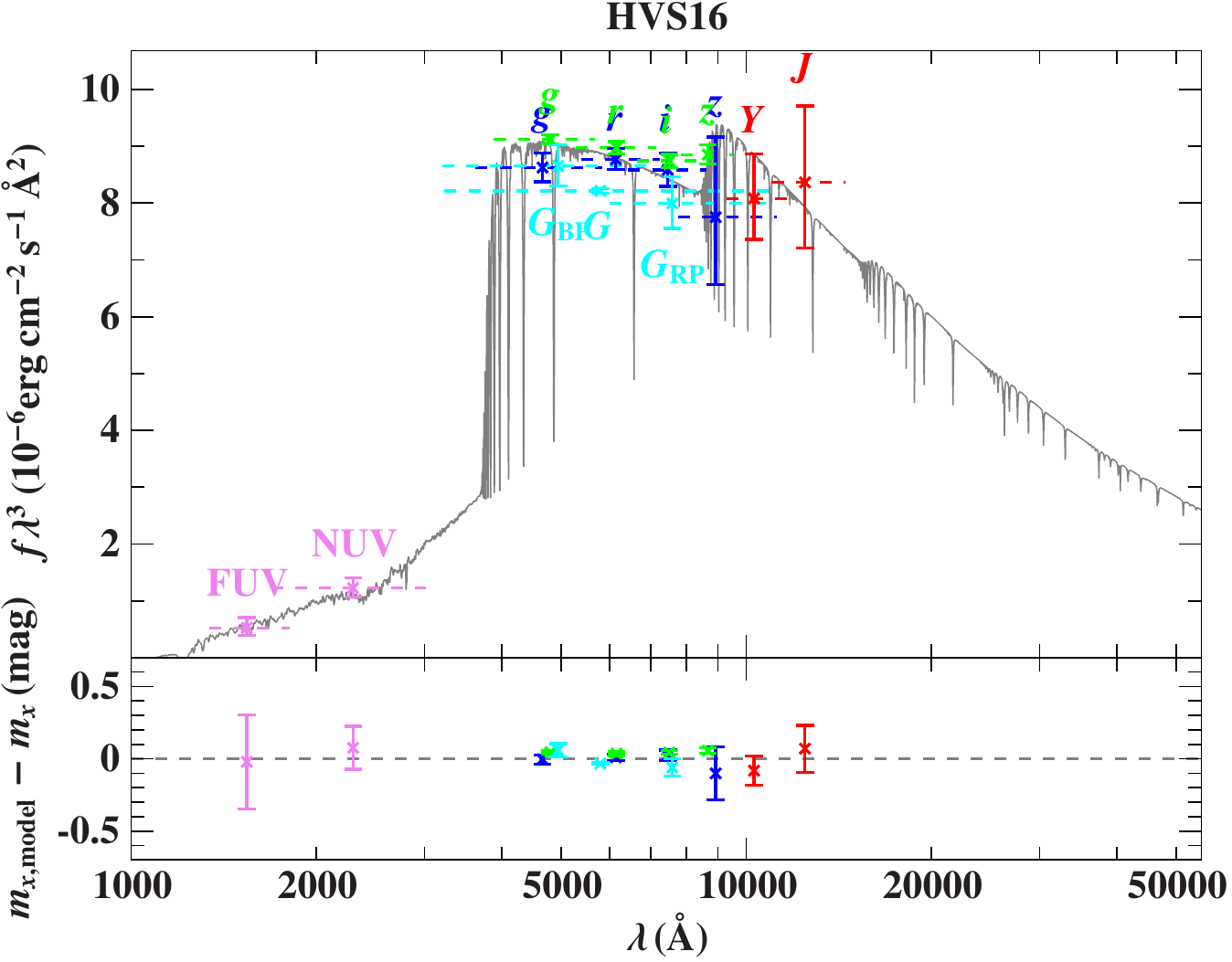}\caption{Same as Fig.~\ref{fig:SED-HVS1-text}. SDSS: blue; Pan-STARRS: green; UKIDSS: red; {\it Gaia}: cyan; GALEX: violet.}\label{fig:SED-HVS16}
\end{figure}\begin{figure}
\centering
\includegraphics[width=\hsize]{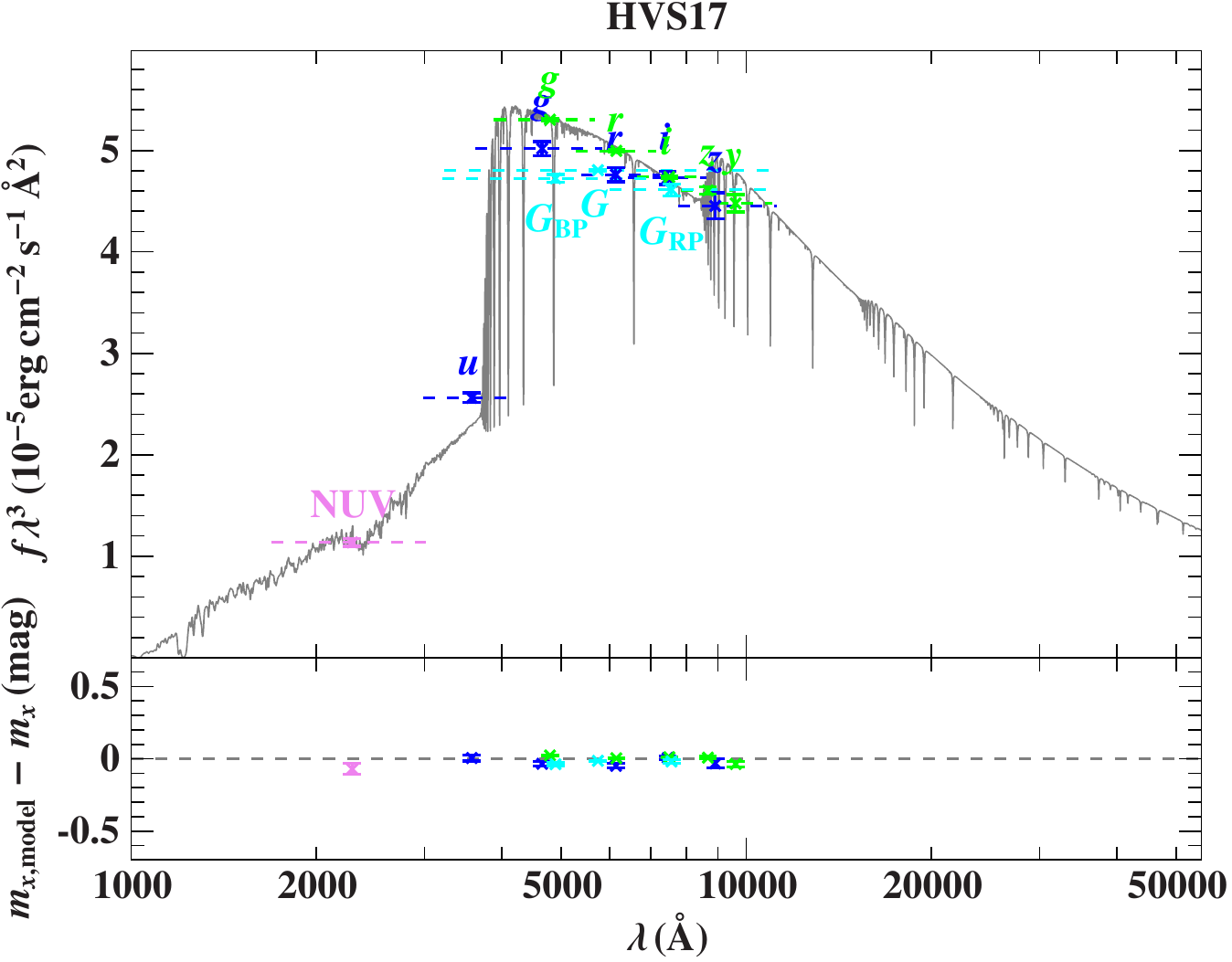}\caption{Same as Fig.~\ref{fig:SED-HVS1-text}. SDSS: blue; Pan-STARRS: green; {\it Gaia}: cyan; GALEX: violet.}\label{fig:SED-HVS17}
\end{figure}\begin{figure}
\centering
\includegraphics[width=\hsize]{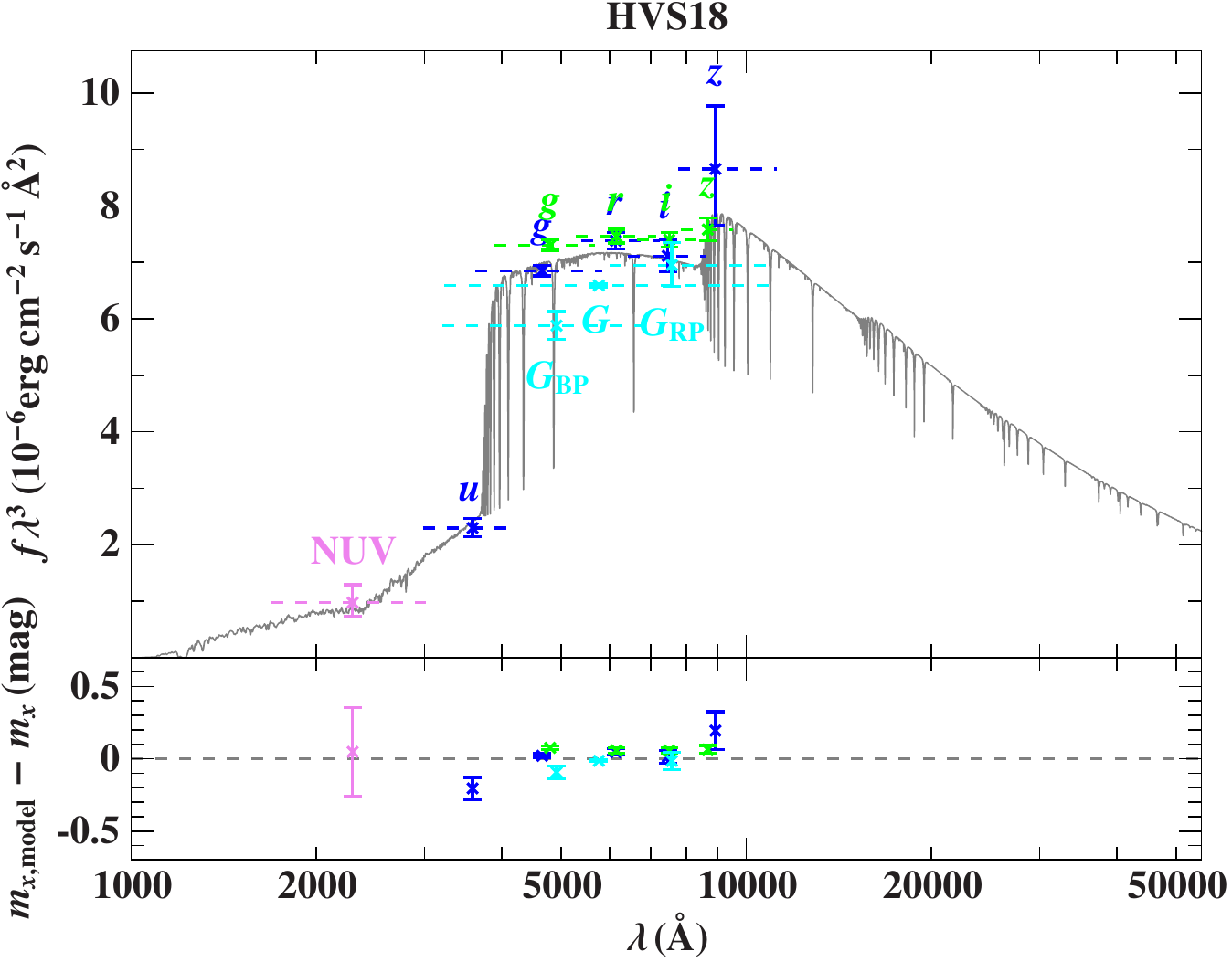}\caption{Same as Fig.~\ref{fig:SED-HVS1-text}. SDSS: blue; Pan-STARRS: green; {\it Gaia}: cyan; GALEX: violet.}\label{fig:SED-HVS18}
\end{figure}\begin{figure}
\centering
\includegraphics[width=\hsize]{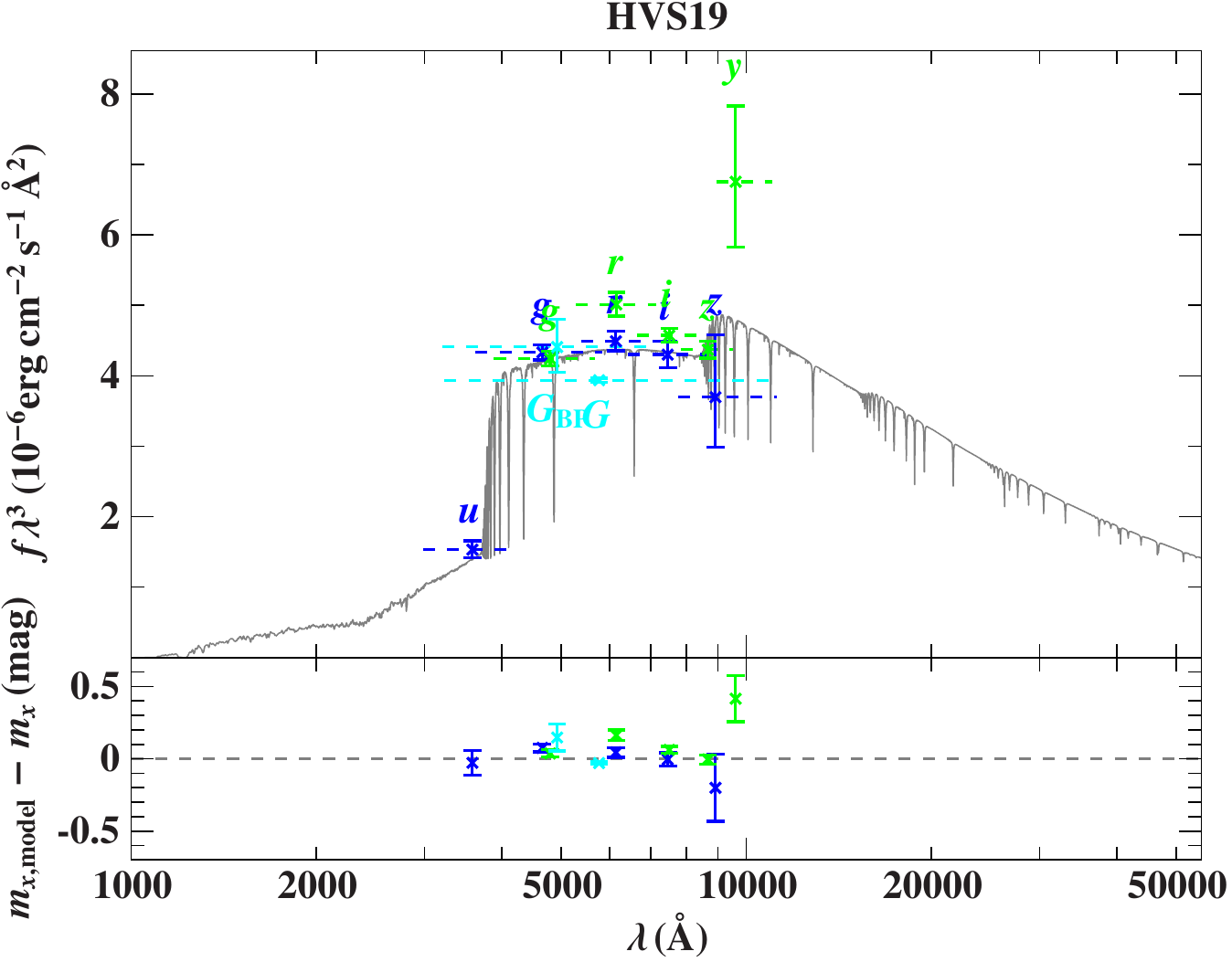}\caption{Same as Fig.~\ref{fig:SED-HVS1-text}. SDSS: blue; Pan-STARRS: green; {\it Gaia}: cyan.}\label{fig:SED-HVS19}
\end{figure}\begin{figure}
\centering
\includegraphics[width=\hsize]{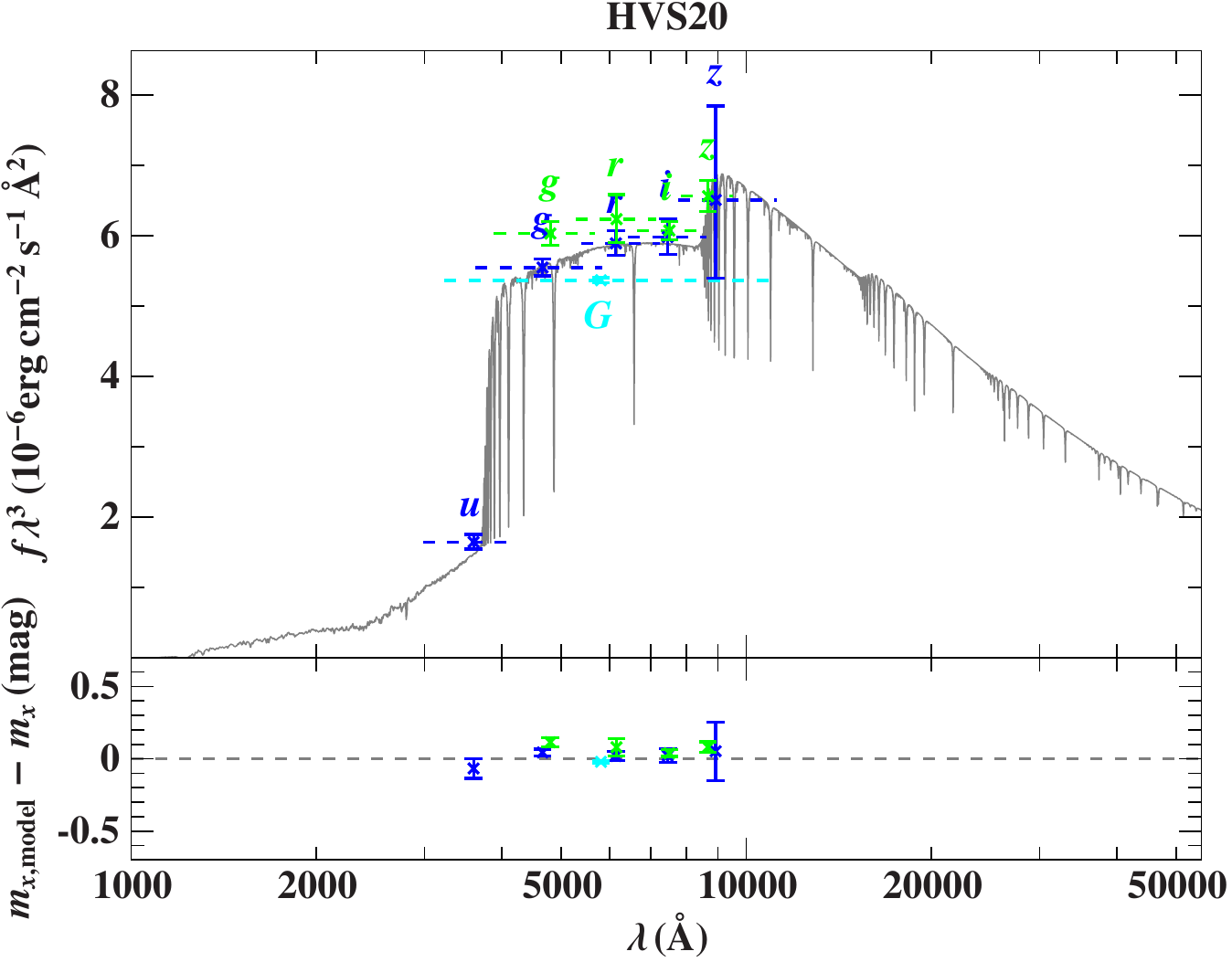}\caption{Same as Fig.~\ref{fig:SED-HVS1-text}. SDSS: blue; Pan-STARRS: green; {\it Gaia}: cyan.}\label{fig:SED-HVS20}
\end{figure}\begin{figure}
\centering
\includegraphics[width=\hsize]{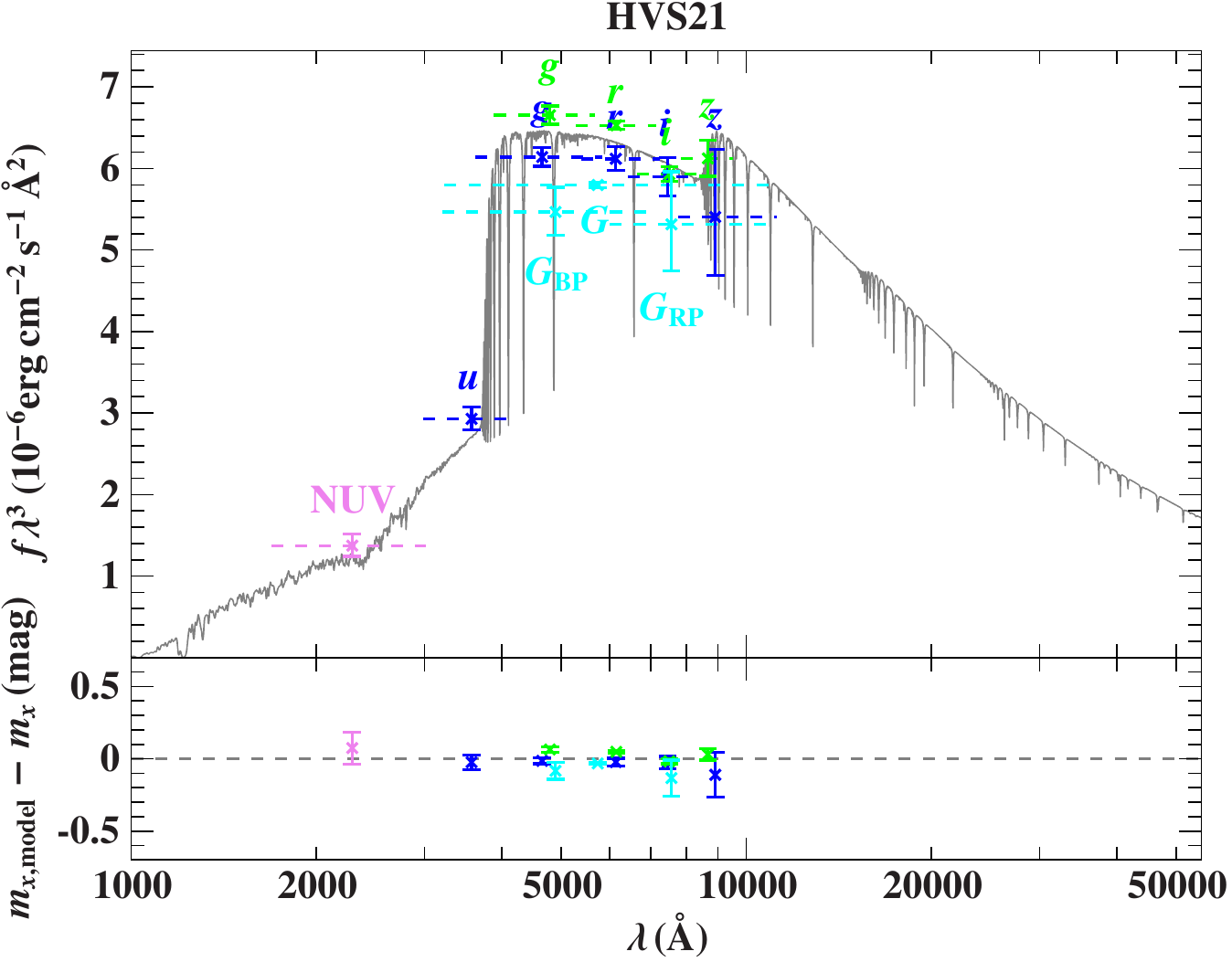}\caption{Same as Fig.~\ref{fig:SED-HVS1-text}. SDSS: blue; Pan-STARRS: green; {\it Gaia}: cyan; GALEX: violet.}\label{fig:SED-HVS21}
\end{figure}\begin{figure}
\centering
\includegraphics[width=\hsize]{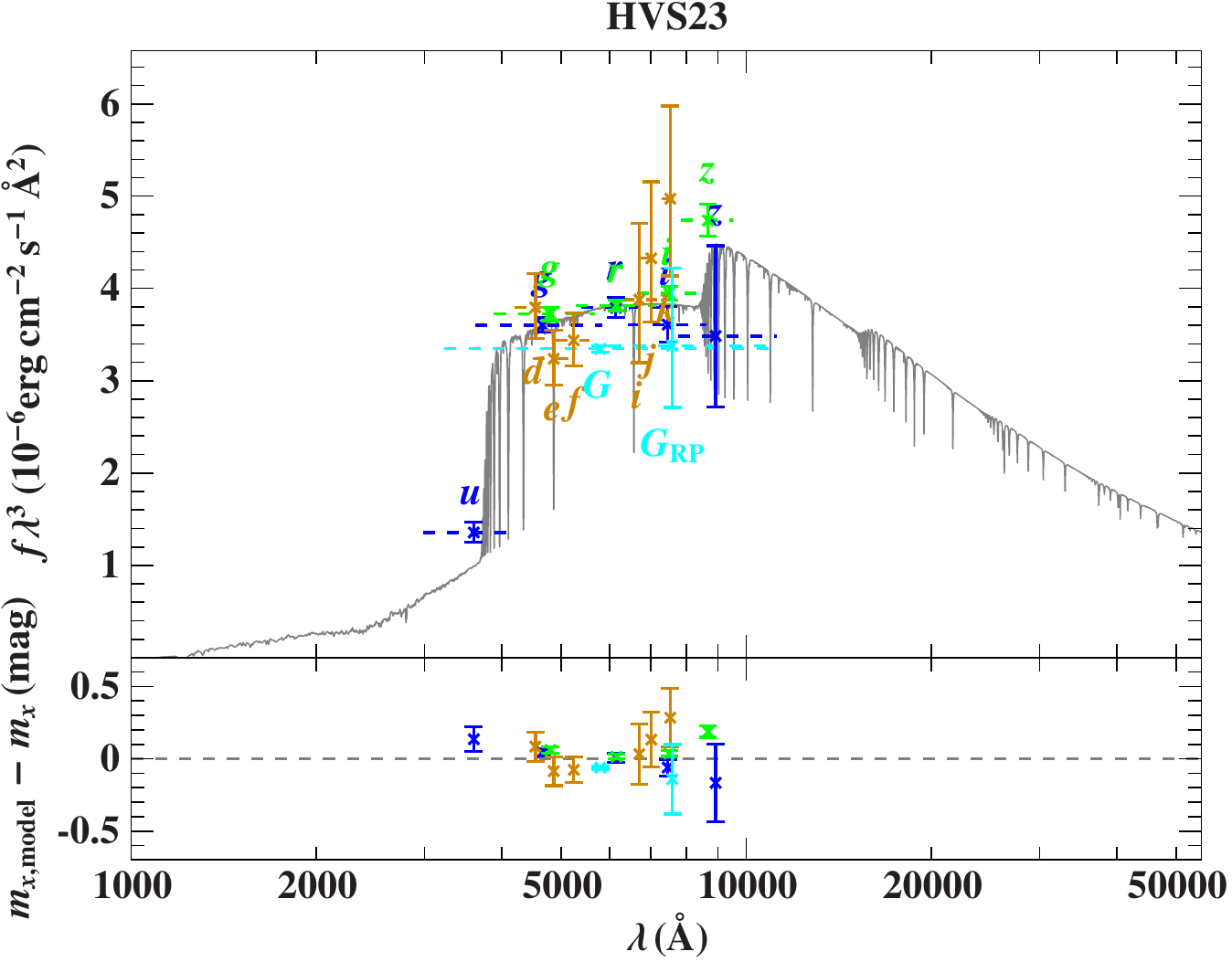}\caption{Same as Fig.~\ref{fig:SED-HVS1-text}. SDSS: blue; Pan-STARRS: green; {\it Gaia}: cyan; BATC: orange.}\label{fig:SED-HVS23}
\end{figure}\begin{figure}
\centering
\includegraphics[width=\hsize]{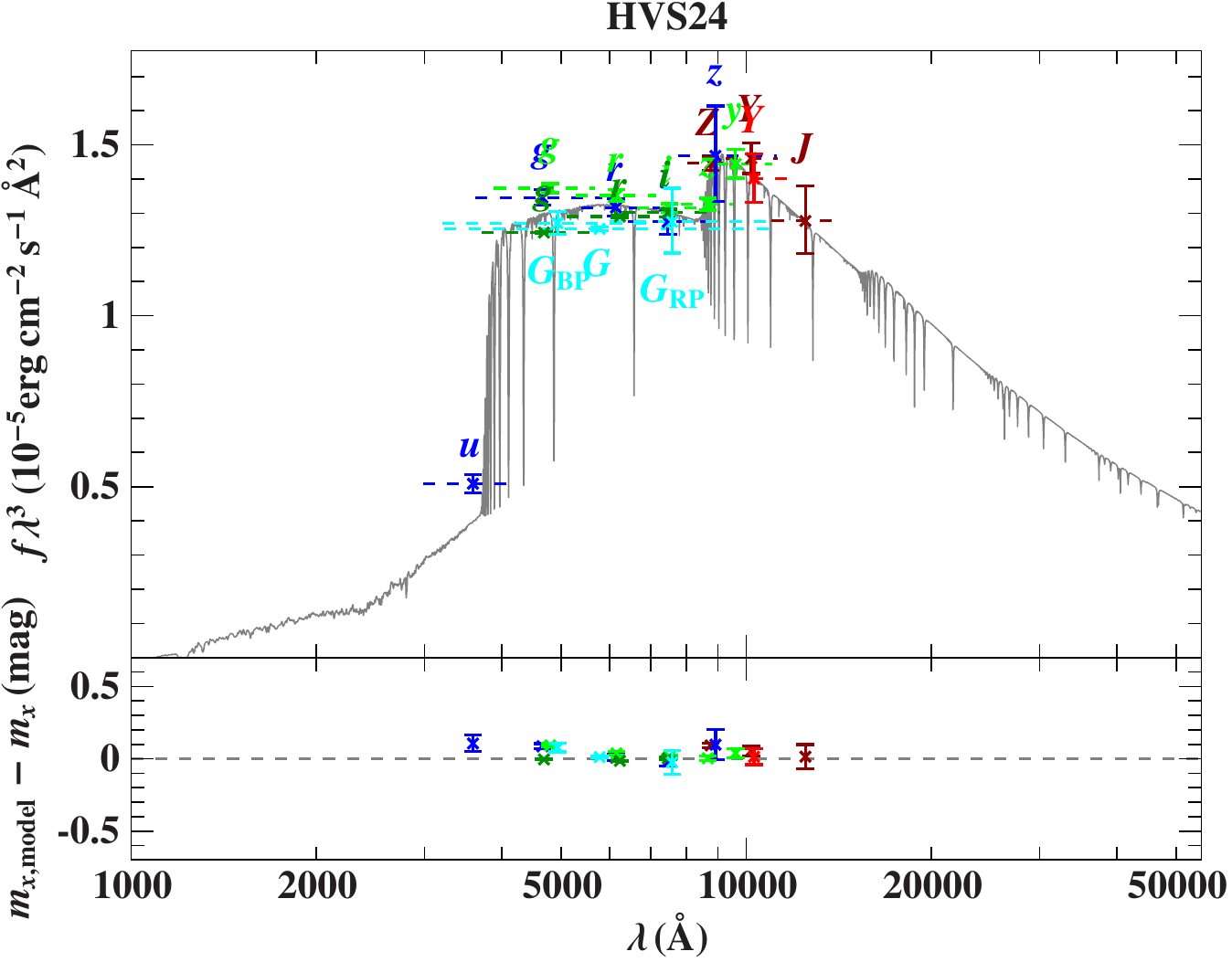}\caption{Same as Fig.~\ref{fig:SED-HVS1-text}. SDSS: blue; Pan-STARRS: green; UKIDSS: red; {\it Gaia}: cyan; VISTA: dark red; VST-KiDs: dark green.}\label{fig:SED-HVS24}
\end{figure}\begin{figure}
\centering
\includegraphics[width=\hsize]{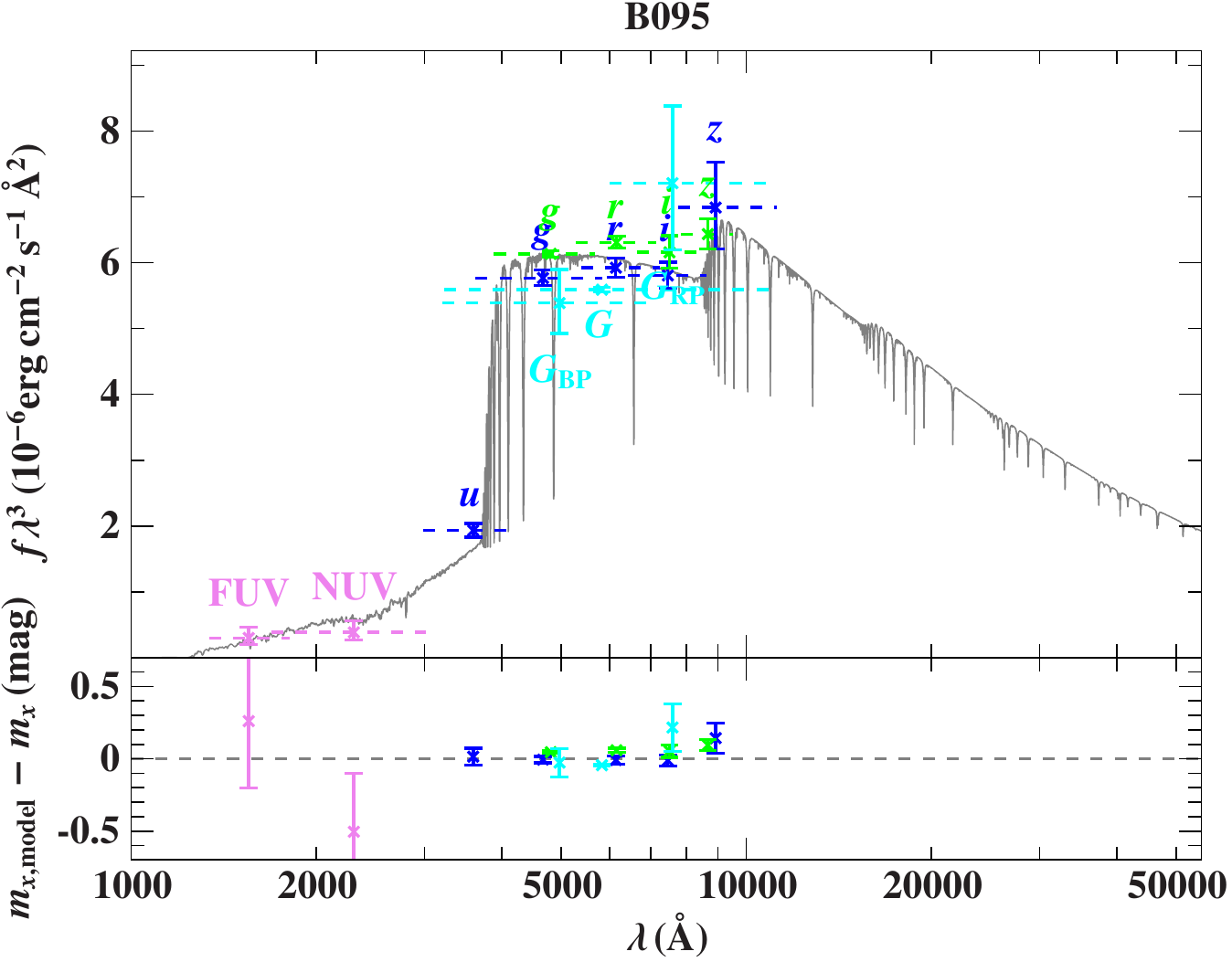}\caption{Same as Fig.~\ref{fig:SED-HVS1-text}. SDSS: blue; Pan-STARRS: green; {\it Gaia}: cyan; GALEX: violet.}\label{fig:SED-B095}
\end{figure}\begin{figure}
\centering
\includegraphics[width=\hsize]{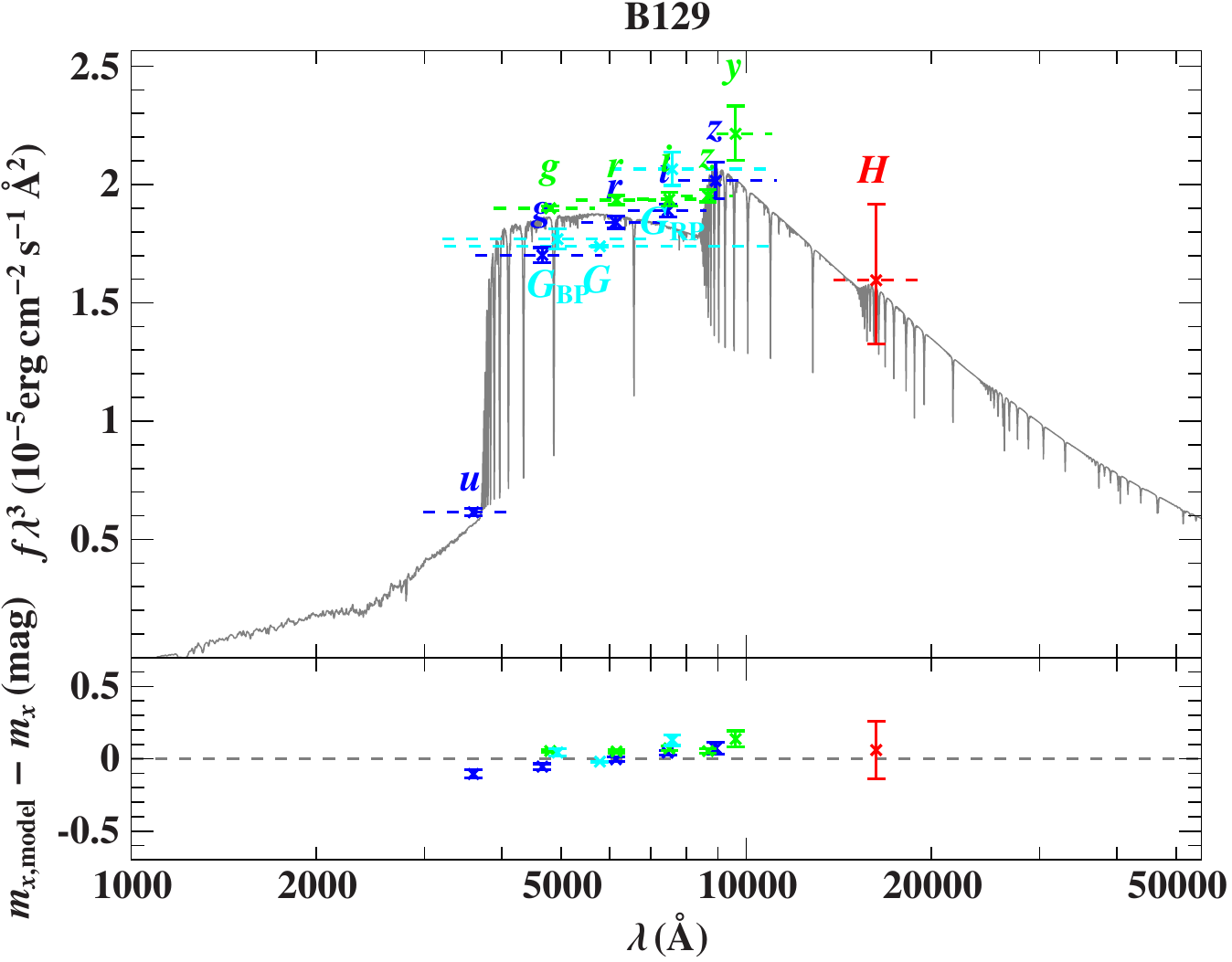}\caption{Same as Fig.~\ref{fig:SED-HVS1-text}. SDSS: blue; Pan-STARRS: green; UKIDSS: red; {\it Gaia}: cyan.}\label{fig:SED-B129}
\end{figure}\begin{figure}
\centering
\includegraphics[width=\hsize]{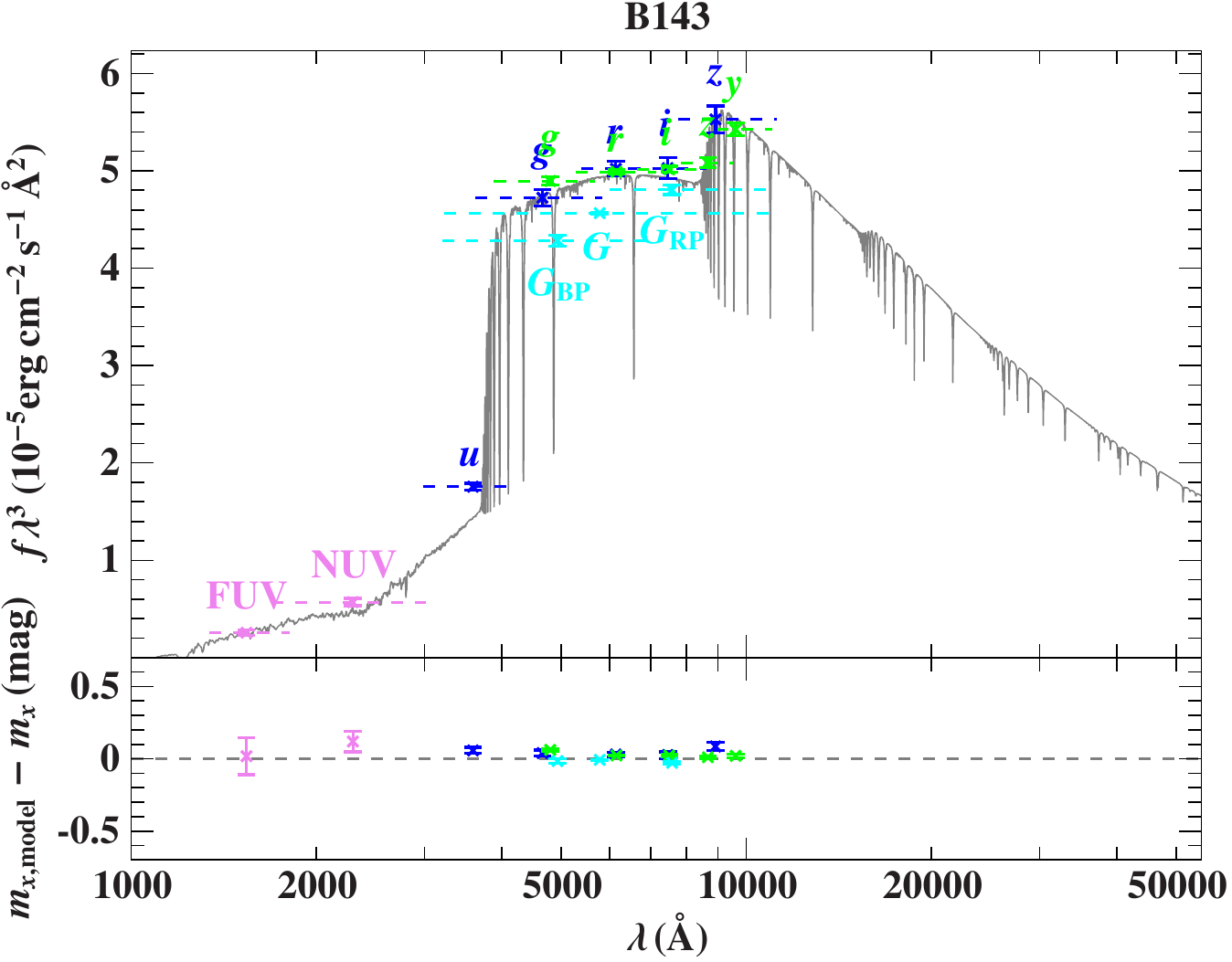}\caption{Same as Fig.~\ref{fig:SED-HVS1-text}. SDSS: blue; Pan-STARRS: green; {\it Gaia}: cyan; GALEX: violet.}\label{fig:SED-B143}
\end{figure}\begin{figure}
\centering
\includegraphics[width=\hsize]{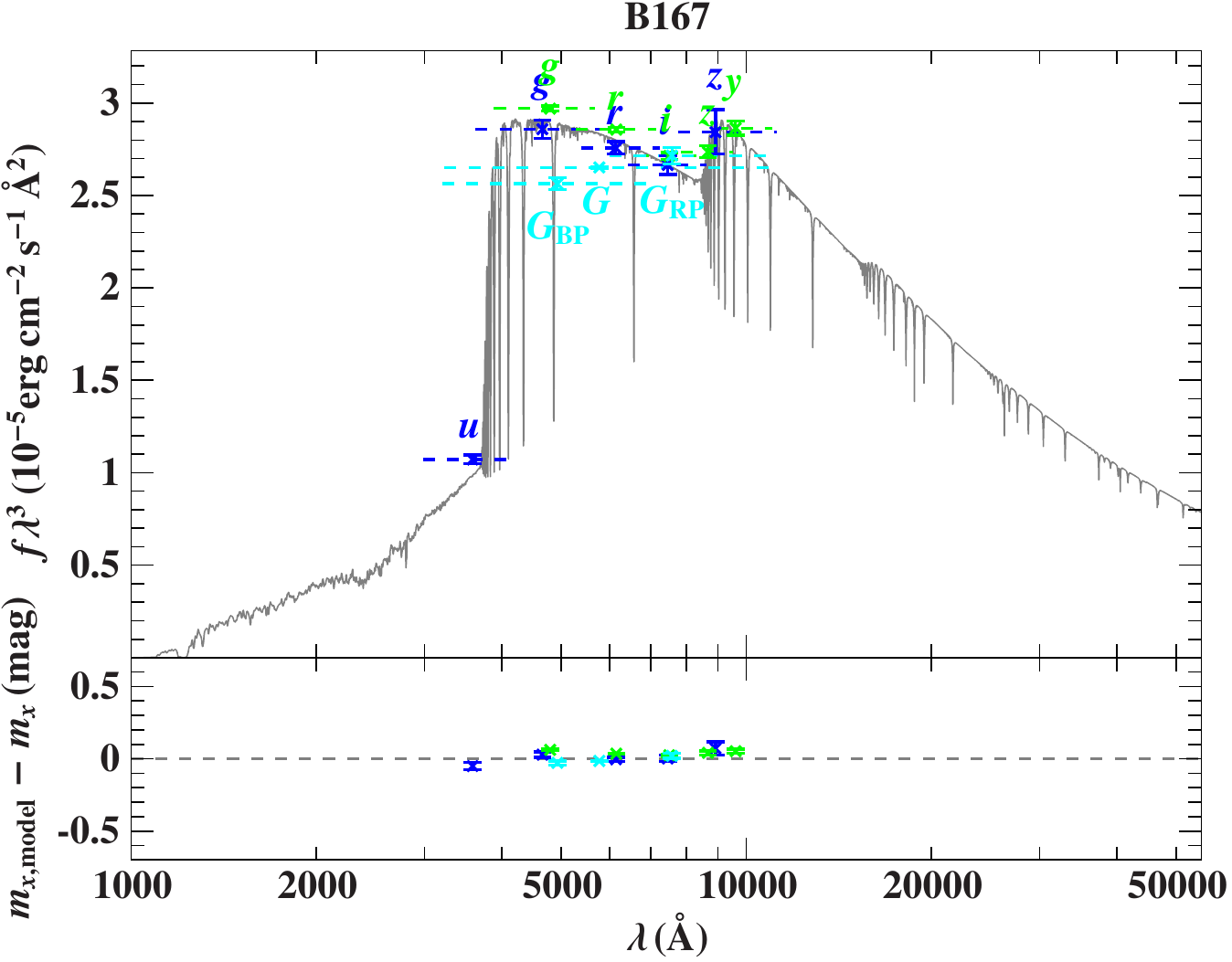}\caption{Same as Fig.~\ref{fig:SED-HVS1-text}. SDSS: blue; Pan-STARRS: green; {\it Gaia}: cyan.}\label{fig:SED-B167}
\end{figure}\begin{figure}
\centering
\includegraphics[width=\hsize]{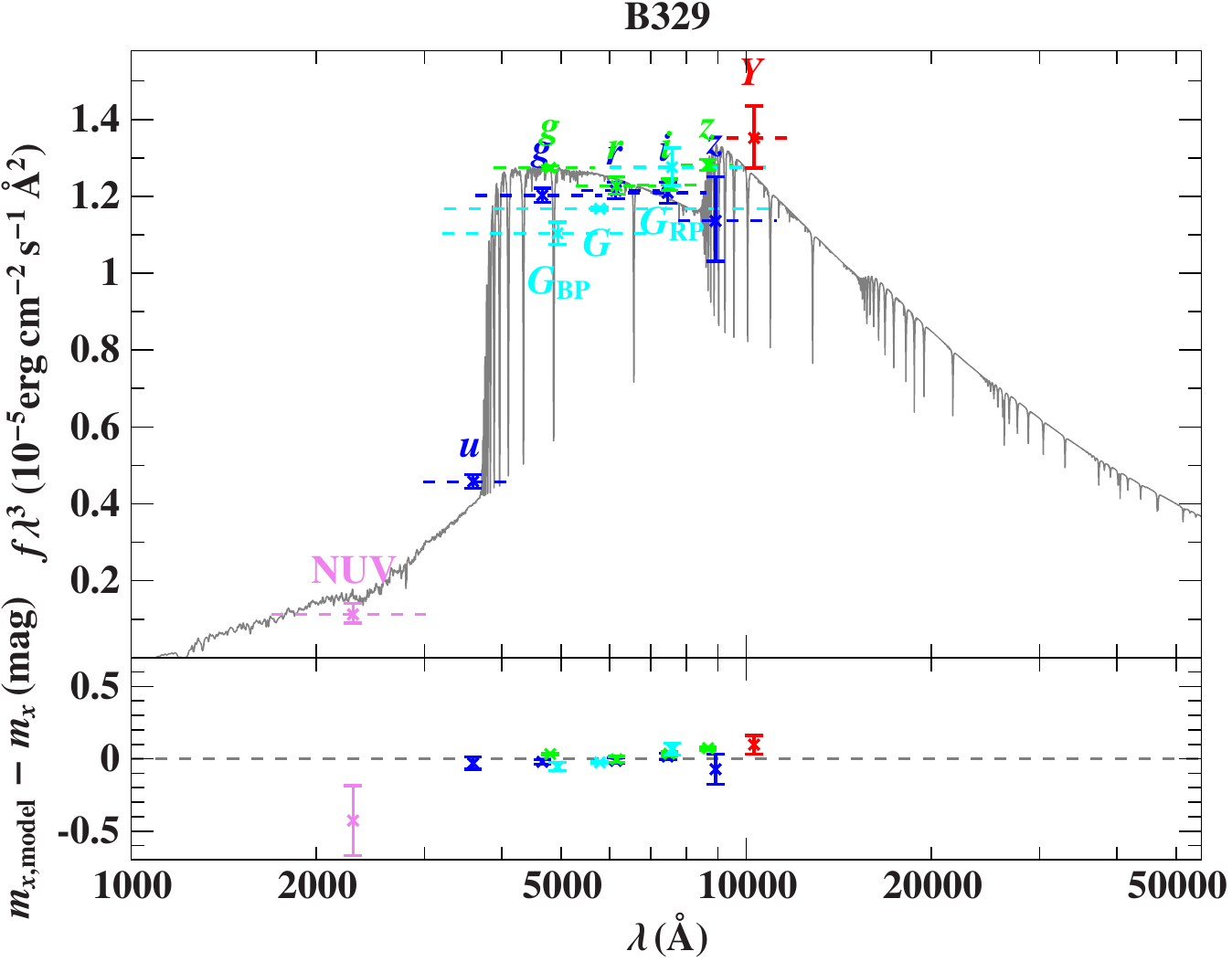}\caption{Same as Fig.~\ref{fig:SED-HVS1-text}. SDSS: blue; Pan-STARRS: green; UKIDSS: red; {\it Gaia}: cyan; GALEX: violet.}\label{fig:SED-B329}
\end{figure}\begin{figure}
\centering
\includegraphics[width=\hsize]{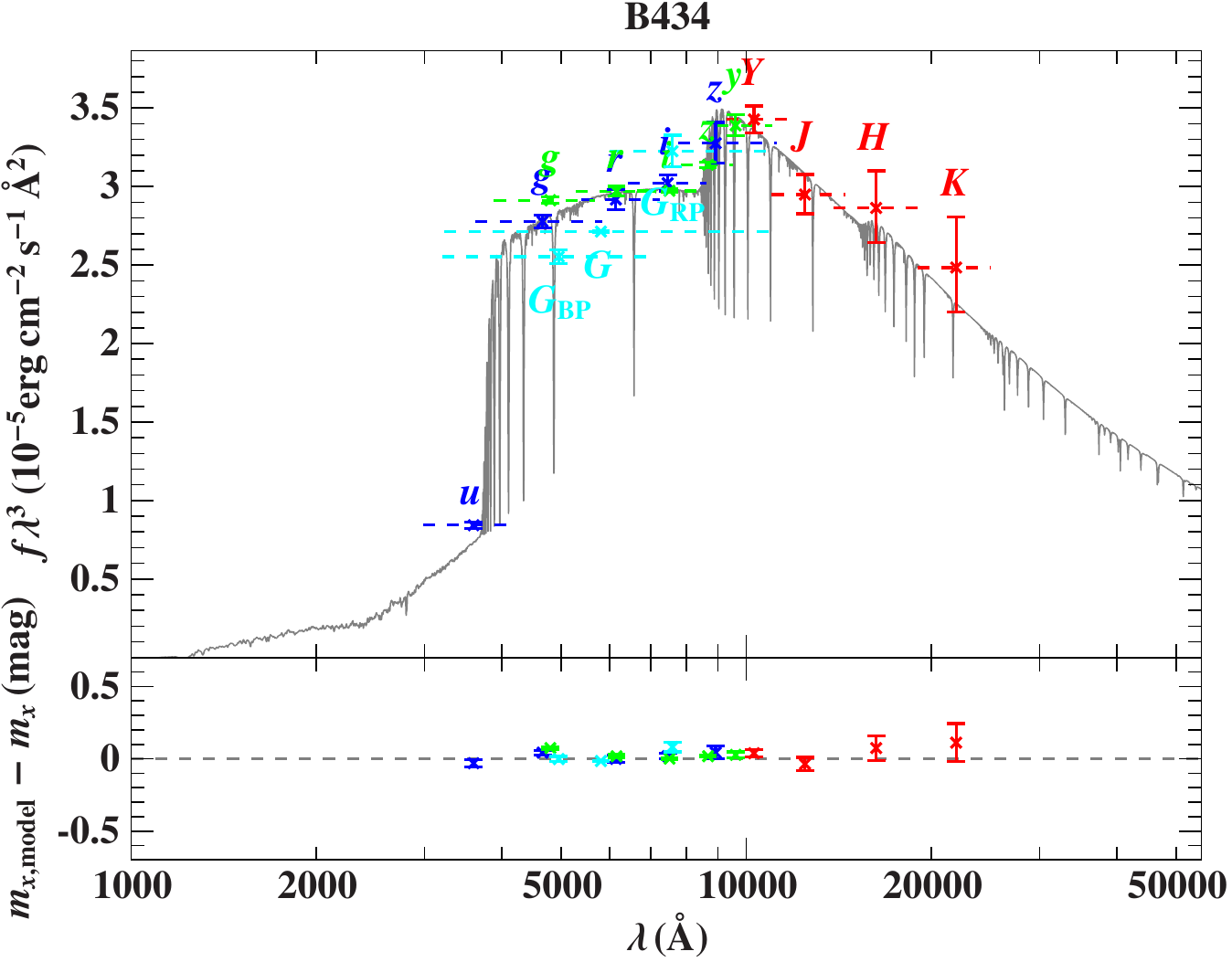}\caption{Same as Fig.~\ref{fig:SED-HVS1-text}. SDSS: blue; Pan-STARRS: green; UKIDSS: red; {\it Gaia}: cyan.}\label{fig:SED-B434}
\end{figure}\begin{figure}
\centering
\includegraphics[width=\hsize]{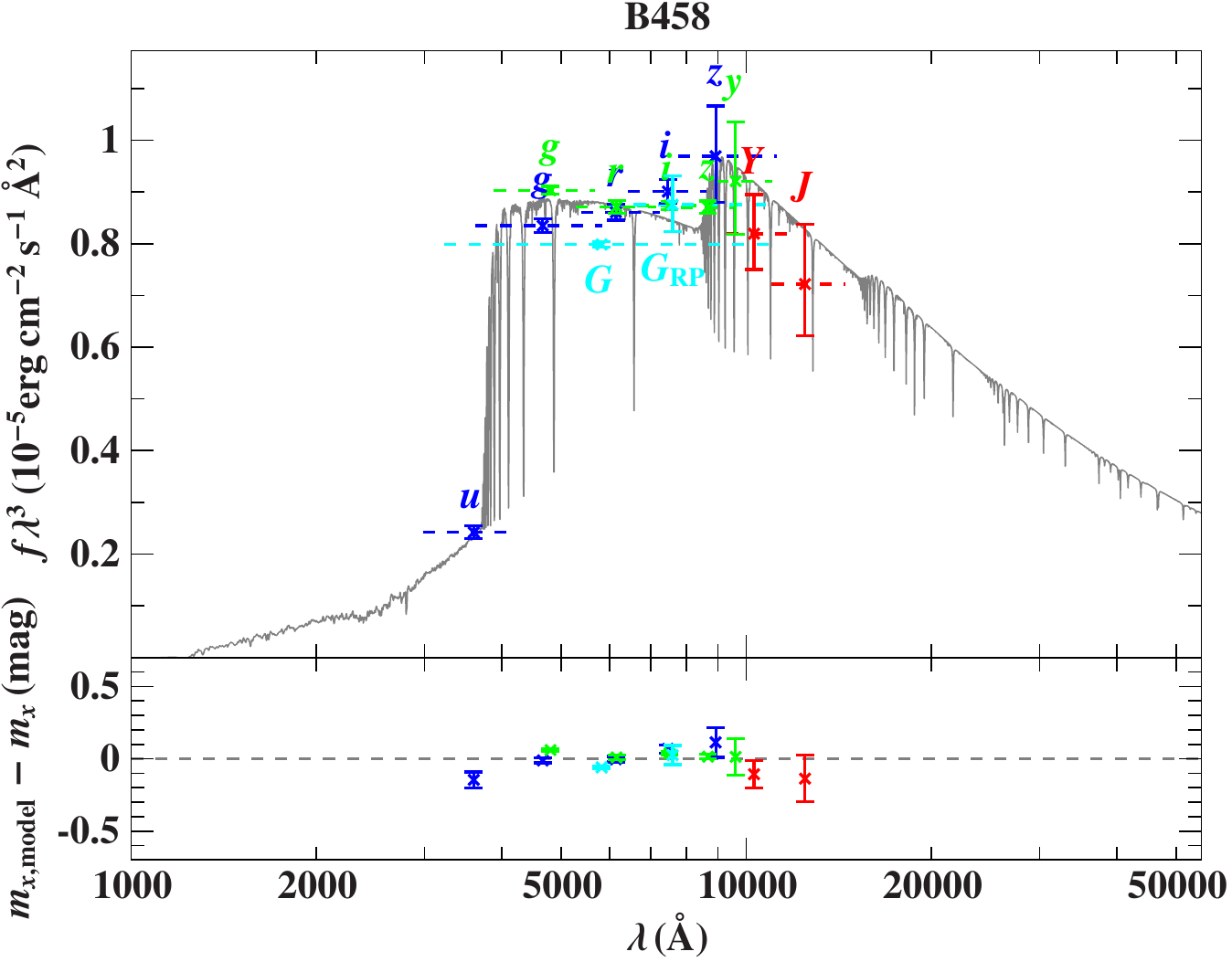}\caption{Same as Fig.~\ref{fig:SED-HVS1-text}. SDSS: blue; Pan-STARRS: green; UKIDSS: red; {\it Gaia}: cyan.}\label{fig:SED-B458}
\end{figure}\begin{figure}
\centering
\includegraphics[width=\hsize]{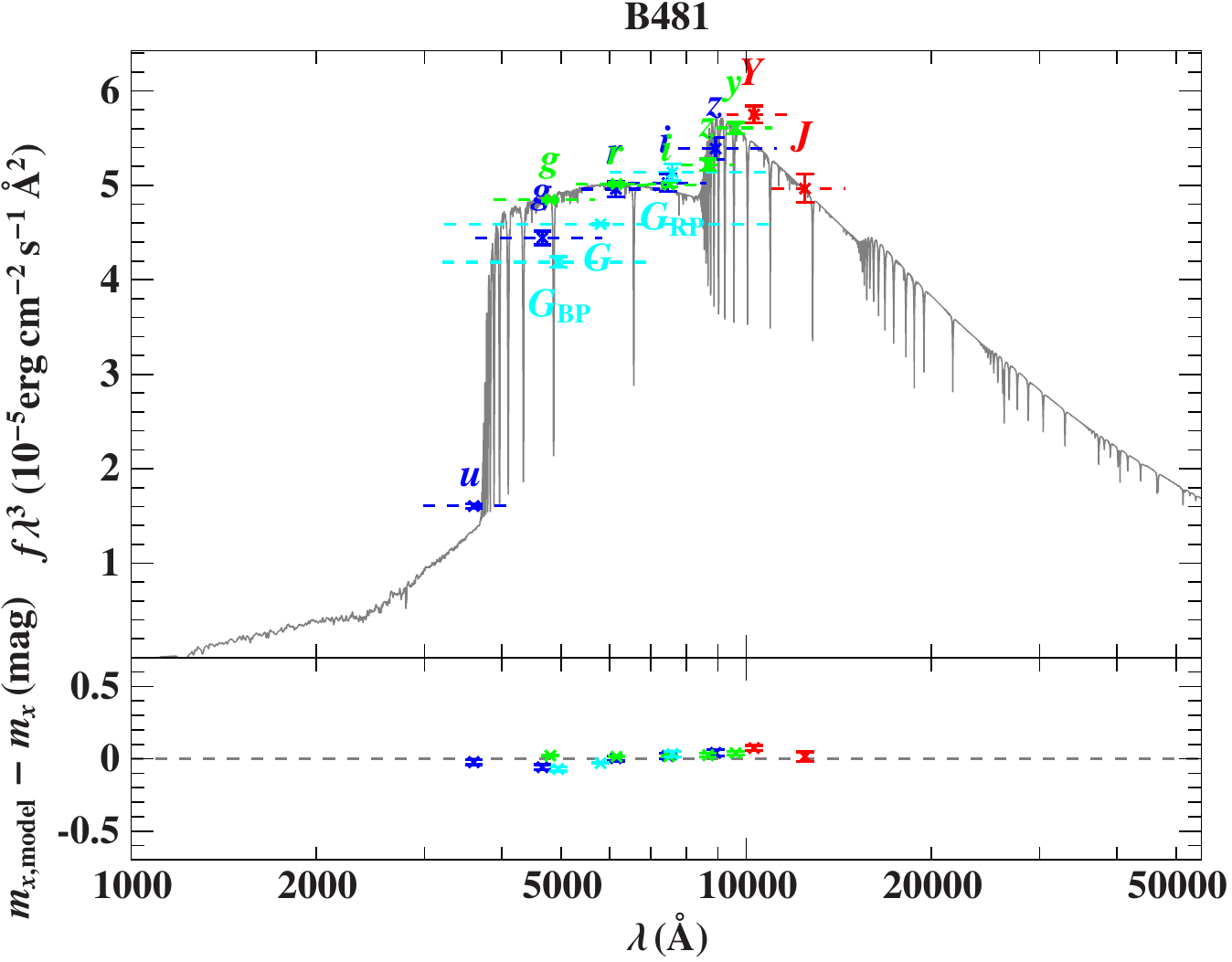}\caption{Same as Fig.~\ref{fig:SED-HVS1-text}. SDSS: blue; Pan-STARRS: green; UKIDSS: red; {\it Gaia}: cyan.}\label{fig:SED-B481}
\end{figure}\begin{figure}
\centering
\includegraphics[width=\hsize]{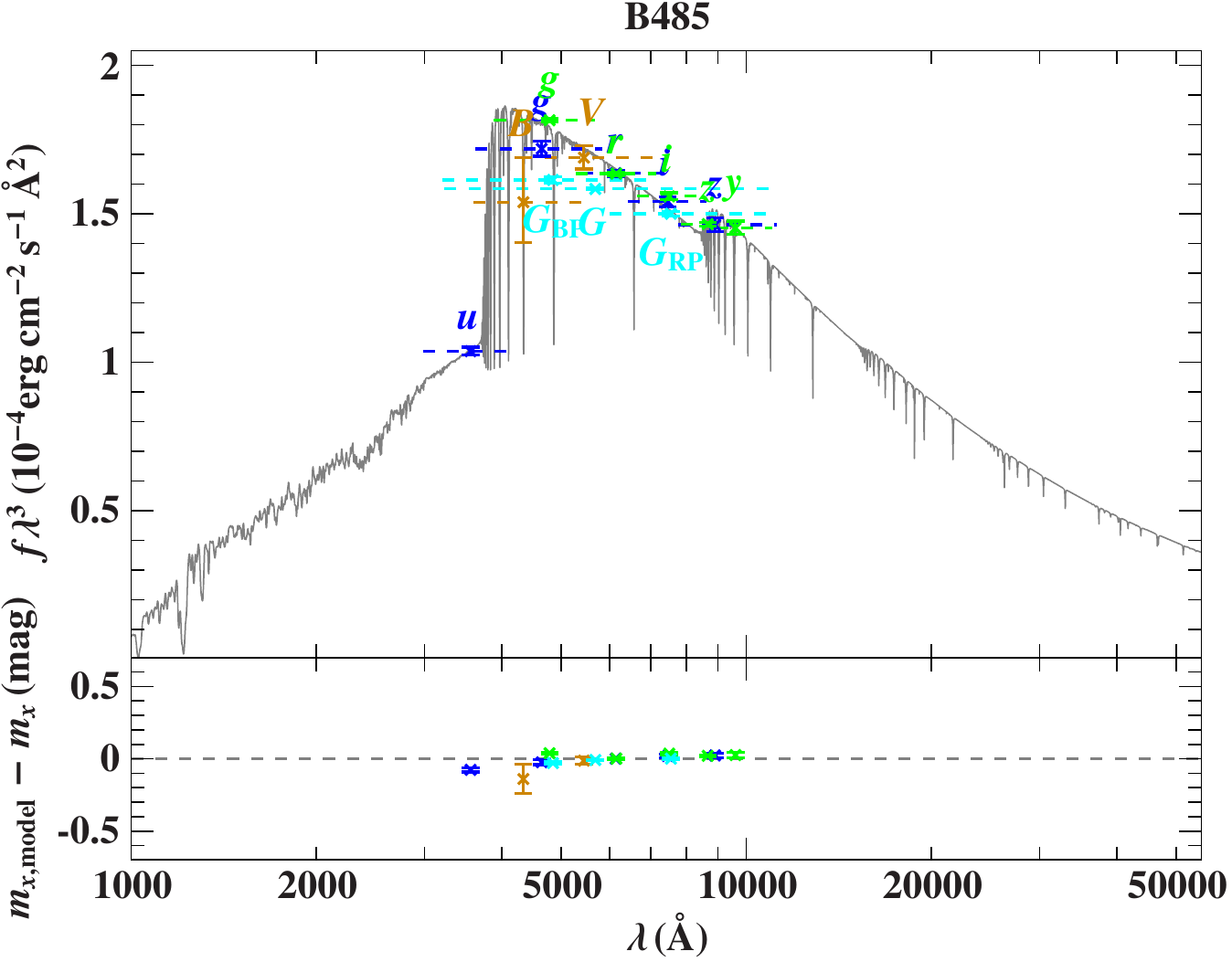}\caption{Same as Fig.~\ref{fig:SED-HVS1-text}. SDSS: blue; Pan-STARRS: green; {\it Gaia}: cyan; Johnson: orange.}\label{fig:SED-B485}
\end{figure}\begin{figure}
\centering
\includegraphics[width=\hsize]{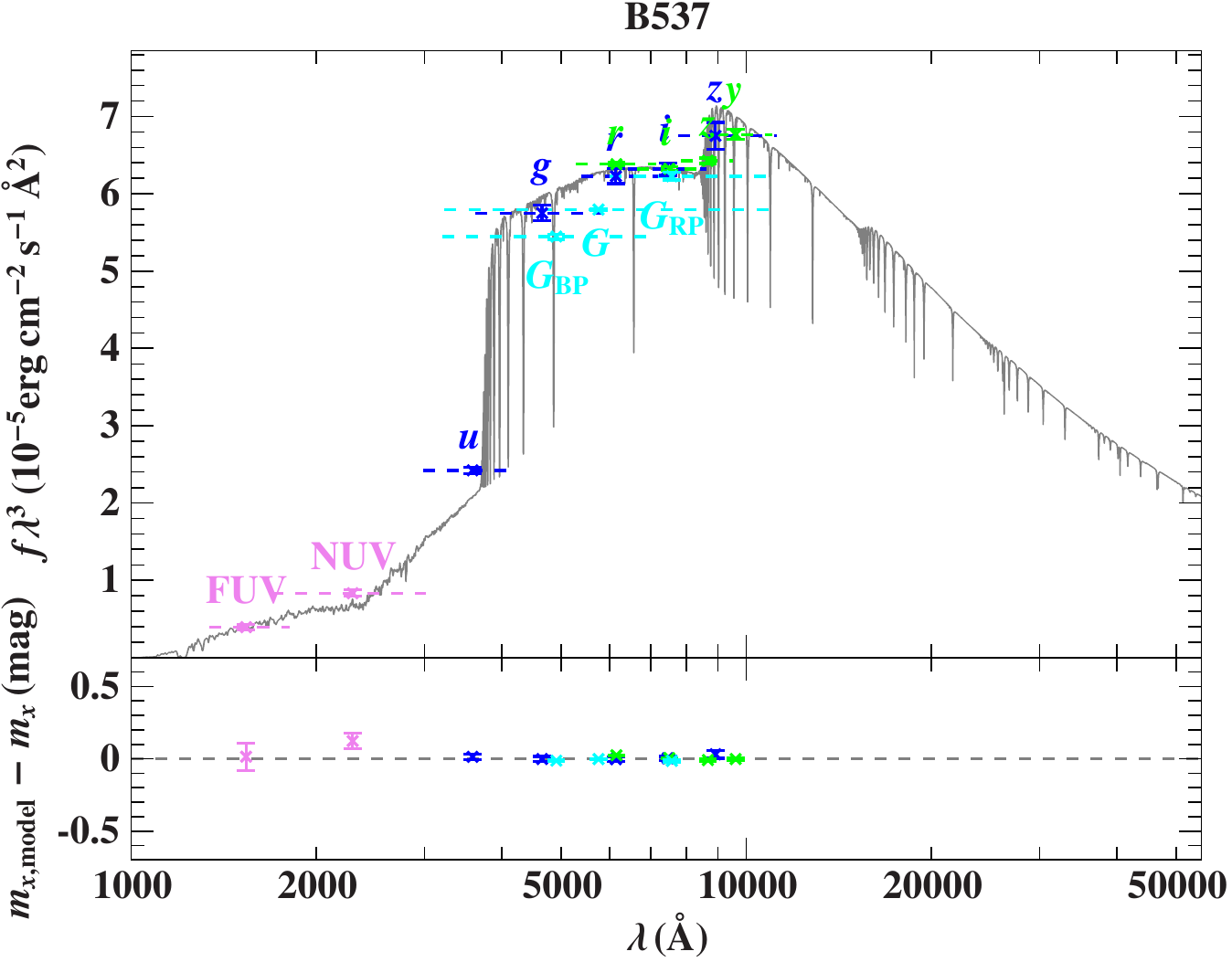}\caption{Same as Fig.~\ref{fig:SED-HVS1-text}. SDSS: blue; Pan-STARRS: green; {\it Gaia}: cyan; GALEX: violet.}\label{fig:SED-B537}
\end{figure}\begin{figure}
\centering
\includegraphics[width=\hsize]{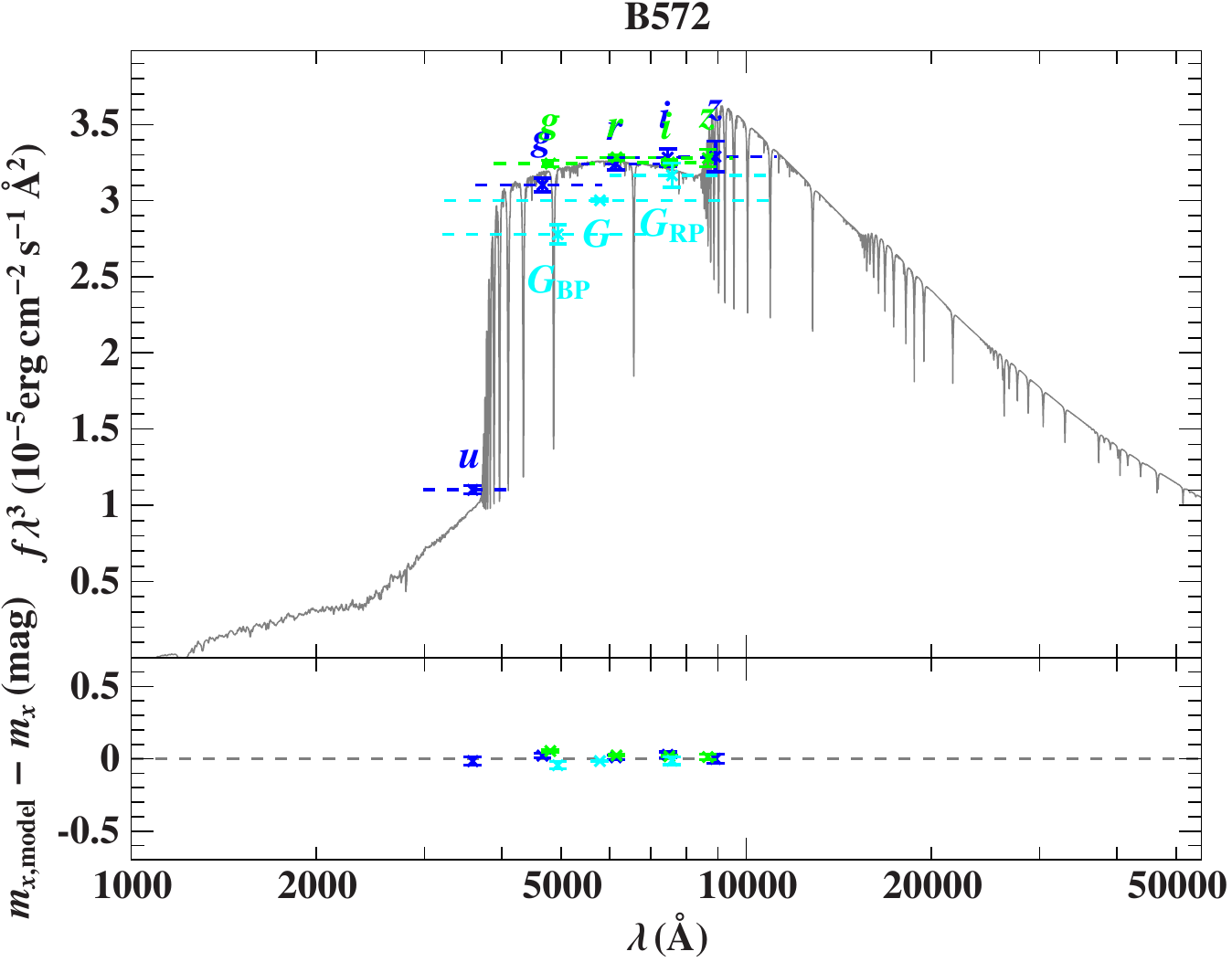}\caption{Same as Fig.~\ref{fig:SED-HVS1-text}. SDSS: blue; Pan-STARRS: green; {\it Gaia}: cyan.}\label{fig:SED-B572}
\end{figure}\begin{figure}
\centering
\includegraphics[width=\hsize]{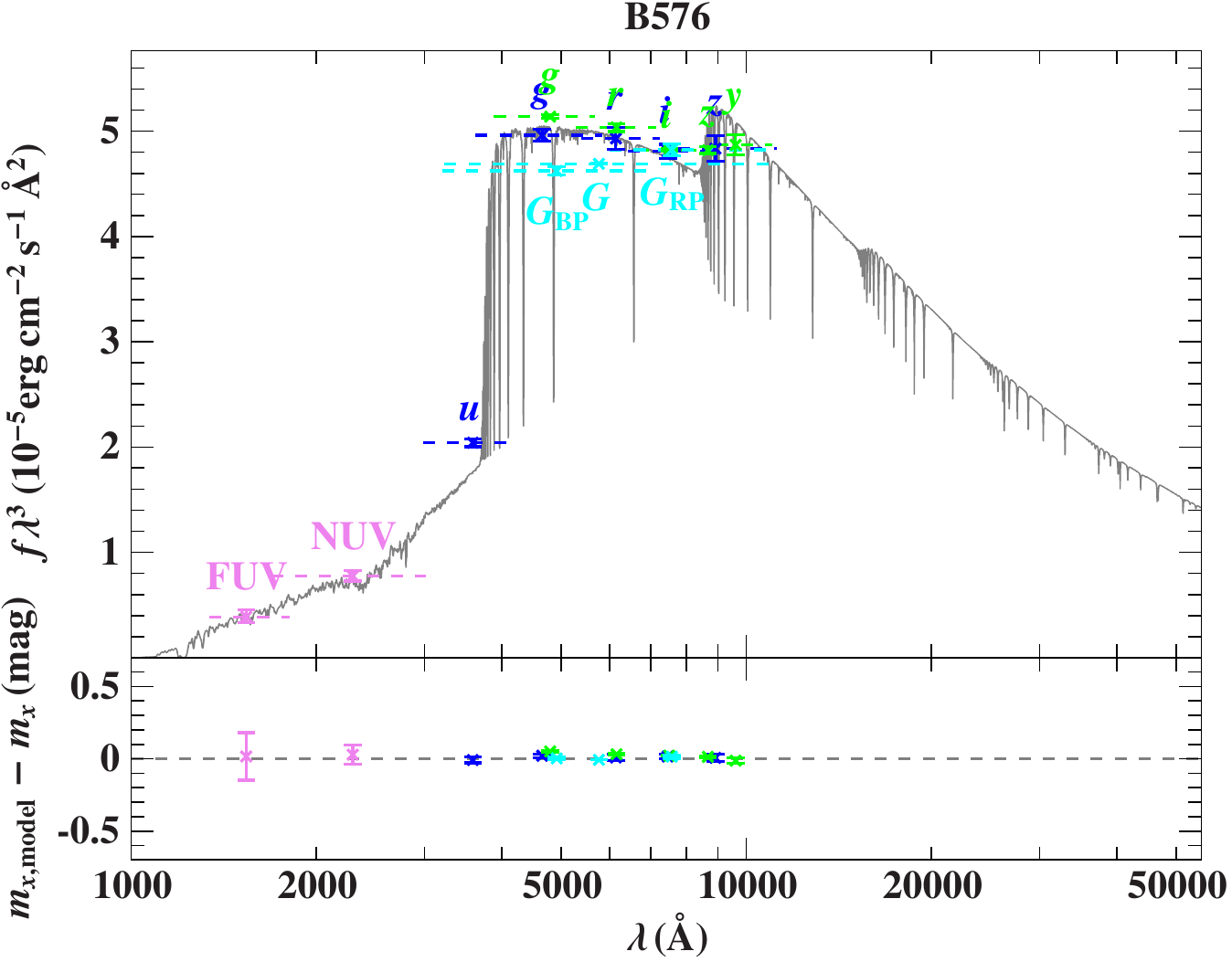}\caption{Same as Fig.~\ref{fig:SED-HVS1-text}. SDSS: blue; Pan-STARRS: green; {\it Gaia}: cyan; GALEX: violet.}\label{fig:SED-B576}
\end{figure}\begin{figure}
\centering
\includegraphics[width=\hsize]{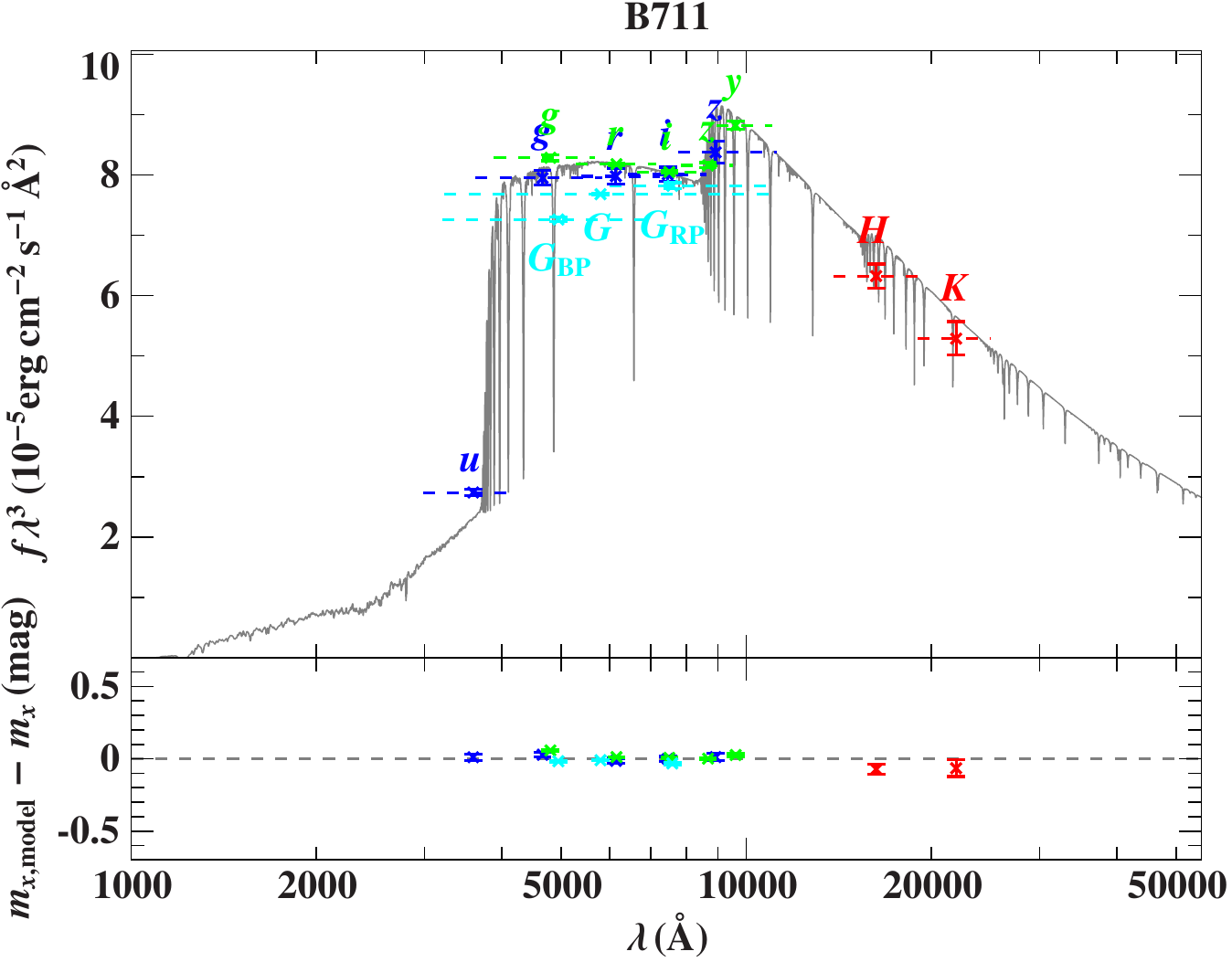}\caption{Same as Fig.~\ref{fig:SED-HVS1-text}. SDSS: blue; Pan-STARRS: green; UKIDSS: red; {\it Gaia}: cyan.}\label{fig:SED-B711}
\end{figure}\begin{figure}
\centering
\includegraphics[width=\hsize]{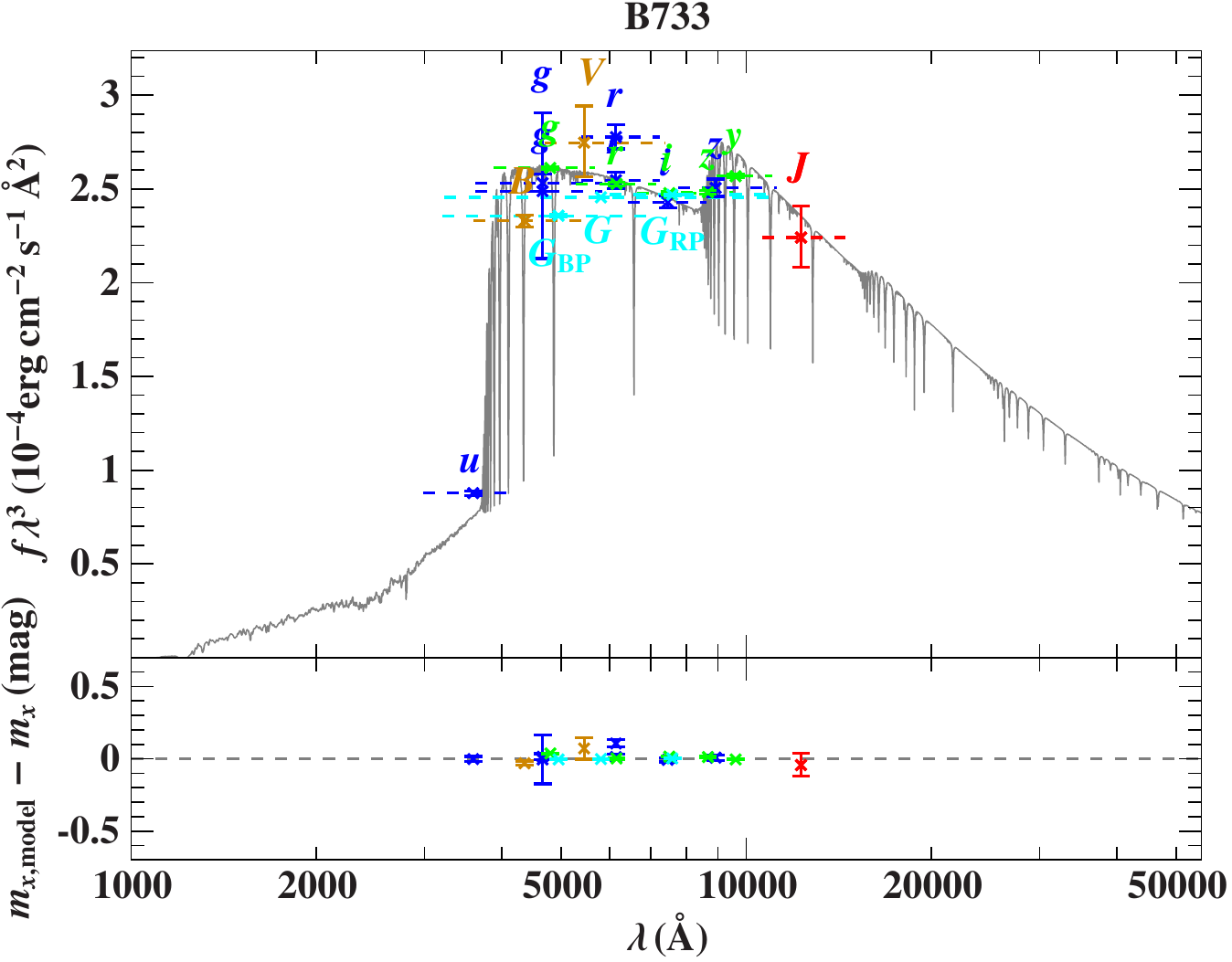}\caption{Same as Fig.~\ref{fig:SED-HVS1-text}. SDSS: blue; Pan-STARRS: green; 2MASS: red; {\it Gaia}: cyan; Johnson: orange.}\label{fig:SED-B733}
\end{figure}\begin{figure}
\centering
\includegraphics[width=\hsize]{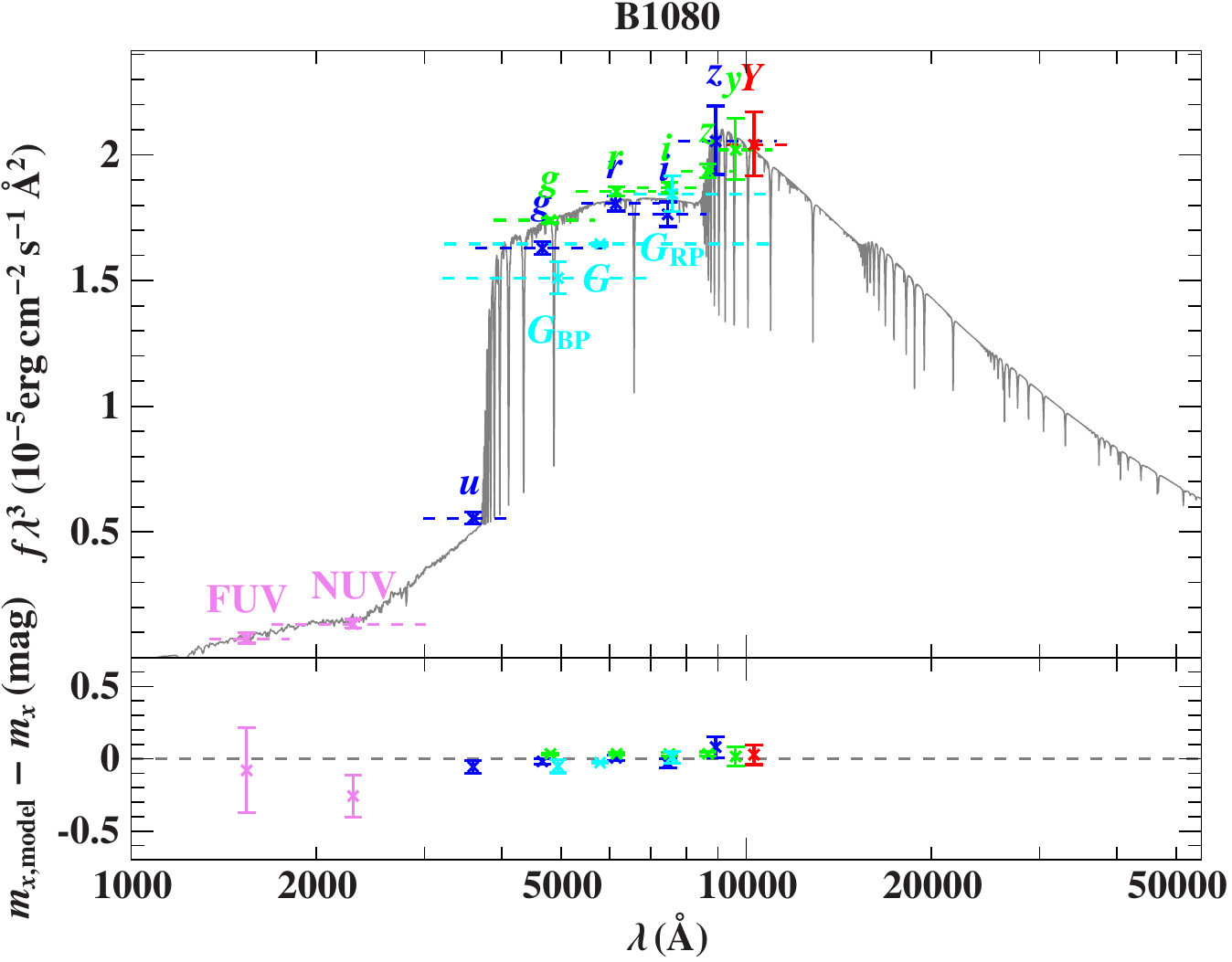}\caption{Same as Fig.~\ref{fig:SED-HVS1-text}. SDSS: blue; Pan-STARRS: green; UKIDSS: red; {\it Gaia}: cyan; GALEX: violet.}\label{fig:SED-B1080}
\end{figure}\begin{figure}
\centering
\includegraphics[width=\hsize]{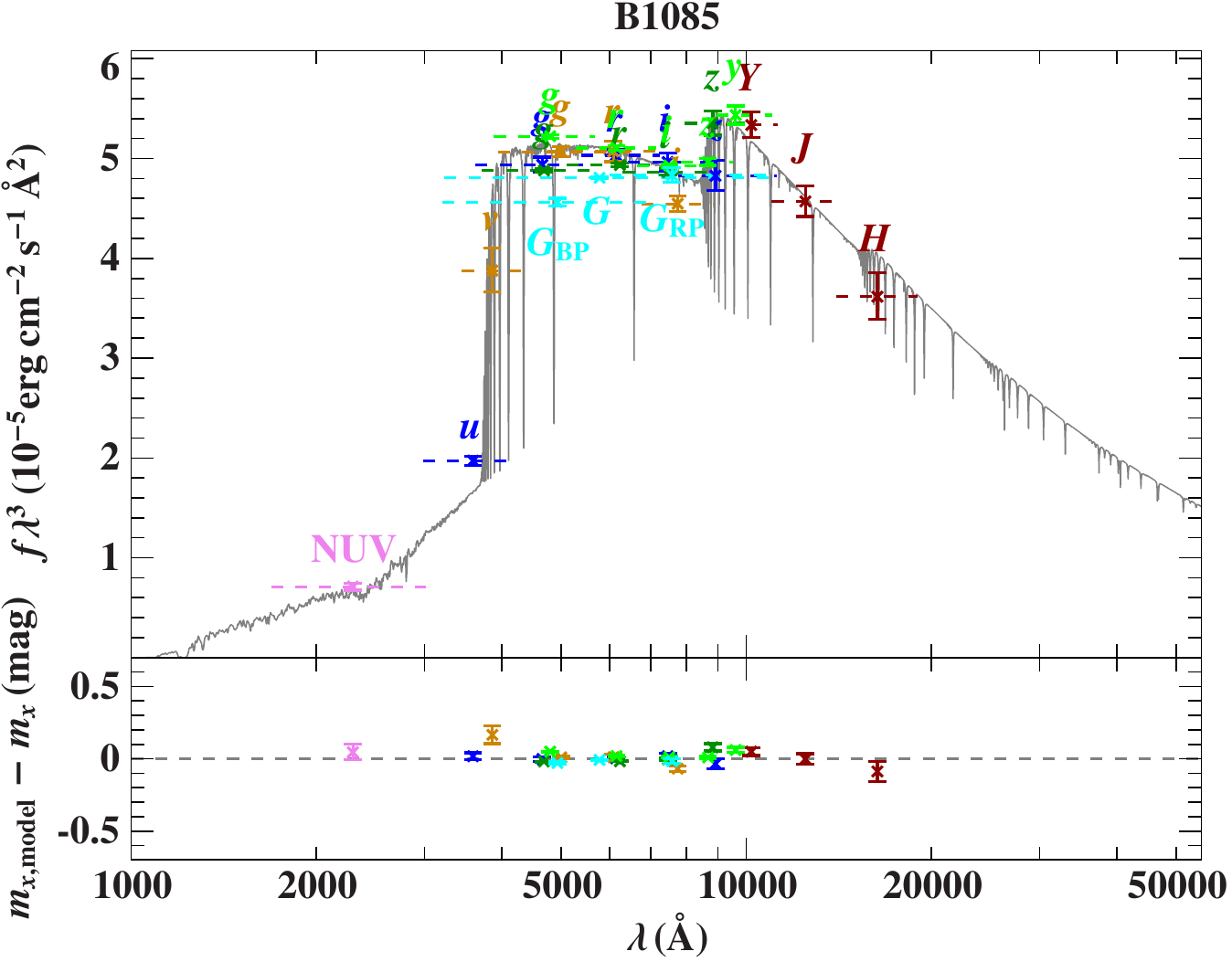}\caption{Same as Fig.~\ref{fig:SED-HVS1-text}. SDSS: blue; Pan-STARRS: green; {\it Gaia}: cyan; GALEX: violet; VISTA: dark red; SkyMapper: orange; VST-KiDs: dark green.}\label{fig:SED-B1085}
\end{figure}\begin{figure}
\centering
\includegraphics[width=\hsize]{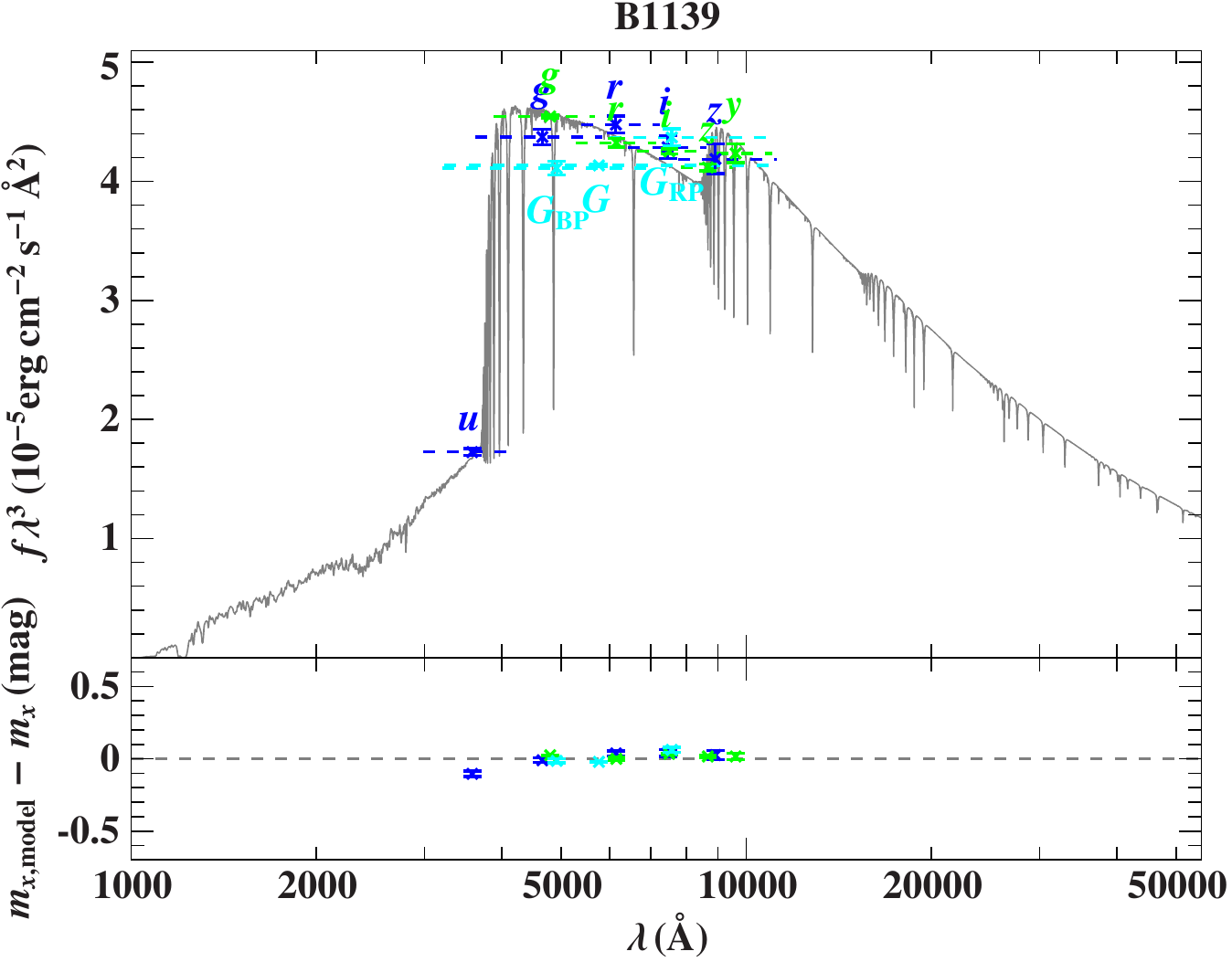}\caption{Same as Fig.~\ref{fig:SED-HVS1-text}. SDSS: blue; Pan-STARRS: green; {\it Gaia}: cyan.}\label{fig:SED-B1139}
\end{figure}

\end{appendix}


\end{document}